\RequirePackage{fix-cm}

\documentclass[numbook]{svjour3} 
\journalname{Empirical Software Engineering}
\counterwithin*{section}{part}
\usepackage[a4paper,bindingoffset=0mm,left=30mm,right=30mm,top=20mm,bottom=20mm]{geometry}

\usepackage{array}
\newcolumntype{L}{>{\arraybackslash}m{12cm}}

\newcolumntype{P}{>{\arraybackslash}m{6cm}}
\newcolumntype{C}{>{\centering\arraybackslash}m{12cm}}

\def\BibTeX{{\rm B\kern-.05em{\sc i\kern-.025em b}\kern-.08em
    T\kern-.1667em\lower.7ex\hbox{E}\kern-.125emX}}

\makeatletter
\renewcommand\paragraph{\@startsection{paragraph}{4}{\z@}%
  {0.5ex \@plus 1ex \@minus .2ex}%
  {-0.5em}%
  {\normalfont\normalsize\itshape}}
\makeatother

\usepackage{amsmath,amssymb,amsfonts} 
\usepackage{times}
\usepackage{url}
\usepackage{cleveref}
\usepackage{relsize}
\usepackage{lscape}
\usepackage{multirow}
\usepackage{ragged2e}
\usepackage{adjustbox, tabularx,booktabs}
\usepackage{dcolumn}
\usepackage{float}
\floatstyle{plaintop}
\usepackage{adjustbox}
\restylefloat{table}
\usepackage{geometry}
\usepackage{graphicx}
\usepackage{rotating}

\usepackage{algorithm}
\usepackage{algpseudocode}
\usepackage{tcolorbox}

\usepackage[utf8]{inputenc}
\usepackage{listings}

\usepackage{booktabs} 
\usepackage{array} 
\usepackage{paralist} 
\usepackage{verbatim} 
\usepackage{enumitem}
\usepackage{ifthen}
\usepackage{xspace}
\usepackage{fancybox} 
\usepackage{soul}
\usepackage{color, colortbl}
\usepackage{balance}

\usepackage{graphicx}
\usepackage{titlesec}
\usepackage{longtable}
\usepackage{tabularx}
\usepackage[comma,authoryear]{natbib}
\usepackage{amsmath,amssymb,amsfonts}
\usepackage{cleveref}
\usepackage{longtable}
\usepackage{cite}
\usepackage{fancybox} 
\usepackage{amsmath,amssymb,amsfonts}
\usepackage{geometry}
\usepackage{graphicx}
\usepackage{titlesec}
\usepackage{longtable}
\usepackage{algorithm}
\usepackage{algpseudocode}
\usepackage{tabularx}
\usepackage{blindtext}
\usepackage{textcomp}
\usepackage{booktabs}
\usepackage{chngcntr}
\usepackage{framed}
\usepackage{mdframed}
\usepackage{amsmath}
\counterwithout{table}{section}
\counterwithout{figure}{section}
\counterwithout{equation}{section}

\definecolor{custom-gray}{cmyk}{0,0,0,0.7,1.0}

\newtcolorbox{Summary}[2][]{
  colback=gray!5,
  colframe=gray!40!black,
  fonttitle=\bfseries,
  coltitle=white,
  title=#2,
  boxrule=0.5mm,
  arc=2mm,
  left=2mm,
  right=2mm,
  top=1mm,
  bottom=1mm,
  #1
}

\usepackage{url}
\usepackage{relsize}
\usepackage{lscape}
\usepackage{multirow}
\usepackage{ragged2e}
\usepackage{adjustbox, tabularx,booktabs}
\usepackage{dcolumn}
\usepackage{float}
\floatstyle{plaintop}
\usepackage{adjustbox}
\restylefloat{table}
\usepackage{geometry}
\usepackage{graphicx}
\usepackage[multiple]{footmisc}

\usepackage{booktabs} 
\usepackage{array} 
\usepackage{paralist} 
\usepackage{verbatim} 

\usepackage{ifthen}
\usepackage{xspace}

\usepackage{fancybox} 
\usepackage{soul}
\usepackage{color, colortbl}
\algnewcommand\algorithmicforeach{\textbf{for each}}
\algdef{S}[FOR]{ForEach}[1]{\algorithmicforeach\ #1\ \algorithmicdo}

\newcommand{\head}[1]{\par\smallskip\noindent\textit{\underline{#1.}}}

\def\RQa{ Which configuration keys and HF repository tags most reliably correlate with reuse relationships between parent and child models on Hugging Face?}
\def\RQb{How effective is semantic fingerprinting based on configuration keys and repository tags at detecting missing model metadata?}
\def\RQc{How does the imputation of missing model metadata reshape our understanding of lineage complexity and license evolution in the PTLM ecosystem?}

\begin{document}


\title{Towards Imputation of Pre-Trained Language Model Metadata using Semantic Fingerprinting}

\author{Adekunle Ajibode  \and
        Oussama Ben Sghaier 
        \and Keheliya Gallaba
        \and Bram Adams
        \and Ahmed E. Hassan}


\institute{Adekunle Ajibode \at
              School of Computing, Queen’s University, Kingston, ON, Canada\\
              \email{ajibode.a@queensu.ca}
           \and
           Oussama Ben Sghaier \at
              School of Computing, Queen’s University, Kingston, ON, Canada\\
              \email{oussama.sghaier@queensu.ca}
              \and
            Keheliya Gallaba \at
              School of Computing, Queen’s University, Kingston, ON, Canada\\
              \email{gallabak@sigsoft.org}
              \and
           Bram Adams \at
               School of Computing, Queen’s University, Kingston, ON, Canada\\
              \email{bram.adams@queensu.ca}
              \and
           Ahmed E. Hassan \at
               School of Computing, Queen’s University, Kingston, ON, Canada\\
              \email{hassan@queensu.ca}}

\date{Received: date / Accepted: date}

\maketitle

\begin{abstract}
Pre-trained language models (PTLMs) hosted on platforms such as Hugging Face form complex lineage structures similar to software dependency graphs. However, unlike traditional software ecosystems, PTLM repositories often lack reliable provenance due to missing metadata, such as licenses, reuse methods, pipeline tags, model types, and training libraries. To address this gap, we introduce Semantic Fingerprinting (SemFin), a lightweight approach that combines Hugging Face (HF) configuration files with model repository tags to automatically impute missing model metadata fields and reconstruct model lineage chains. We evaluate SemFin on a large-scale dataset of 317,133 PTLMs. Our results show that configuration files typically encode the technical requirements necessary to instantiate and reuse models, enabling them to serve as a structural blueprint for model reuse, particularly for transformer-based architectures. By combining these configuration files with model repository tags, SemFin significantly outperforms the existing propagation-based imputation approaches, improving prediction accuracy by up to 31.4\% and 26.6\% compared to Graph Avg and Hub Avg baselines. Importantly, SemFin also imputes metadata for 16.6\% of isolated models where propagation-based methods fail. Applying SemFin to impute missing reuse-method and license metadata for 167,089 unlabeled models reveals that traceable reuse method chains expand by 75.9\% and license lineage chains by 53.6\%, uncovering 86 previously invisible reuse method patterns, while the proportion of incompatible license patterns only increases from 34.8\% to 36.8\%. These findings demonstrate how automatically derived structural signals can support the automated construction of AI Bills of Materials (AIBOMs), helping transform metadata from an error-prone manual declaration into information inferred directly from model artifacts.

\end{abstract}

\section{Introduction}\label{introduction}
In traditional software engineering, centralized repositories like Maven Central\footnote{https://central.sonatype.com/} and PyPI\footnote{https://pypi.org/} serve as critical infrastructure for software dependencies. Package Management Systems (PMS) such as Maven or npm automatically enforce versioning and dependency resolution~\citep{decan2019empirical}, ensuring that every artifact is traceable to its source code, version, and license via explicit manifests such as \texttt{pom.xml} or \texttt{package.json}. This automated tracking plays an essential role in generating and validating Software Bills of Materials (SBOMs), which are verifiable records of a product's supply chain required to check its safety and legal compliance.

The widespread adoption of pre-trained language models (PTLMs) has shifted this paradigm toward centralized model hubs, most notably Hugging Face (HF), which currently hosts over two million models~\citep{laufer2025anatomy}. While other repositories exist, such as ONNX\footnote{https://github.com/onnx/models}, PyTorch Hub\footnote{https://pytorch.org/hub/}, Model-Zoo\footnote{https://modelzoo.co/}, and ModelHub\footnote{http://app.modelhub.ai/}, HF remains the largest and most prominent. The presence of multiple repositories further reflects the rapid growth and distribution of PTLMs across platforms. This massive scale is driven by PTLMs becoming foundational building blocks for applications ranging from code generation to complex reasoning~\citep{zhao2023survey, min2023recent}. 

These new hubs face a significant issue because they lack the automated package managers found in traditional software. Instead, the PTLM ecosystem relies on manual metadata provided by users, which are often left empty or inconsistent. Although HF allows uploaders to manually specify metadata such as \texttt{license}, \texttt{reuse method}, \texttt{pipeline tags}, \texttt{training libraries}, and \texttt{parent model}, our analysis shows that more than 50\% of license and reuse method metadata are left empty. For example, reuse methods, such as fine-tuning, quantization, and model merging, define the specific technical process used to derive a child model from its parent  (i.e., the original source model, crucial for tracing model lineage and reuse). Even well-downloaded models, including hmellor/tiny-random-BambaForCausalLM\footnote{https://huggingface.co/hmellor/tiny-random-BambaForCausalLM}, state-spaces/mamba-130m-hf\footnote{https://huggingface.co/state-spaces/mamba-130m-hf}, and peft-internal-testing/opt-125m\footnote{https://huggingface.co/peft-internal-testing/opt-125m}, simultaneously lack both license and reuse method information.

This metadata sparsity results in a transparency debt and creates ``lineage mirages" where the true change history of a model is obscured~\citep{horwitz2025we}. This opacity is a critical barrier to the adoption of AI Bills of Materials (AIBOMs), a concept analogous to SBOMs but focused on models, which should provide verifiable records of a model's supply chain, provenance, and licensing. Without these records, users might unknowingly adopt models that violate licenses or inherit biases from undocumented parents. Until we have a reliable way to recover this missing data, AIBOM standards will remain theoretical concepts that stakeholders cannot use to audit their AI supply chains~\citep{rajbahadur2025building}. 

Existing approaches that attempt to reconstruct model lineage, particularly graph-based methods, are inherently constrained by this data sparsity and the presence of isolated models where lineage information is incomplete. Furthermore, these methods become unreliable when faced with model merging, a practice of fusing multiple model weights into a single artifact~\citep{wortsman2022model}, which obscures traditional lineage paths. As a result, an artifact-driven approach becomes increasingly necessary.

To address the transparency debt, we propose a lightweight approach that fuses models' HF repository tags with the readily available configuration files packaged with the models. Current approaches to mitigate this debt have significant gaps because they rely on external heuristics, for example analyzing how metadata propagates across declared parent child links in repositories like Hugging Face \citep{horwitz2025we}. However, these methods treat models as black boxes, ignoring the structural blueprints of models provided within a HF repository. \citep{laufer2025anatomy} explicitly acknowledge these limitations, arguing that future work must move beyond manual metadata provided on the platform to analyze the actual model artifacts. That is why we concretely operationalized by leveraging a model's \texttt{config.json} file to extract configuration keys (i.e., only the parameter names, without their values).

Our proposed imputation technique for model metadata is based on \textit{Semantic Fingerprinting} (SemFin), a novel methodology that combines model configuration keys with repository tags to reconstruct model identity and lineage. Unlike previous methods that rely on other model metadata, SemFin fuses models' configuration keys and HF repository tags to infer missing metadata. This approach is based on a preliminary analysis of the configuration files of 317,133 PTLMs, which demonstrate that model reuse leaves a distinct fingerprint: derived models systematically remove generic inference configuration keys while introducing task-specific keys. This is very important for our imputation method because these predictable changes in configuration keys provide simple structural signals that our classifiers can use to distinguish between different reuse methods and tasks. We leverage these lightweight signals to train machine learning classifiers that each can recover a missing metadata field with high accuracy. While our empirical evaluation of SemFin currently targets five core model metadata fields, specifically licenses, reuse methods, pipeline tags, model types, and training libraries, it highlights the feasibility of a generalized, artifact-driven approach to constructing comprehensive AIBOMs.

Specifically, we address the following research questions:
\begin{itemize}
    \item[\textbf{$RQ_1.$}] \textbf{\RQa}

    \noindent \textit{\underline{Motivation}}: Understanding which parts of a model's configuration change during reuse is crucial for distinguishing between model copies and genuine derivatives. While our eventual SemFin approach fuses both configuration keys and repository tags, establishing how models' set of configuration keys evolve through model reuse requires isolating the signals within the configuration files. Currently, configuration files are often treated as static descriptors, whereas repository tags are often inconsistent or manually assigned. We hypothesize that as new models are derived down the supply chain, the specific configuration keys included in their configuration files change by adding new keys or removing old ones to reflect their new purpose. Identifying how configuration keys are added or removed, and how they correlate with repository tags, is the first step toward automating lineage tracking and identifying how a child model was derived from its parent.

    \noindent \textit{\underline{Findings}}: We find that configuration files act as a stable core with 68 invariant structural keys surviving across all models. These files signal reuse through the systematic removal of 2,405 generic configuration keys upon certain types of model reuse. Specifically, child models frequently strip away generic inference keys such as \texttt{top\_k} and \texttt{temperature} while simultaneously adding 751 more specific keys. These additions include task-specific keys like \texttt{problem\_type} and various quantization settings. We also observe that repository tags and configuration keys frequently appear together in child models. This indicates that repository tags and architectural configuration keys are often added or removed simultaneously during the same reuse step.

    \item[\textbf{$RQ_2.$}] \textbf{\RQb}

    \noindent \textit{\underline{Motivation}}: Building on $RQ_1$, which showed that configuration files contain specific keys that reliably signal reuse (i.e., the general process of using a parent model to produce a child model, which changes the configuration keys and repository tags) and co-change with repository tags, we hypothesize that these signals can reconstruct missing metadata. Because graph-based heuristics fail for isolated or weakly connected models, we investigate whether machine learning models trained on fingerprints that fuse configuration keys and repository tags can accurately recover a wide range of missing model metadata fields. This includes metadata labels such as licenses, pipeline tags, and reuse methods, even when reuse signals from the parent model or other related models are absent. 

    \noindent \textit{\underline{Findings}}: SemFin significantly outperforms existing graph- and hub-based heuristics, achieving near-perfect recovery for specific model metadata, such as Model Type (97.7\% accuracy), and strong performance for others, such as Pipeline Tag (91.3\% accuracy). Crucially, SemFin achieves full coverage by design, as it always produces a prediction from available configuration keys, whereas baselines abstain on 7.5 to 16.6\% of models due to missing graph connectivity or metadata. This full coverage is possible because our method relies on configuration files, which are mandatory for models to function within HF's popular transformers framework. In contrast, the baseline imputation approaches are limited to the 83--92\% of models that have explicit graph connections. McNemar's test confirms that these improvements are statistically significant across all metadata ($p < 0.001$), characterized by a predominantly large effect size (Cohen's $g > 0.25$).

    \item[\textbf{$RQ_3.$}] \textbf{\RQc}

    \noindent \textit{\underline{Motivation}}: Having shown in $RQ_2$ that SemFin can recover missing reuse method and license metadata, we next examine how such incompleteness distorts our understanding of model reuse across the Hugging Face ecosystem. While parent-child dependency edges between models exist for all models in Hugging Face, missing reuse method metadata obscures the specific technical transformations from a parent model into a child model (e.g., fine-tuning, quantization, merging), and missing license metadata obscures licensing constraints. As a result, the AI supply chain appears fragmented, hiding how models evolve and masking potential license incompatibilities such as non-commercial to commercial transitions or share-alike violations. By imputing missing labels with SemFin, we uncover reuse method lineage patterns and license lineage patterns that were previously unobservable.

   \noindent \textit{\underline{Findings}}: Metadata imputation increases the number of traceable reuse method lineage chains from 31,795 to 131,788, which in turn reveals more than twice as many unique reuse method lineage patterns (155 vs. 76). This recovery shows that Finetune reuse method metadata is substantially under-reported, becoming the dominant reuse method pattern (56.62\%) after imputation , while many apparently standalone Merge operations are revealed to be part of broader multi-step patterns. Similarly, license lineage pattern reconstruction more than doubles the number of observable license lineage chains from 60,965 to 131,356, uncovering a broader range of unique license patterns (419 vs. 250) , in which the single-step license lineage pattern apache-2.0 remains the dominant pattern (34.58\%) after imputation. Importantly, although the wider structural diversity is exposed, the overall proportion of incompatible license patterns remains relatively stable, rising only slightly from 34.8\% to 36.8\%. Within these chains, the most common issues consistently involve Non-commercial $\rightarrow$ Commercial transitions (44.7\% in incomplete, 40.4\% in complete) , followed by AI-Restricted $\rightarrow$ Non-AI License (19.3\%) and ShareAlike $\rightarrow$ Different License (14.0\% in incomplete, 18.7\% in complete) violations. Overall, these results indicate that while the baseline percentage of licensing conflicts remains stable, missing metadata significantly underestimates the absolute structural complexity and the total volume of hidden legal risks propagating through the ecosystem. 
\end{itemize}

Our findings demonstrate that PTLM configuration files and repository tags act as socio-technical artifacts that can transform model metadata from a manual burden into an automatically generated property, or at minimum, provide meaningful semi-automated suggestions to users during model upload. This capability directly supports the automated creation of AIBOMs by scanning model artifacts at ingestion time to reliably populate missing provenance data.  For practitioners, SemFin offers a robust mechanism to audit model supply chains and ensure compliance in an increasingly complex ecosystem.

Specifically, our study provides the following contributions:
\begin{itemize}
    \item We provide the first empirical taxonomy of how configuration keys change (i.e., addition/removal/remaining of keys) when a model is modified. By fusing configuration keys with repository tags, we show how these combined signals correlate with the work done to create a child model from its parent.
    \item We introduce SemFin, a novel approach that uses configuration keys and repository tags to reconstruct model lineage, serving as a technical foundation for future automated AIBOM generation.
    \item We uncover a massive hidden ecosystem of model modification, showing that the true landscape of how models are derived is five times larger than visible metadata suggests.
    \item We provide an open-source replication package, including the curated dataset, SemFin source code, and our trained machine learning models, to facilitate reproducibility and future research on automated metadata imputation.
\end{itemize}

\section{Background and Related Work}\label{background}
\subsection{Pre-Trained Language Models}\label{pretrained}
Pre-trained language models (PTLMs) are general-purpose models trained on large-scale corpora to learn representations that can be transferred across downstream tasks~\citep{devlin2019bert}. Unlike traditional models developed independently for specific objectives, modern architectures adopt a pre-training paradigm that captures broad linguistic regularities prior to task-specific reuse~\citep{williams2018broad}. Their effectiveness stems from the combination of high-capacity architectures, large training datasets, and advances in training methodologies~\citep{mao2020survey}. Prominent PTLMs, including BERT~\citep{devlin2018bert}, GPT~\citep{openai2023gpt}, and RoBERTa~\citep{liu2019roberta}, serve as the foundation for many NLP applications, enhancing capabilities such as text classification, machine translation, and question answering~\citep{raffel2020exploring}. 

In this study, we examine PTLMs as reusable software artifacts. Specifically, we conceptualize the ecosystem as a lineage of reuse in which a \textbf{parent model} serves as an initialization point, and a \textbf{child model} represents the outcome of one or more reuse methods. A reuse method refers to a distinct transformation applied to a pre-trained model to produce a derived model~\citep{ajibode2025towards}. While terms such as ``fine-tuning'' are often used generically, the contemporary PTLM ecosystem employs a diverse set of reuse methods involving architectural and weight-level transformations. In this study, we categorize the following reuse methods:

\begin{itemize}
    \item \textbf{Finetuning:} Further training a pre-trained model on a task-specific or domain-specific dataset, resulting in updated model weights to adapt the parent model to new domains or tasks~\citep{severyn2015unitn,howard2018universal}.
    \item \textbf{Parameter-Efficient Fine-Tuning (PEFT):} Adapting a pre-trained model by updating only a small subset of its parameters (e.g., adapters or LoRA), while keeping the majority of the parent model's weights fixed, thereby reducing computational overhead~\citep{hu2022lora,houlsby2019parameter}.
    \item \textbf{Quantization:} Reducing the numerical precision of a model's weights (e.g., from 16-bit floating point to 8-bit or 4-bit integers) to lower memory usage and improve inference efficiency~\citep{jacob2018quantization}.
    \item \textbf{Model Merging:} Combining the weights of multiple parent models into a single model to blend capabilities, typically through parameter-space operations such as weight averaging~\citep{wortsman2022model}.
    \item \textbf{Distillation:} Training a smaller child model to replicate the behavior of a larger parent model, producing compact models suitable for resource-constrained environments~\citep{hinton2015distilling}.
    \item \textbf{Pruning:} Removing structurally redundant or less important components from a neural network, such as individual weights, neurons, or entire attention heads, to reduce model size and computational cost without significantly degrading performance. For example, in transformer models, many attention heads can be pruned with minimal impact on accuracy \citep{michel2019sixteen}.
    \item \textbf{Deduplication:} A data-centric reuse that removes duplicate data or redundant representations to improve training efficiency and downstream model quality \citep{lee2107deduplicating}.
\end{itemize}

To maintain consistency, we define these techniques as ``reuse methods" for establishing parent-child relationships, though we recognize that a child model may result from multiple, undocumented reuse steps.

\subsection{Model Configuration on Hugging Face}\label{config_file}

\begin{figure*}[t]
\centering
\includegraphics[width=\textwidth]{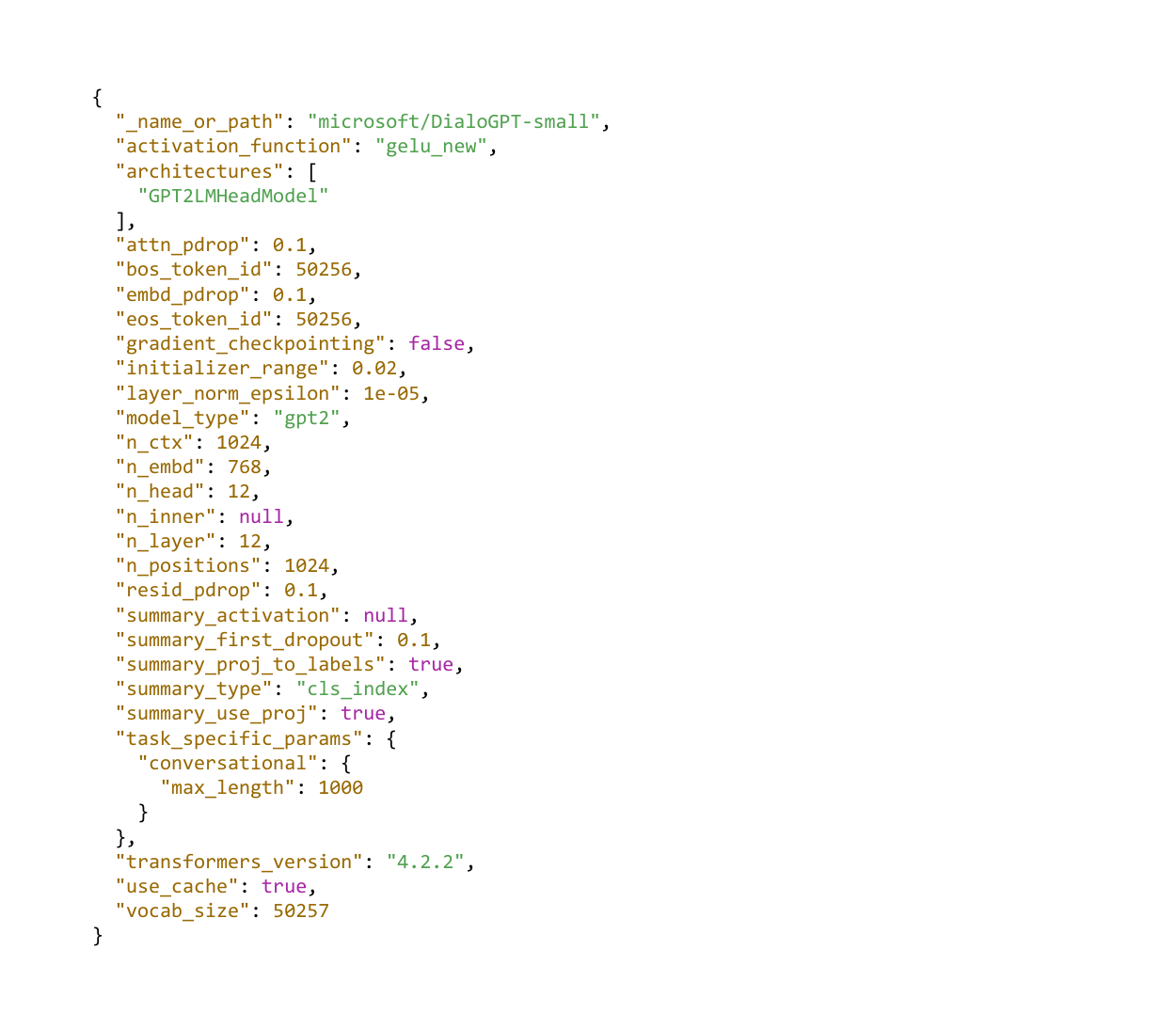}
\caption{Example of configuration file for bigjoedata/rockchatbot model}
\label{ex_config}
\end{figure*}

Within the Transformers ecosystem on Hugging Face, model configuration files (managed by classes derived from \texttt{PreTrainedConfig}) serve as the canonical store for critical architectural attributes, such as the number of hidden layers, attention heads, and vocabulary size. The framework requires these specifications to programmatically reconstruct the model's architecture prior to loading weights and initializing the tokenizer. These settings are serialized into a JSON artifact typically named \texttt{config.json}\footnote{https://huggingface.co/docs/transformers.js/en/api/configs} and located at the root of a model's repository. For example, the \texttt{config.json} for \texttt{bigjoedata/rockchatbot}\footnote{https://huggingface.co/bigjoedata/rockchatbot/blob/main/config.json} is shown in \Cref{ex_config} , while a similar configuration file exists for the \texttt{zai-org/GLM-4.7} model\footnote{https://huggingface.co/zai-org/GLM-4.7/blob/main/config.json}.

This file is a fundamental prerequisite for model instantiation: when a user invokes AutoConfig.from\_pretrained(), the Transformers library relies on this artifact to identify the appropriate model class and infer tensor shapes and architectural constraints (see Hugging Face documentation). In practice, users encounter problems\footnote{https://huggingface.co/vidore/colqwen2-v0.1/discussions/5}\footnote{https://stackoverflow.com/questions/75626974/is-it-possible-to-load-huggingface-model-which-does-not-have-config-json-file} with models that lack a configuration file. Absence or inconsistency of config.json often causes initialization failures, rendering the model unusable regardless of the quality of its learned weights. 

As detailed in the Hugging Face configuration documentation\footnote{\url{https://huggingface.co/docs/transformers/main_classes/configuration}}, unlike the model’s learned weights, which are stored as large binary artifacts, the configuration file (typically \texttt{config.json}) acts as a mandatory structural blueprint\footnote{https://docs.aws.amazon.com/glue/latest/dg/developing-blueprints-code-config.html}. The Transformers framework strictly requires this file to construct the empty neural network graph prior to loading any weights~\citep{wolf2020transformers}. Without it, the framework cannot deduce essential architectural dimensions (such as hidden layers or vocabulary size), causing initialization to fail. Furthermore, the framework's \texttt{from\_pretrained} pipeline relies on this configuration as the ultimate authority on structural metadata, allowing the users and practitioners to programmatically override default parameters, such as modifying dropout probabilities or the number of classification output, during initialization, superseding the original structural settings.

From a software engineering perspective, the config.json file functions analogously to package manifests in traditional software development, such as package.json in Node.js\footnote{https://nodejs.org/api/packages.html\#packagejson-and-file-extensions}, package.json in npm\footnote{https://docs.npmjs.com/cli/v11/configuring-npm/package-json}, and .ini configuration files in system administration. Just as a software package requires a manifest to declare dependencies, entry points, and environment settings in order to build and execute correctly, a pre-trained language model requires its configuration file to ensure reproducibility and predictable execution. This parallel highlights a critical dependency in the Machine Learning Operations (MLOps) lifecycle: model artifacts are not merely raw collections of learned weights, but complex software packages that rely on precise configuration management to function within larger systems.

The absence of configuration files in some HF models is particularly problematic given the recent push for formal AI documentation standards. Frameworks such as SPDX\footnote{https://spdx.dev/understanding-spdx-profiles/} (Software Package Data Exchange) and CycloneDX\footnote{https://cyclonedx.org/} have recently extended their specifications to support AI Bill of Materials (AIBOMs), attempting to standardize fields for model architecture, training data, and upstream lineage \citep{rajbahadur2025building}. However, these standards currently rely on data that does not reliably exist in the model repositories. Without an automated method to extract missing metadata from model artifacts, AIBOM schemas lack the lineage data necessary to be functional.

\subsection{Related Work}
\subsubsection{AI Supply Chains \& Transparency Debt}

The structural opacity of AI development has prompted a growing body of research into the dependencies governing model co-change. \citep{hopkins2025ai} formalize this environment as an \textit{AI Supply Chain} modeled via directed graphs, where upstream components (models, datasets) are reused to create downstream applications. Crucially, their analysis reveals that these supply chains are \textit{non-modular}; unlike traditional software, upstream design choices (e.g., fairness constraints) propagate through the network in complex ways, creating systemic risks of ``hidden interactions'' and dispersed control \citep{hopkins2025ai}.

Previous efforts to trace AI supply chain relationships have primarily focused on conceptualizing the ecosystem using existing, external documentation, rather than computationally reconstructing missing lineage data. For instance, \citep{bommasani2023ecosystem} propose ``Ecosystem Graphs" to link assets (datasets, models, applications) via metadata, highlighting the transparency deficit inherent in pre-trained models. Similarly, from a socio-technical perspective, \citep{widder2023dislocated} demonstrate how this supply chain fragmentation leads to ``dislocated accountability,'' where engineers perceive critical ethical responsibilities as residing elsewhere in the chain.

However, these existing frameworks face a practical limitation: they rely heavily on manual curation or incomplete extrinsic metadata (e.g., model cards). While the literature defines the \textit{theoretical} risks of opaque supply chains, practical methods to map them remain limited by data sparsity. Our study addresses this gap by proposing a lightweight approach, i.e., SemFin, that leverages intrinsic configuration files and repository tags to automatically reconstruct the ``actual'' supply chain, recovering the structural lineage that theoretical works argue is essential for governance.

\subsubsection{Mining model ecosystems}

A nascent body of literature has applied Mining Software Repositories (MSR) techniques to the Hugging Face (HF) ecosystem, establishing it as a critical yet volatile data source for understanding machine learning software evolution. Foundational studies by \citep{laufer2025anatomy} and \citep{ait2025suitability} have characterized the ecosystem's macro-structure, quantifying the sheer volume of models and datasets while validating HF's utility for empirical research. However, both studies identify a critical methodological barrier: the reliance on user-provided metadata tags, which are often inconsistent or absent. \citep{laufer2025anatomy} explicitly argue for the necessity of moving ``beyond metadata'' to analyze internal model artifacts, a recommendation that motivates the configuration-centric approach adopted in this study.

Building on this structural analysis, subsequent research has audited the development and documentation practices within the ecosystem. \citep{ajibode2025towards} and \citep{adekunle2025synchronization} uncovered significant fragmentation in model release patterns, revealing that a small set of base architectures is reused to derive a vast proportion of models, but lacks consistent versioning or lineage tracking. Parallel audits of model documentation by \citep{pepe2024hugging} expose severe gaps in transparency, particularly regarding training data, bias disclosures, and licensing. Collectively, these findings paint a picture of a ``transparency debt,'' where the essential supply chain metadata required for governance is systematically missing. 

More broadly, prior MSR research has shown that structured non-code artifacts, such as configuration files and manifests, encode valuable signals about software evolution and reuse \citep{kula2015trusting, hejderup2018software}. For instance, dependency constraints in package manifests have been used to study library adoption latencies and ecosystem-wide co-evolution, motivating their use beyond traditional source code analysis.

Despite these insights, existing work remains primarily \textit{descriptive}, diagnosing the opacity of the Hugging Face ecosystem without providing automated mechanisms to resolve it. Prior studies are constrained by the very user-provided metadata they identify as flawed, limiting their ability to analyze models where such information is missing. Our work addresses this limitation by shifting from description to \textit{prediction}. By integrating the internal \texttt{config.json} file with available metadata tags, we establish a lightweight and reliable mechanism to overcome sparse documentation, effectively automating the recovery of missing metadata that previous studies could only identify as absent.

\subsubsection{Lineage inference \& metadata imputation}
Automatically recovering missing metadata for machine learning models is an emerging research challenge. The most closely-related prior work by Horwitz et al. \citep{horwitz2025we} proposes graph-based neighbor-voting methods (e.g., Graph Avg, Hub Avg) to infer metadata by propagating known labels across a repository’s model dependency network. However, prior research suggests that graph-based inference is inherently constrained by the presence and quality of observable reuse relationships; if explicit lineage information is incomplete or missing, propagation fails.

To address this limitation, our work explores a complementary inference strategy. Rather than relying solely on the external structure of reuse graphs, we investigate how internal technical artifacts, specifically the \texttt{config.json} file, can be combined with available repository-level metadata to recover missing metadata. Because these intrinsic configuration signals remain available even in the absence of explicit graph edges, they provide a robust, additional source of structured evidence. Accordingly, we evaluate our configuration-centric approach against the methods of Horwitz et al. as our primary baselines to examine the extent to which internal features can complement external graph-based recovery.

In traditional software engineering, ``fingerprinting" is a well-established technique for identifying code provenance \citep{davies2011software, alrabaee2022survey}, detecting plagiarism \citep{naik2015review}, and managing license compliance \citep{rubella2012fingerprint} by extracting lightweight structural signatures rather than performing exhaustive source-level comparisons. Techniques ranging from token-based hashing \citep{kamiya2002ccfinder} to abstract syntax tree (AST) comparison allow engineers to assess code similarity, detecting reused fragments that span a well-defined spectrum: from identical copies (Type-1) and structurally modified snippets (Type-2 and Type-3) to functionally equivalent semantic clones (Type-4) \citep{roy2009comparison}. Just as a hash of a function's bytecode can serve as a unique identifier for a software library, we argue that the config.json file serves as a distinct signature for a PTLM.

While traditional fingerprinting targets source code or binary executables, SemFin adapts this concept to the structural configuration of machine learning models. By treating configuration keys, such as architectures or quantization parameters, as semantic tokens, we apply the principles of clone detection to the AI supply chain. This allows us to identify derived models and reconstruct complex reuse relationships even when the model repository tags are missing or misleading.

\section{Empirical Study Design}\label{methodology}
In this study, we leverage a dataset mined from Hugging Face. It will first be used by RQ1 for a feasibility study, and later to evaluate the SemFin approach (RQ2) and study the reconstructed model supply chain (RQ3). For RQ1, we constructed a dataset of parent–child model pairs and extracted their configuration files to identify configuration settings that change between parent and child models, indicating that a parent was adapted to produce a child model. The following sections detail the data collection, filtering, and metadata extraction procedures developed to support these analyses.

\subsection{Data Collection Methodology}\label{data_collection}
In this study, we selected Hugging Face\footnote{https://huggingface.co/models} (HF) as our primary platform due to its widespread adoption and central role in the distribution of PTLMs. Compared to other platforms, such as \textit{ONNX}\footnote{https://github.com/onnx/models}, \textit{PyTorch Hub}\footnote{https://pytorch.org/hub/}, \textit{Model-Zoo}\footnote{https://modelzoo.co/}, and \textit{ModelHub}\footnote{http://app.modelhub.ai/}, HF provides broader model coverage, richer metadata, and stronger community engagement \citep{ait2025suitability}. This prominence is evident in the platform hosting over two million models spanning a wide range of tasks. To construct our dataset, we followed a multi-stage curation process designed to capture the comprehensive lineage of PTLMs by systematically aggregating parent-child lineages and their corresponding architectural configuration files.

\begin{figure*}[t]
\centering
\includegraphics[width=\textwidth]{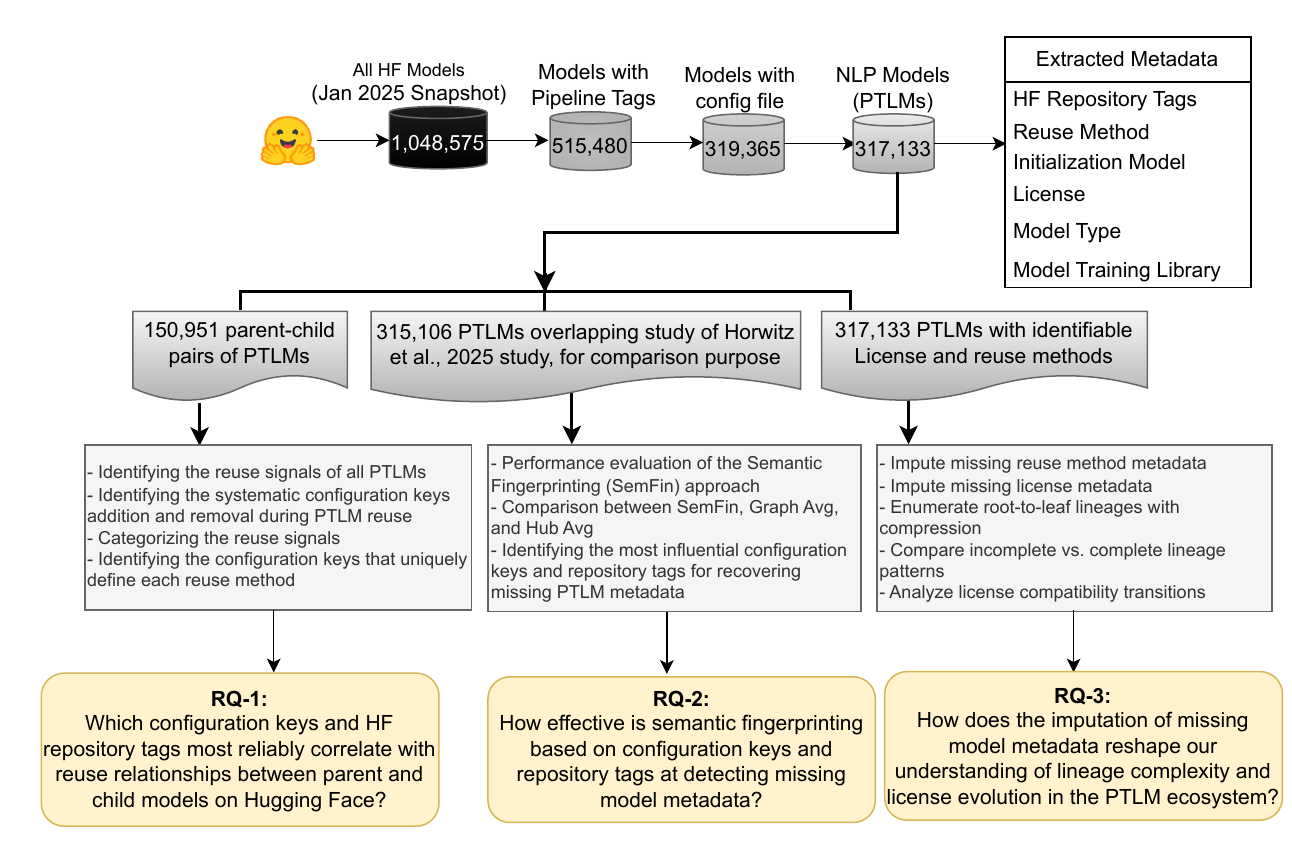}
\caption{Data collection procedure}
\label{framework}
\end{figure*}

\begin{itemize}
\item \textbf{Step 3.1.1: Extracting all models from Hugging Face:}
To retrieve all models hosted on Hugging Face, we used the Hugging Face Hub API (\texttt{HfApi})\footnote{\url{https://huggingface.co/docs/huggingface_hub/package_reference/hf_api}}. We developed a Python script, available in our replication package \citep{SemFin}, to collect a snapshot of the platform as of January~2025. We established this fixed cut-off to ensure dataset compatibility with the baselines used by \citet{horwitz2025we} for our comparative evaluation (RQ2) and to exclude extremely recent uploads, which often have incomplete or rapidly changing metadata. This collection process resulted in a total of 1{,}048{,}575 model entries (see \Cref{framework}). For each model, we extracted the platform-assigned \texttt{pipeline\_tag}, which is provided by the Hugging Face infrastructure and indicates the primary task category associated with the model (e.g., distinguishing an image classifier from a text generator). We observed that 533,095 (50.8\%) models lacked these tags. Since our study strictly focuses on PTLMs, we excluded these untagged entries to prevent the accidental inclusion of non-language models (e.g., Vision or Audio models). We note that this exclusion primarily filters out low-impact artifacts, as the included tagged models exhibit a mean download count of 2,168 compared to just 168 for the excluded untagged entries. This significant disparity suggests that the untagged group largely comprises experimental or inactive artifacts that do not meaningfully contribute to the active reuse ecosystem. For the remaining 515,480 models (49.2\%), we categorized each model into a high-level task domain based on its \texttt{pipeline\_tag}, following the official Hugging Face taxonomy\footnote{\url{https://huggingface.co/tasks}}. This categorization yielded the following distribution: Natural Language Processing (355,952), Computer Vision (71,427), Reinforcement Learning (48,168), Audio (32,126), Multimodal (7,322), Tabular (420), and Other (65). Subsequent filtering focused exclusively on the 515,480 models with verified pipeline tags.

\textbf{Step 3.1.2: Extracting configuration files for models with identifiable pipeline tags:} Among the 515,480 models associated with identifiable pipeline tags, we retrieved configuration files by directly accessing the model repositories via standard HTTP requests. For each model, we programmatically constructed the URL to its default configuration file (following the pattern https://huggingface.co/{model\_name}/resolve/main/config.json). We retained only those models where the configuration file was publicly accessible and successfully retrieved. Instances where the file was missing, the repository was private, or the model had been deleted were collectively categorized as having no accessible configuration. Similar to our filtering in Step 3.1.1, we investigated whether the exclusion of these models introduced bias. We found that models with valid configuration files exhibited a significantly higher mean download count of 2,272 compared to just 665 for models lacking them. This disparity demonstrates that models without a config.json file experience substantially lower community engagement, making our retained subset highly representative of the actively reused ecosystem. 

In total, 319,365 models (61.9\%) yielded usable configuration files, while the remainder did not. Their distribution across tasks is as follows: Natural Language Processing (317,133), Multimodal (1,000), Audio (428), Reinforcement Learning (392), Computer Vision (366), Tabular (28), and Other (18). These configuration files formed the basis for subsequent analyses, serving as the foundation for constructing the parent-child lineage.

\item \textbf{Step 3.1.3: Selecting PTLMs:} Among the 319,365 models with configuration files, Natural Language Processing (NLP) models exhibited the highest proportion of configuration availability. This aligns with our primary focus on PTLMs. We selected PTLMs because they serve as the foundational components for widely adopted AI technologies, such as chat interfaces and code assistants. This central role in the advancement of AI \citep{zhao2023survey} is reflected in our dataset, where PTLMs account for 99.3\% (317,133) of the configured models. Furthermore, PTLMs tend to be larger, better documented, and more widely adopted than models from other domains \citep{castano2024analyzing}, making them a suitable basis for capturing reuse pathways.

\end{itemize}

\subsection{Extracting Repository Tags} While the \texttt{pipeline\_tag} was used in Step 3.1.1 as a taxonomic filter to define our study's scope, it represents only a single dimension of model metadata. For the development of SemFin, we further extracted the complete set of Repository Tags for each PTLM via the \texttt{HfApi} to serve as additional predictive features. Unlike the standardized pipeline tag discussed in the previous section, repository tags are an unstructured collection of user-provided descriptors. We identified 54,211 unique entries in our dataset, including tags such as \texttt{language:en}, \texttt{license:apache-2.0}, and \texttt{dataset:wikipedia}. These tags are integrated into our approach because the metadata categories they represent, such as linguistic focus or licensing, are expected to carry over along the model supply chain during reuse. By tracking these repository tags, our study evaluates whether these user-provided descriptors can act as viable signals for model provenance. In our downstream predictive analysis (RQ2), these user-declared tags complement (RQ1) the intrinsic technical signals extracted from configuration files.

\subsection{Extracting the reuse method of the PTLMs}\label{reuse_extraction} 
As discussed in \Cref{pretrained}, a reuse method is any distinct reuse process (e.g., fine‑tuning) applied to a pre‑trained model able to create a derived model. To construct a reliable ground truth dataset for training our predictive classifiers (RQ2) and to characterize the evolving nature of model reuse (RQ3), it is essential to categorize these methods accurately. To achieve this, we adopted the keyword-based extraction strategy established in \citep{ajibode2025towards}, which utilizes specific keywords (e.g., \textit{ft}, \textit{4bit}, \textit{dedupe}) extracted from multiple repository sources, such as model names, repository tags, and model cards, to identify variant types. However, given the rapid change of the ecosystem, we hypothesized that new reuse methods would employ nomenclatures not covered in prior work. We addressed this through a three-step process.

\begin{itemize}
    \item \textbf{Step 3.3.1: Taxonomy Expansion via Multi-Source Manual Analysis.} To ensure that our taxonomy covered emerging reuse techniques, we performed an exploratory manual analysis. We randomly selected 384 PTLMs (confidence level: 95\%, margin of error: 5\%) to serve as a data source for discovering new indicators. The first two authors independently inspected the \textit{model names, configuration files, and repository tags} of these 384 models to extract potential reuse keywords. Unlike a classification task where inter-rater agreement (e.g., Cohen’s Kappa) is the primary objective, our goal was completeness. Consequently, we treated the findings of both authors as complementary rather than competing. We consolidated the extracted keywords into a unified set, resolving ambiguities by cross-referencing candidate configuration keys against official framework documentation, including the Hugging Face Transformers\footnote{https://huggingface.co/docs/transformers}, PEFT\footnote{https://huggingface.co/docs/peft}, and Optimum specifications\footnote{https://huggingface.co/docs/optimum}. This process allowed us to identify emerging indicators absent in previous studies. For instance, while both authors identified common reuse method indicators, the independent review ensured that rarer indicators such as \texttt{ia3} (PEFT) and \texttt{exl2} (quantization) were captured. We also discovered entirely new indicator sets for Model Merging\footnote{https://huggingface.co/docs/peft/en/developer\_guides/model\_merging} (5 keys), PEFT\footnote{https://huggingface.co/docs/peft/en/index} (7), and Pruning\footnote{https://huggingface.co/docs/optimum/v1.2.1/en/intel/pruning} (4). Additionally, we enriched existing categories, identifying 4 new indicators for Distillation\footnote{https://huggingface.co/docs/setfit/en/how\_to/knowledge\_distillation}, 8 for Quantization\footnote{https://huggingface.co/docs/transformers/en/main\_classes/quantization}, and 8 for Fine-tuning\footnote{https://huggingface.co/docs/transformers/en/training}. 
    
    During this enrichment process, it was critical to distinguish between semantically similar keywords were critical.  For example, we note that the indicator ``compression" under Fine-tuning refers to task- or objective-level compression (e.g., token or sequence compression, as in \textit{infgrad/Jasper-Token-Compression-600M}), rather than model compression techniques such as quantization or pruning. Conversely, we classified models with a ``compressed" indicator under Quantization. After manually inspecting all 44 model cards in this compressed category, we confirmed that these models are indeed quantized but neither pruned nor distilled (e.g., \textit{royleibov/granite-7b-instruct-ZipNN-Compressed} and \textit{royleibov/solar-pro-preview-instruct-ZipNN-Compressed}). We explain each of these reuse methods in \Cref{reuse_methods}.

    \item \textbf{Step 3.3.2: Automated Multi-Source Extraction.} Leveraging the expanded taxonomy of step 3.3.1, we developed an automated extraction pipeline to classify the reuse method of every PTLM in our dataset. This automated phase scans the same three metadata artifacts (i.e., configuration file, repository tags, and model name) to maximize detection accuracy. We implemented a \textbf{Source Priority Protocol} to rank these artifacts based on their reliability. First, we scan the \texttt{config.json} file (Priority 1), considering it the ``observed metadata" of the model's architecture. If no signal is found, we scan the repository tags (Priority 2), and finally, we parse the free-text model name (Priority 3) as a last resort.

    \item \textbf{Step 3.3.3: Hybrid Conflict Resolution.} A critical challenge in this process is ``method stacking," where a single model exhibits signals for multiple reuse methods (e.g., \textit{unsloth/DeepSeek-R1-Distill-Qwen-7B-bnb-4bit\footnote{https://huggingface.co/unsloth/DeepSeek-R1-Distill-Qwen-7B-bnb-4bit}}), which contains indicators for Distillation (``Distill") and Quantization (``bnb", ``4bit"). To resolve these conflicts, we implemented a \textbf{hybrid resolution strategy} tailored to the structural nature of the metadata source:
    \begin{itemize}
        \item \textbf{For Configuration File and Repository Tags (Unordered):} Since these sources are unordered sets, we rank the reuse methods based on how invasive the model transformations are, from most to least invasive. This priority ranking places methods that combine distinct model lineages (merging) or fundamentally alter the model structure or training objective (architectural changes) over those that primarily optimize model storage. Accordingly, we rank Merge, PEFT, Distillation, and Pruning as high priority, while Quantization, Deduplication, and Finetuning are classified as low priority. This priority scheme resolves multi-method conflicts by selecting the single primary reuse method that reflects the most significant transformation, rather than merely a final optimization step.
        
        \item \textbf{For Model Names (Ordered):} Since naming conventions on Hugging Face typically follow a chronological pattern where new modifications are appended as suffixes, we employ a temporal heuristic, which posits that the right-most indicator in a model name represents the final, usable state of the artifact. For example, in the \textit{unsloth/DeepSeek-R1-Distill-Qwen-7B-bnb-4bit\footnote{https://huggingface.co/unsloth/DeepSeek-R1-Distill-Qwen-7B-bnb-4bit}}, although ``Distill" appears earlier, the final token ``4bit" confirms the artifact is deployed as a quantized model. Thus, in cases where we rely on the model name (i.e., when higher-priority Config/Tag signals are absent), we classify it as Quantization based on this final indicator.
    \end{itemize}
\end{itemize}

Consequently, we successfully identified reuse methods for 150,044 (47.31\% of 317,133) PTLMs in our study, with the following distribution: Finetune (101,595), Quantization (29,510), Merging (13,641), Peft (2,276), Distillation (1,892), Pruning (601), and Deduplication (529).

\begin{table}[t]
\centering
\caption{Taxonomy of Model reuse methods and Indicators with Ecosystem Distribution}
\label{reuse_methods}
\renewcommand{\arraystretch}{1.2}
\begin{tabular}{p{0.15\linewidth} p{0.25\linewidth} p{0.10\linewidth} p{0.40\linewidth}}
\toprule
\textbf{Reuse Method} & \textbf{Key Indicators} & \textbf{Count} & \textbf{Description} \\
\midrule
\textbf{Merge} & merge, merged, fusion, fused, combined, passthrough & 6 & Combining the weights of multiple models into a single architecture to blend their capabilities. \\
\midrule
\textbf{PEFT} & lora, qlora, adapter, adapters, peft, ia3, prefix & 7 & Adapting a pre-trained model by updating only a small subset of parameters (Parameter-Efficient Fine-Tuning). \\
\midrule
\textbf{Quantization} & quantized, quantization, 4bit, 8bit, int4, int8, q4, q8, awq, gptq, exl2, gguf, ggml, bitsandbytes, bnb, qat, compressed & 17 & Reducing the numerical precision of model weights (e.g., from float to int) to lower memory usage and latency. \\
\midrule
\textbf{Distillation} & distilled, distillation, teacher, student, tinybert, mobilebert & 6 & Training a smaller ``student'' model to mimic the outputs and behavior of a larger ``teacher'' model. \\
\midrule
\textbf{Pruning} & pruned, pruning, sparse, sparsity & 4 & Removing redundant network parameters or neurons to reduce model size while maintaining performance. \\
\midrule
\textbf{Deduplication} & deduped, dedupe & 2 & Eliminating duplicate data or parameters to improve dataset quality and training efficiency. \\
\midrule
\textbf{Finetune} & finetuned, finetune, ft, tuned, tuning, instruct, chat, rlhf, dpo, sft, compression (task-level) & 11 & Updating the parameters of a pre-trained model on a specific dataset or instruction-following task. Keywords such as \texttt{instruct}, \texttt{chat}, and \texttt{compression} were observed in the model name, indicating reuse type or target usage. \\
\bottomrule
\end{tabular}
\end{table}

\subsection{Extracting Additional Metadata}
We followed the procedure outlined below to extract additional metadata utilized in this study, specifically the parent model (initialization model), license, and training library.

\subsubsection{Initialization Model Extraction}\label{parent_extraction}
To determine the provenance of PTLMs in our dataset, we extracted for each PTLM the raw text string that specifies the original model used as a starting point for training, such as the \textit{\_name\_or\_path} value in a \textit{config.json} file. Formally, we define this raw extracted string as the \textit{Initialization Model}. We explicitly treat this raw string as a candidate reference rather than an assumed valid parent model because configuration manifests are often filled manually or automatically exported with environment-specific references that do not correspond to public repository paths. Treating these extractions as candidates allows us to establish a raw baseline that we systematically filter and validate against our active model corpus during lineage reconstruction experiment.

We extracted initialization model strings from two primary sources: the model's configuration file and the README.md. We combined these two sources to maximize coverage, as documentation practices on the Hugging Face Hub vary significantly. For instance, some models, such as \textit{Stickmu/HailBERT-de-v1}, specify initialization information only in the README metadata, while others, such as \textit{interneuronai/real\_estate\_listing\_analysis\_bart}, rely solely on the configuration file. For models that provided information in both sources (e.g., \textit{fakespot-ai/roberta-base-ai-text-detection-v1}), we observed no discrepancies, confirming consistency between our extraction methods.

To extract initialization model strings, we first parsed the \textit{config.json} file and extracted the value of the \textit{\_name\_or\_path} field. We observed that this field frequently contains local filesystem paths or temporary directories rather than canonical model identifiers. For example, \textit{PrunaAI/bigscience-bloomz-560m-bnb-4bit-smashed} specifies \textit{/tmp/tmp9vq9eg\_x}, \textit{PrunaAI/mosaicml-mpt-7b-chat-bnb-4bit-smashed} reports \textit{/tmp/tmp2l8uulmf}, and \textit{PrunaAI/mistralai-Mistral-7B-Instruct-v0.2-bnb-4bit-smashed} lists \textit{/ceph/hdd/staff/charpent/.cache/modelsnnadc2ao0skejqu6}. These examples illustrate the diversity of raw initialization strings obtained through configuration parsing.

When the \textit{\_name\_or\_path} field was missing or incomplete, we parsed the YAML metadata header at the top of the README.md file to extract the \textit{base\_model} tag. This step was particularly important for merge models, which often declare multiple initialization models explicitly in this structured field.

We deliberately excluded two additional sources due to low reliability. First, we did not mine the unstructured free text of Model Cards, as this approach introduces substantial noise \citep{ajibode2025towards,oreamuno2024state}, with many cards referencing multiple unrelated models for comparison. Second, we excluded repository tags that contain dataset names or other identifiers that resemble model names, making automated extraction unreliable. Accordingly, we restricted our extraction to structured configuration and metadata fields.

At the end of this step, we identified initialization model strings for 93.46\% of the dataset. These raw extractions serve as unverified parent identifiers that we subsequently validate and filter to reconstruct the verified parent-child links used to answer our research questions.

\subsubsection{License Extraction}
To identify the license associated with each PTLM repository, we developed a Python script that extracts information from four primary sources: repository tags, the YAML metadata header in the README, the model’s configuration file, and the model card text. We implemented a hierarchical extraction strategy that prioritizes structured metadata over unstructured text.

Our algorithm proceeds as follows. First, we examine repository tags (e.g., Hugging Face metadata), specifically searching for tags prefixed with \texttt{license:} (e.g., \texttt{license:apache-2.0}), and extract the value following the colon. If no license tag is present, we parse the YAML metadata header at the top of the README file to retrieve the value associated with the \texttt{license} field.

If the license remains unidentified, we inspect the model’s configuration file (e.g., \texttt{config.json}) for a \texttt{license} key. Finally, as a last resort, we perform a keyword search in the raw text of the model card for the string \texttt{license:} and extract the immediately following text. If none of these four sources yields a license, we assign the placeholder value \texttt{unknown}.

At the end of this extraction step, we successfully extracted explicit license information for 129,295 (40.77\% of 317,133) models. While the remaining models lack explicit license definitions, a sparsity consistent with prior findings \citep{jewitt2025hugging, pepe2024hugging}, the 129,295 labeled instances provide a substantial and representative ``observed metadata" dataset. This labeled subset is sufficiently large to train and validate our machine learning classifiers during the evaluation phase (RQ2), serving as the technical foundation to eventually impute the missing license metadata for the unlabeled models across the broader ecosystem.

\subsubsection{Model Type Extraction}
To identify the underlying model architecture on which each PTLM is based, we relied on the configuration file associated with each repository. Unlike licensing information, which is often optional or dispersed across multiple sources, the model architecture is a fundamental technical specification required for a model to function within the Hugging Face transformers library.
We developed a Python script to parse the configuration file (typically config.json) and extract the value of the model\_type field. This field explicitly specifies the model architecture (e.g., bert, gpt2, llama).
At the end of this extraction step, we successfully identified model types for all 317,133 PTLMs in our dataset (100\% coverage). This complete coverage is expected because the transformers framework strictly requires the model\_type within the configuration file to identify the appropriate model class and construct the neural network graph.

\subsubsection{Model Training Library Extraction} To identify the software library or framework used to train or fine-tune each PTLM, we utilized the Hugging Face API, specifically targeting the \texttt{library\_name} metadata associated with each repository. This metadata indicates the primary library (e.g., \texttt{transformers}, \texttt{adapter-transformers}, \texttt{spacy}) required to load and run the model.

Our extraction process identified 80 unique libraries across the dataset. In total, we successfully retrieved library information for 311,828 repositories, representing 98.32\% of our dataset.

\section{\textbf{RQ$_1$:} \RQa}\label{RQ1_result}
Although configuration files define model behavior and repository tags provide descriptive metadata, prior work treats them as static artifacts rather than evolving signals of model reuse \citep{di2024automated, schlegel2023management}. In practice, tags are often inconsistent or manually assigned, as detailed in our preliminary ecosystem analysis showing high metadata sparsity, and there is limited understanding of how they change alongside configuration files during reuse. For metadata to be useful in tracking model changes, it must be both readily available, such as from configuration files or repository tags, and able to change as models are modified along the supply chain. Metadata that is missing or remains unchanged across reuse provides little value for distinguishing between models. Therefore, we focus on identifying lightweight reuse signals, such as the addition or removal of configuration keys, to differentiate between models that simply inherit from a parent from those that introduce meaningful modifications. These signals allow us to relate structural changes in configurations to repository tags and support the recovery of missing metadata and hidden reuse pathways in RQ2 and RQ3.

To this end, we empirically analyze how configuration files and repository tags co evolve between parent and child models on Hugging Face. Our goal is to determine whether observable structural changes align with tag annotations, and to identify the minimal set of features required to characterize different reuse patterns, such as fine-tuning and quantization. If decent correlations are found, this analysis will help us design a metadata field imputation technique, and in that case, our focus on configuration keys and tags will ensure that the technique will be lightweight.
\medskip

\begin{figure*}[t]
    \centering
    \includegraphics[width=\textwidth]{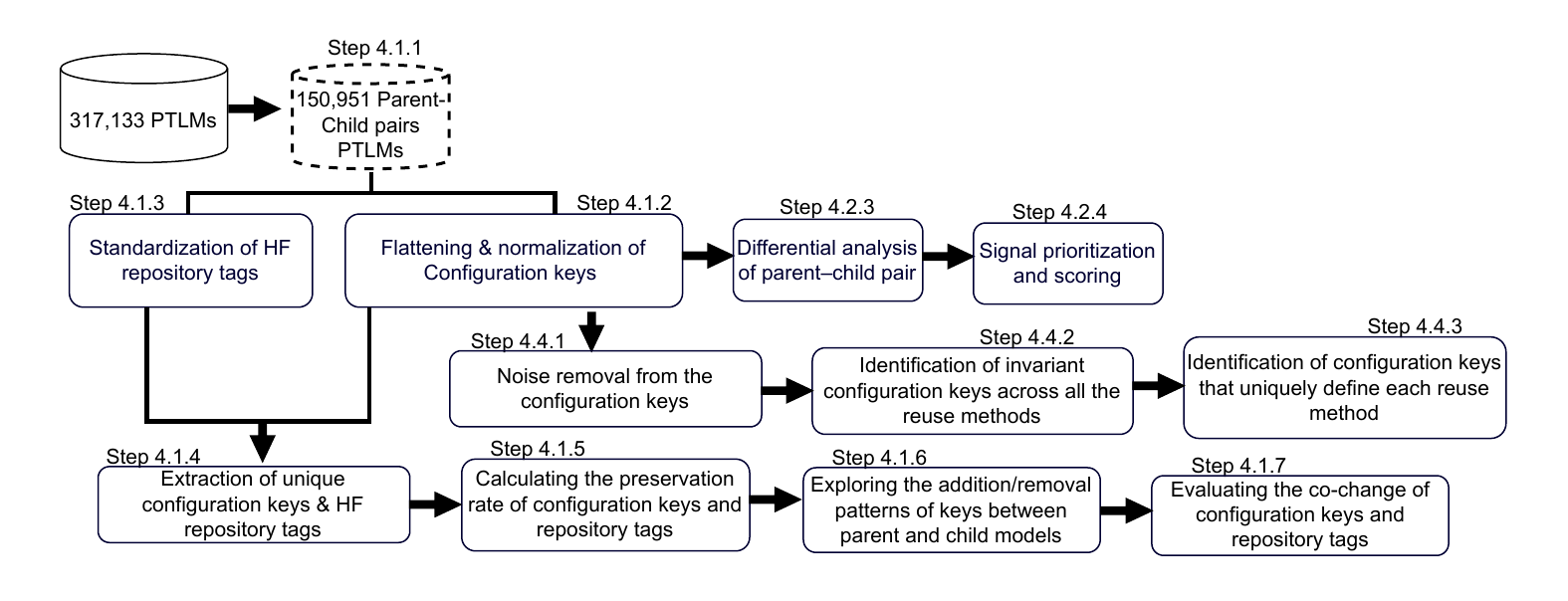} 
    \caption{Overview of the $RQ_1$ methodology for exploring the co-change of configuration keys and repository tags, extracting structural reuse signals, and identifying reuse-method-specific configuration keys.}
    \label{rq1_approach}
\end{figure*}

\noindent\emph{Approach and Results}\label{RQ1_approach}\\
To address this research question, we employ a multi-stage workflow to identify reuse signals, analyze their potential co-change, and extract method-specific configuration fingerprints. We depict the complete methodological flow for $RQ_1$ in \Cref{rq1_approach}. Given the distinct analytical techniques required for each phase, ranging from statistical scoring to qualitative manual coding, we structure this section by presenting, for each sub-question, the specific \textbf{Approach} (a subset of \Cref{rq1_approach}) followed immediately by the corresponding \textbf{Results}.

\subsection{Exploring the co-change of configuration keys and repository tags during PTLM reuse}\label{section_4.1}
To explore the potential of a lightweight method for metadata imputation, we investigate the feasibility of fusing configuration files and repository tags as joint predictive features. We hypothesize that if repository tags systematically co-evolve with configuration keys, they encode aligned reuse semantics rather than isolated metadata annotations. This structured alignment would indicate that a subset of configuration keys and repository tags capture complementary aspects of the same reuse process, confirming that they can be reliably fused for metadata prediction.
\head{Approach}\\
To identify how repository tags co-evolve with configuration files and to measure the extent of structural inheritance during model reuse, we follow the following steps:

\begin{itemize}

    \item \textbf{Step 4.1.1 – Identify valid parent–child model pairs:}
    Building on the extracted initialization model strings from \Cref{parent_extraction}, we define a valid \textit{parent model} as one whose initialization string can be resolved to a public repository on the Hugging Face Hub and is present within our collected dataset. This ensures that both the parent and child models have accessible configuration data required for co-change analysis. Initialization strings that refer to local filesystem paths (e.g., \texttt{/home/yakovelm/.cache/...}), temporary directories, or otherwise unresolvable identifiers are excluded, as they cannot be linked to verifiable upstream models.
    
    To construct the final set of valid parent–child pairs, entries containing multiple parents were expanded into individual parent–child relationships. We then retained only those pairs where the parent model exists in our verified dataset. This filtering step removed 263 additional initialization models that could not be cross-referenced within our PTLM corpus.
    
    This refinement reduced the sample from 317,133 models to 150,951 parent–child pairs. To assess whether this subset remains representative of the active ecosystem, we examined its community engagement. Models in this refined set exhibit a mean download count of 4,664, substantially higher than the 3,448 mean downloads observed for child models without valid parent models. Consequently, this active and verified subset provides a robust basis for analyzing reuse signals in configuration keys.
    
    \item \textbf{Step 4.1.2 – Configuration flattening and key normalization:}
    To enable granular comparison across heterogeneous model architectures, we flatten configuration files into dot-notated key–value pairs and normalize index-specific keys. Configuration files of many PTLMs on Hugging Face frequently contain indexed and nested structures such as:
    
    \begin{verbatim}
    "layers": {
     "0": {"bias": true, "hidden_size": 768},
     "1": {"bias": true, "hidden_size": 768}
    }
    \end{verbatim}
    
    Flattening this structure initially yields indexed keys including \texttt{layers.0.bias}, \texttt{layers.0.hidden\_size}, \texttt{layers.1.bias}, and \texttt{layers.1.hidden\_size}. Because direct comparison is hindered by architectural depth differences (e.g., a parent model having 12 layers and a child having 6), we normalize these keys by programmatically stripping layer indices using regular expressions. This results in generalized, representative keys such as \texttt{layers.bias} and \texttt{layers.hidden\_size}.

    Because the values for these structural keys are fundamentally identical across layers within a given model, our normalization script collapses these redundant indexed entries into a single representative key. In the rare event that a value varies between layers, our script is designed to let the final layer determine the value used for the entire model. This dimension reduction allows us to perform a singular, direct comparison of the overall architectural characteristics (e.g., parent \texttt{layers.hidden\_size} vs. child \texttt{layers.hidden\_size}) without being obstructed by differing layer counts. Furthermore, we removed non-ASCII characters to guarantee uniform key representation. This process ensures that configuration keys are compared based on their functional role rather than their position within the layer stack, enabling consistent analysis across models with varying architectural depths.

    \item \textbf{Step 4.1.3 - Standardize HF repository tags:}
    To standardize the comparison of parent and child HF repository tags metadata, we clean these tags by isolating the categorical keys from structured key-value pairs. Specifically, for tags containing both a key and a value (e.g., \texttt{license:mit}), we strip away the specific value and retain only the underlying key (e.g., \texttt{license}). This reduction ensures that our analysis tracks the inheritance of broader metadata categories rather than model-specific values during model reuse.

    \item \textbf{Step 4.1.4 - Extracting unique child configuration keys and repository tags:}
    To understand the relationship between newly added configuration keys and repository tags, we retrieved the specific keys and tags that are unique to each child model (compared to its parent). We achieve this by subtracting the inherited parent repository tags and configuration keys from the child's respective sets. Retrieving these novel additions allows us to analyze the specific configuration keys and repository tags introduced during reuse and evaluate the correlation between them.

    \item \textbf{Step 4.1.5 - Calculating the preservation rate of configuration keys and repository tags:}
    To quantify the extent of a parent model's keys and/or tags inherited by a given child model when a parent model is adapted to produce a child model, we calculate the preservation rate for both configuration keys and repository tags across all valid parent-child pairs. While Step 4.1.4 focused on the subtraction of inherited keys or tags to identify new additions, this step uses the intersection of parent and child metadata to measure what is retained. The preservation rate is computed by dividing the number of retained configuration keys and repository tags by the total number of respective elements originally present in the parent model. We also identify pairs where a child model retains the entire set of configurations and tags of metadata profile of its parent without modification. Finally, we aggregate these pairs globally to compare the total volume of unique configuration keys and repository tags that are systematically removed by child models against the total volume of keys and tags that are newly introduced during the reuse step.

    \item \textbf{Step 4.1.6 - Exploring the addition/removal patterns of keys between parent and child models:} To identify configuration keys that models frequently add or remove during reuse, we analyzed keys that appear or disappear in child models relative to their parents. We first removed administrative metadata keys (e.g., name\_or\_path, transformers\_version, and \_commit\_hash) from our key statistics. For added keys, we selected all configuration keys that are present in child models but absent in their corresponding parents, resulting in 751 keys. We ranked them by the number of parent-child pairs in child models. For removed keys, we selected all configuration keys that are present in parent models but missing from their corresponding children, resulting in 2,405 keys. We ranked them by the number of parent–child pairs in which they were removed.

    \item \textbf{Step 4.1.7 - Evaluating the co-change of configuration keys and repository tags:}
    To quantify the connection between configuration keys and repository tags added during the same reuse step, we measure how frequently specific configuration keys in the child models appear alongside specific repository tags in child models. We focus this analysis on the subset of parent–child pairs where the child model introduces at least one new configuration key and one new repository tag. This ensures we are analyzing models with observable modifications in both metadata types. To ensure the analysis captures meaningful structural reuse, we explicitly exclude administrative artifacts from the configuration keys, such as \texttt{\_name\_or\_path}, and remove metadata identifiers that do not describe the model's technical properties, such as \texttt{region}, from the repository tags. We then retrieve the most frequently added repository tags. For each tag, we calculate the co-occurrence rate by dividing the number of times a specific configuration key is introduced alongside that tag by the total number of times that tag appears as a new addition. This metric identifies which configuration keys are strictly associated, serving as a measure of redundancy versus complementarity, with the introduction of specific repository tags during the reuse process.

\end{itemize}

\head{Result}\\
\noindent \textbf{On average, child models inherit 98.2\% of their parent's configuration keys and 88.4\% of their repository tags, showing that reuse preserves most of the parent's structure and metadata while allowing targeted modifications.} Across roughly 151,000 parent–child pairs, child models retain a mean of 98.2\% of configuration keys (median 100\%) and 88.4\% of repository tags (median 90.5\%). While nearly one-third (31.4\%) of pairs show perfect configuration key overlap, only 5.1\% exhibit perfectly identical tag sets, indicating that users are much more likely to modify a model's repository tags than its core architectural configuration keys.

At the corpus level, we identify 4,851 unique configuration keys in parent models and 5,920 in child models. Among these, 1,077 keys appear exclusively in child models, whereas only 8 are unique to parents. This expansion is significantly more pronounced in the tags metadata: of the 405,098 total unique tags observed across the corpus, an overwhelming 294,826 are introduced exclusively by child models, compared to just 3,735 unique to parents. This stark asymmetry confirms that child models primarily expand upon what they inherit from parents, as key or tag removals are drastically less frequent than additions across the ecosystem.

Taken together, these findings demonstrate a strong inheritance pattern across the corpus. Because the vast majority of keys and tags remain stable across lineages, the small minority of features that do change provide a lean, highly discriminative set of structural signals that our machine learning models can leverage for targeted metadata imputation in RQ2.

\begin{figure*}[t]
\centering
\includegraphics[width=\textwidth]{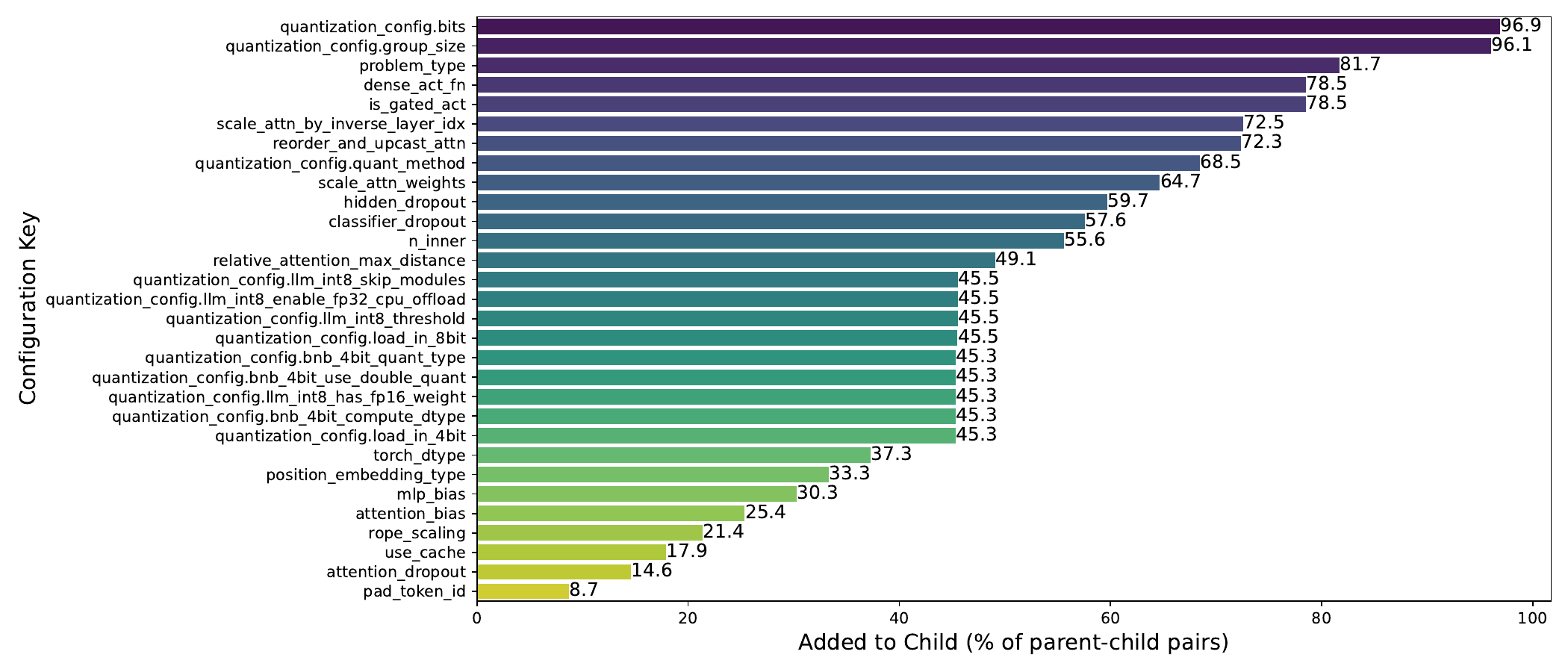}
\caption{Top-30 configuration keys added to child models but absent in parent models on Hugging Face. The x-axis shows the total number of models in which each key was added, while the y-axis lists the configuration keys. Bars are annotated with the percentage of child models containing the added key, highlighting the prominence of quantization e.g., quantization\_config.bits and task-specific additions (e.g., problem\_type.))}
\label{key_added_to_child}
\end{figure*}

\noindent\textbf{Child models extend the parent configuration schema primarily to enable post-training quantization, task specialization, and structural specificity.} \Cref{key_added_to_child} displays the top configuration keys that are absent in parent models but explicitly added to child models. Among the top-added keys, quantization configuration keys dominate: \texttt{quantization\_config.bits} appears in 96.9\% of child models, and \texttt{quantization\_config.group\_size} in 96.1\%. These results indicate that a large portion of model reuse on Hugging Face involves converting parent models into resource-efficient formats (e.g., 4-bit or 8-bit versions). Child models also frequently introduce task-specific definitions that were undefined in the parent, such as \texttt{problem\_type} (added in 81.7\% of child models). In a separate but related pattern, they explicitly define architectural specifiers like \texttt{dense\_act\_fn} (78.5\%) and \texttt{is\_gated\_act} (78.5\%). These results indicate that the reuse process systematically drives configuration towards greater specificity, where child models explicitly resolve operational configuration keys that were left implicit or default in their immediate parents.

\begin{figure*}[t]
\centering
\includegraphics[width=\textwidth]{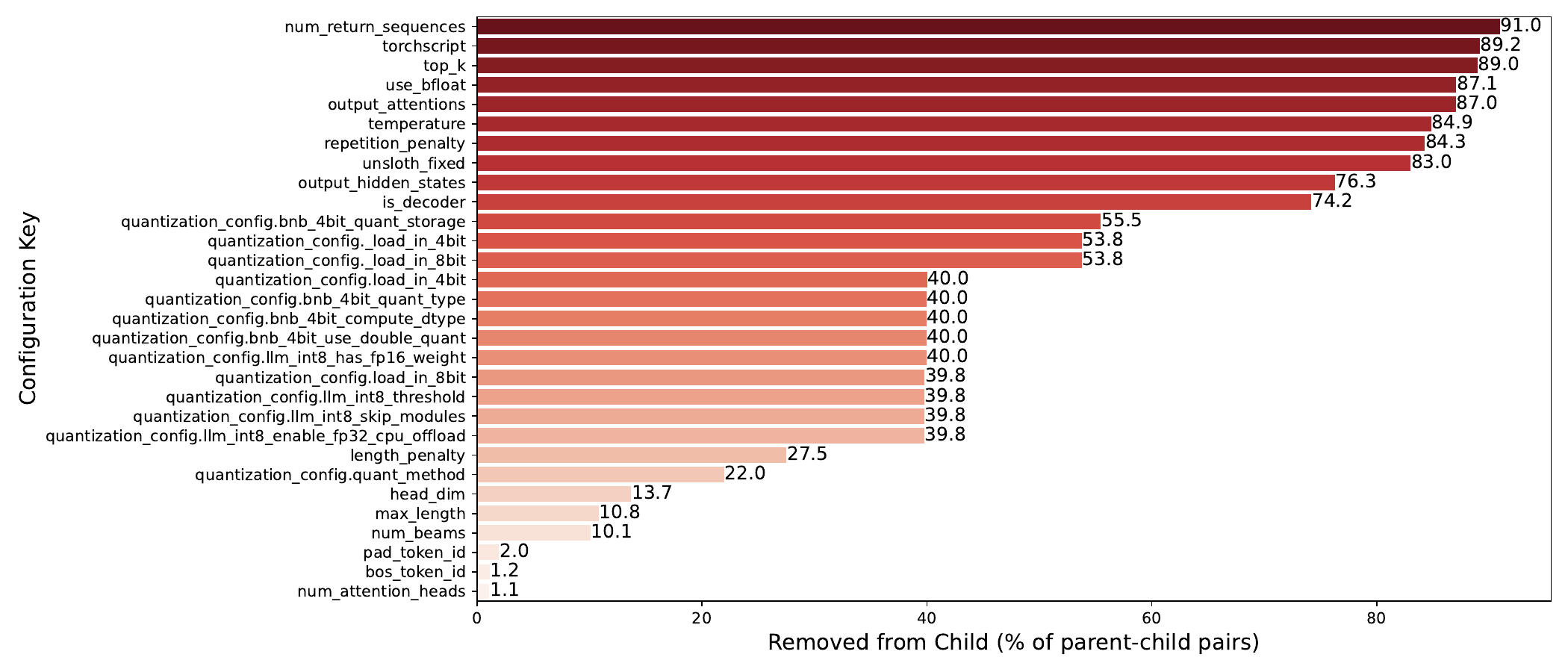}
\caption{Configuration keys frequently present in parent models but absent in child models on Hugging Face. The x-axis shows the number of models in which each key was not included in child models, while the y-axis lists the configuration keys. Bars are annotated with the percentage of child models that no longer include the key, highlighting common removal of runtime generation configuration keys, legacy export options, and quantization-related keys.}
\label{key_only_in_parent}
\end{figure*}

\noindent\textbf{Child models frequently remove runtime generation configuration keys and framework compilation flags.} \Cref{key_only_in_parent} shows the configuration keys most commonly present in parent models but absent in their children. Inference-time configuration keys are often removed: \textit{top\_k} (89.0\%), \textit{temperature} (84.9\%), \textit{repetition\_penalty} (84.3\%), and \textit{do\_sample} (69.3\%). Keys related to model compilation and C++ deployment, such as \textit{torchscript} (89.2\%), are also frequently removed. Additionally, quantization-related keys (e.g., \textit{quantization\_config.bnb\_4bit\_quant\_storage}, 55.5\%) are removed in a substantial fraction of child models. These results indicate that child models streamline the configuration file to focus strictly on architecture and weight definitions. The systematic removal of inference parameters likely occurs because practitioners prioritize architectural and weight definitions over transient runtime settings during model export. For our proposed approach in \Cref{semfin}, this pattern provides a discriminative signal: the absence of inference keys such as \textit{temperature} or \textit{top\_k} reliably indicates a derivative model optimized for specific deployment contexts, whereas the presence of these keys suggests a general-purpose ancestor or unmodified checkpoint. Our approach leverages this signal to distinguish between base models and their adapted descendants without requiring value-level comparisons.

\noindent\textbf{Across all parent-child pairs, we observed 40,791 unique (tag, key) pairs, though the distribution is highly long-tailed: the median co-occurrence rate is only 1.4\%.} These unique pairs were calculated as pairwise combinations, where every unique tag added by a child model is paired with every unique configuration key added in the same step. However, when child models introduce new tags during reuse, these additions strongly correlate with targeted structural modifications, as shown in \Cref{tag_config_coevolution}. For example, quantization tags such as \texttt{bitsandbytes} (99.9\%), \texttt{gptq} (99.1\%), and \texttt{4-bit} (95.7\%) almost perfectly predict the addition of \texttt{quantization\_config} keys. Similarly, task-specific tags like \texttt{text-classification} predict the \texttt{problem\_type} key in 85.7\% of cases. In contrast, administrative tags, such as \texttt{license}, \texttt{tensorboard}, and \texttt{dataset}, demonstrate weak predictive power, ranging from 24.9\% to 36.4\%, for structural changes. 

This high co-occurrence is not redundancy. The configuration key tells us quantization happened, while the tag tells us which quantization method was used. The variation across tags further confirms they convey method-specific semantics. For instance, \texttt{dataset} co-occurs with \texttt{problem\_type}, while \texttt{safetensors} co-occurs with \texttt{use\_cache}. This alignment suggests that repository tags capture reuse semantics that complement intrinsic configuration signals.

\begin{table}[t]
\centering
\caption{Co-change of newly introduced HF repository tags and configuration keys in parent--child model pairs. Co-occurrence rates show that specific repository tags systematically appear alongside newly added configuration keys, establishing repository tags as observable signals of model reuse.}
\label{tag_config_coevolution}
\resizebox{\textwidth}{!}{%
\begin{tabular}{lrlrl}
\toprule
\textbf{New Tag} & \textbf{Total Tag Occurrences} & \textbf{Associated New Config Key} & \textbf{Co-occurrence Count} & \textbf{Co-occurrence Rate} \\
\midrule
license & 52,454 & problem\_type & 13,055 & 24.9\% \\
generated\_from\_trainer & 38,165 & problem\_type & 15,589 & 40.8\% \\
text-classification & 27,501 & problem\_type & 23,561 & 85.7\% \\
tensorboard & 24,569 & problem\_type & 8,944 & 36.4\% \\
safetensors & 16,469 & use\_cache & 4,355 & 26.4\% \\
dataset & 8,183 & problem\_type & 2,538 & 31.0\% \\
endpoints\_compatible & 7,540 & rope\_theta & 1,288 & 17.1\% \\
4-bit & 6,933 & quantization\_config.quant\_method & 6,638 & 95.7\% \\
model-index & 6,514 & problem\_type & 2,336 & 35.9\% \\
token-classification & 5,925 & use\_cache & 1,295 & 21.9\% \\
merge & 5,289 & attention\_dropout & 1,360 & 25.7\% \\
mergekit & 5,250 & attention\_dropout & 1,304 & 24.8\% \\
autotrain\_compatible & 5,238 & architectures & 1,948 & 37.2\% \\
trl & 4,733 & rope\_scaling & 1,173 & 24.8\% \\
bitsandbytes & 3,664 & quantization\_config.llm\_int8\_skip\_modules & 3,662 & 99.9\% \\
sft & 3,265 & rope\_scaling & 794 & 24.3\% \\
gptq & 2,981 & quantization\_config.bits & 2,955 & 99.1\% \\
question-answering & 2,920 & use\_cache & 733 & 25.1\% \\
generated\_from\_keras\_callback & 2,592 & use\_cache & 1,066 & 41.1\% \\
conversational & 1,990 & attention\_dropout & 453 & 22.8\% \\
\bottomrule
\end{tabular}%
}
\end{table}

\subsection{Identifying configuration keys that consistently signal changes between parent and child models} \label{section_4.2}
Unlike \Cref{{section_4.1}}, which focused on whether configuration keys are added or removed during reuse, this section shifts focus to keys that exist in both parent and child but take on different values. Although the overall configuration schema exhibits a strong structural inheritance pattern where the vast majority of keys are preserved in the child model, we hypothesize that the underlying values assigned to these shared keys are frequently altered during model reuse. We therefore examine which specific configuration keys experience systematic value-level modifications, as these differences may carry discriminative signals that reflect intentional functional transformations. To do this, we analyze the value-level variance of configuration keys, evaluating whether these continuous or categorical shifts can serve as distinct semantic fingerprints for metadata field prediction.

\head{Approach}\\
Here, we analyze configuration keys whose values change between parent and child models in order to determine their potential for distinguishing intentional reuse from unmodified duplication. We follow these steps:

\begin{itemize}
    \item \textbf{Step 4.2.1 - Identify valid parent-child pairs:} We utilized the verified dataset of parent--child PTLM pairs established in Step 4.1.1 for this analysis.

    \item \textbf{Step 4.2.2 - Configuration flattening and key normalization:} We applied the same configuration flattening and index-stripping techniques detailed in Step 4.1.2.

    \item \textbf{Step 4.2.3 – Differential analysis of parent–child pairs:}
    We perform a systematic, key-by-key comparison between the normalized configuration files of parent and child models. For each key, we classify its relationship into one of four states:
    \begin{itemize}
        \item \textbf{Same:} The configuration key exists in both models and has an identical value.
        \item \textbf{Different:} The configuration key exists in both models but the value differs, indicating a potential reuse or tuning.
        \item \textbf{Added to child:} The configuration key exists only in the child model.
        \item \textbf{Removed from child:} The configuration key exists only in the parent model.
    \end{itemize}
    
    We explicitly exclude specific administrative and environment metadata keys from this behavioral comparison. While keys such as \texttt{\_name\_or\_path} (utilized earlier for lineage extraction) and \texttt{transformers\_version} provide highly useful provenance and compatibility information, they appear with overwhelmingly high frequency across configuration files and almost always experience automated value modifications during the standard model saving and uploading process. 
    
    Consequently, including them would introduce a disproportionate statistical bias toward routine operations, drowning out the signals of true functional or architectural reuse. Similarly, keys like \texttt{\_commit\_hash} are purely administrative versioning artifacts with no bearing on the model's actual functional behavior. Filtering these three specific keys allows us to eliminate obvious, uninformative noise, focusing our analysis on potential derivation signals rather than the parameter values of administrative keys that do not disclose structural links between parents and children. In total, this step compares the value-level variance of 6,026 unique normalized configuration keys across parent–child model pairs.

    \item \textbf{Step 4.2.4 - Signal prioritization and scoring:} To identify the most significant signals of reuse, we aggregated the occurrence statistics for each unique configuration key. Since raw probability rates can be misleading for a rarely used configuration key, where a key appearing in only 1 parent model and experiencing a value modification in its corresponding child model yields a perfect but statistically insignificant rate of 1.0, we calculated a single \textbf{Weighted Score} ($S_w$) for each key using Equation \ref{eq:weighted_score}:

    \begin{equation}
        S_{w} = R_{\text{change}} \times \ln(N_{\text{total}})
    \label{eq:weighted_score}
    \end{equation}

    where $N_{\text{total}}$ is the total count of parent-child pairs in which the key was relevant (i.e., present in the parent, child, or both), and $R_{\text{change}}$ is the proportion of those instances where the key underwent any functional modification (i.e., its value was changed, added to the child, or removed from the child) rather than remaining identical. The natural logarithmic term balances this scoring: it dampens the influence of raw frequency to prevent common configuration keys from dominating, while still penalizing sparse data to filter out small-sample noise, consistent with standard term-weighting practices in information retrieval \citep{manning2008introduction}. The resulting score $S_w$ ranges from 0 to approximately 11.9 (i.e., $\ln(150,951)$). A high value indicates a key that is both widely adopted in the ecosystem and frequently modified during reuse, serving as a strong overall signal of changing reuse.

    \item \textbf{Step 4.2.5 - Visualization of reuse signals:} Finally, we ranked the configuration keys by their unified weighted score ($S_w$) in descending order. We visualized the top 30 keys using a bar chart generated with \texttt{matplotlib} and \texttt{seaborn}. This visualization highlights the configuration keys most prone to modification during the reuse process.
\end{itemize}

\begin{figure*}[t]
\centering
\includegraphics[width=\textwidth]{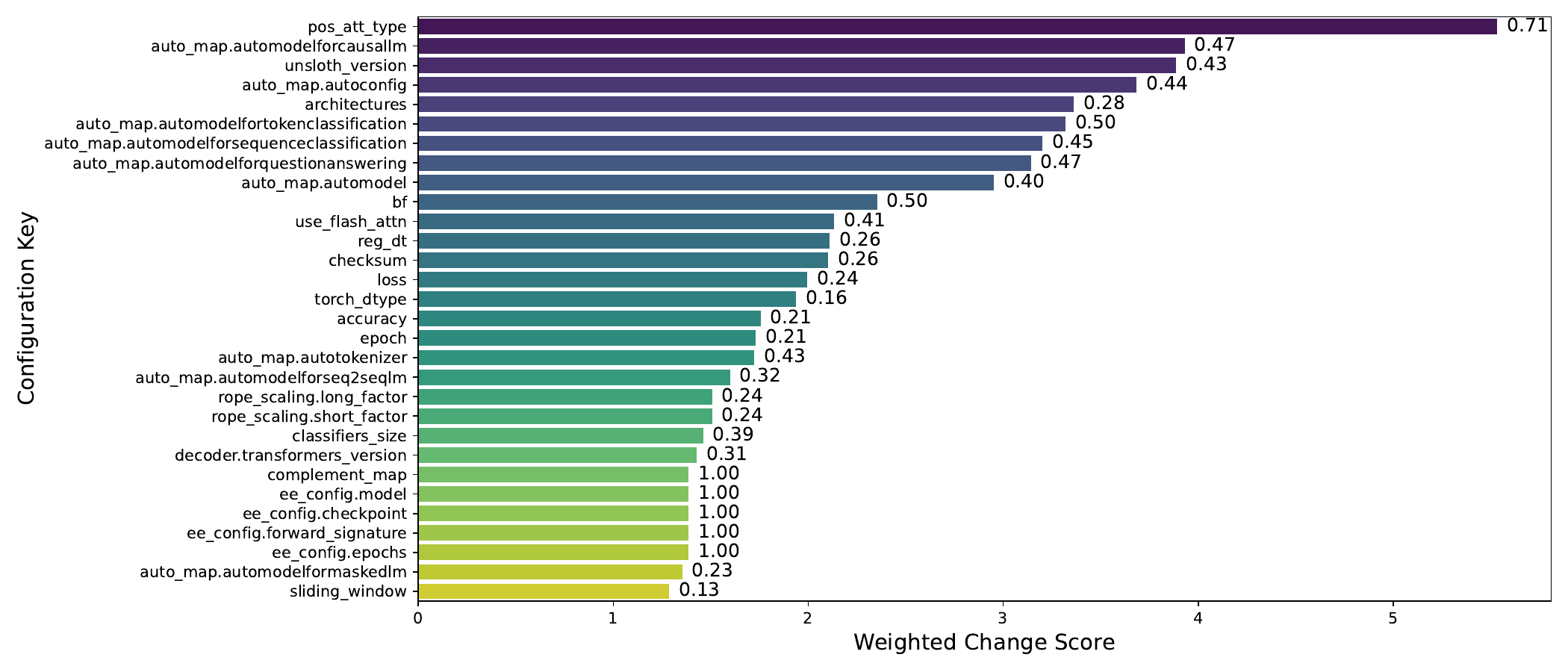}
\caption{Top 30 configuration keys with the highest weighted change scores ($S_w$) between parent and child PTLMs on Hugging Face. The x-axis represents the weighted change score, while the y-axis lists the configuration keys. Bars are annotated with the corresponding overall change rates ($R_{change}$) observed across parent–child model pairs, accounting for all modification states (i.e., different, added, and removed).}
\label{key_value_changes}
\end{figure*}

\head{Result}\\
\textbf{Configuration keys with the highest weighted change scores ($S_w$) during PTLM reuse include those controlling internal learning behavior (e.g., \texttt{pos\_att\_type}), optimization framework versions (e.g., \texttt{unsloth\_version}), and model identity (e.g., \texttt{architectures}).} Of the 6,026 unique normalized configuration keys that we analyzed, only 324 keys (5.4\%) have any value change between parent and child models. Among these 324 changing keys, we rank them by $S_w$ and report the top 30 in \Cref{key_value_changes}. This ranking prioritizes the configuration keys with the most significant value divergence, meaning keys that are both frequent and prone to modification. Our analysis confirms that although 31.4\% of parent and child models on Hugging Face inherently share the same configuration schema, \Cref{key_value_changes} reveals that they frequently assign different values to these shared keys. Specifically, parent and child model relationships show substantial value divergence in attention-related settings such as \texttt{pos\_att\_type} in 71.2\% of pairs and framework integration metadata like \texttt{auto\_map}, which changes in approximately 40\% to 50\% of cases. We also see high weighted scores for \texttt{unsloth\_version} and \texttt{architectures}, which indicate that practitioners frequently update optimization tools or redefine model classes during the reuse process. We further observe systematic value shifts in training and inference configuration keys, including numerical precision (\texttt{torch\_dtype}: 16.3\%, \texttt{bf}: 49.6\%) and training duration (\texttt{epoch}: 21.0\%). 

Together, these results indicate that model reuse commonly involves specific key-level additions and removals rather than value modifications or unmodified duplication. This structural schema variance reveals that value changes are highly restricted across the ecosystem. Because the vast majority of shared parameters remain stable and parameter formats vary from continuous floats to arbitrarily nested lists, incorporating raw configuration values would cause an unsustainable explosion in feature space complexity. Consequently, this restricted empirical threshold establishes that the simple presence or absence of specific configuration keys provides the precise, lightweight technical signals necessary to automatically impute missing metadata, allowing us to classify and trace functional reuse across the ecosystem.

\subsection{Identifying configuration keys that uniquely define and differentiate model reuse methods} \label{section_4.3} 
So far, our analysis has established that configuration changes are predominantly driven by learning settings and hyperparameter optimization. However, distinguishing the specific reuse method is important, particularly because 52.7\% of the pre-trained language models in our study lack explicit metadata describing their reuse method. Identifying the configuration keys that never change is equally essential for understanding which structural properties remain constant across all forms of model reuse. While these invariant core keys carry no value changes, mapping them allows us to systematically filter them out as shared architectural noise, allowing our lightweight approach to focus on more discriminative configuration signals. 

While \Cref{section_4.1} established that child models primarily add new configuration keys to what they inherit from parents, \Cref{section_4.2} revealed that only 5.4\% of unique normalized configuration keys have any value change between parent and child models. This empirical finding justifies our decision to exclude the corresponding values of those configuration keys to prevent an explosion in the feature space. Therefore, this section identifies both the unique configuration keys that fingerprint specific reuse paradigms and the invariant core keys that persist across all reuse methods, isolating the configuration keys to serve as the lightweight feature space for SemFin.

\head{Approach}\\
To identify the configuration keys that uniquely characterize and differentiate each model reuse method (i.e., a method used to adapt the parent model), we first preprocess and normalize model configuration files to retain only semantically meaningful information. Specifically, we perform the following steps.
\begin{itemize}
    \item \textbf{Step 4.3.1 Preprocessing the configuration file:}  
    Unlike the general preprocessing used for identifying parent--child pairs, this phase filters the configuration files to isolate keys that uniquely identify specific reuse methods. First, we remove uninformative metadata fields from the configuration file, including commit hashes (e.g., \texttt{sha}), URLs, Git repositories, homepages, DOIs, and licenses. Next, following the same flattening and key normalization procedure defined in Step 4.1.2, we extract only configuration keys from nested configuration dictionaries and discard all associated values. Finally, we identify and remove universal architectural keys, such as \texttt{num\_heads}, \texttt{num\_layers}, and \texttt{attention\_dropout}, that are present across all reuse methods on Hugging Face. Because these universal keys, along with model-size indicators (e.g., \texttt{22B}, \texttt{34B}), represent the baseline architecture rather than the reuse behavior, they provide no discriminative signal for classifying how a model was modified. Stripping these universal keys ensures our analysis strictly captures the unique semantic fingerprinting of each reuse method.

    \item \textbf{Step 4.3.2 Identifying platform-level configuration keys across all reuse methods:} After preprocessing and normalizing configuration files using \emph{level-1 key normalization} (retaining only the first hierarchical level, e.g., ``model" from ``model.layers.attention"), we identify keys that are common across all model reuse methods (i.e., techniques used on parent model to produce the child model). We employ Level-1 normalization to ensure a robust intersection of features across diverse architectures. For example, while Level-2 keys such as \texttt{rope\_scaling.low\_freq\_factor} and \texttt{rope\_scaling.high\_freq\_factor} provide granular configuration details, they fundamentally represent the presence of the same configuration object. If we were to use Level-2 keys at this stage, the intersection would effectively fracture; a model defining only \texttt{low\_freq\_factor} and another defining only \texttt{high\_freq\_factor} would appear to have no commonality. This fragmentation would defeat our goal of establishing a common keys (i.e., keys that are common across all the reuse methods), causing us to falsely retain standard architectural components instead of successfully filtering them out. For each reuse method, we collect the set of normalized configuration keys appearing in its associated models. We consider a key present in a category if it appears at least once in a model configuration file belonging to that category. We then compute the intersection of these key sets across all reuse methods. Any key that appears in every reuse method is classified as a platform-level configuration key. 
    
    Finally, the first and second authors collaboratively performed a manual thematic analysis of these 68 platform-level configuration keys. Using the functional definitions provided in the official Hugging Face configuration documentation\footnote{https://huggingface.co/docs/transformers/main\_classes/configuration} as ground truth, we reached a consensus to group functionally similar keys into three synthesized categories: \textit{Structural Dimensions} (keys defining the physical shape, depth, and tensor sizes of the network graph, such as layer counts and hidden dimensions), \textit{Architectural Behaviors} (keys governing internal computational dynamics and mathematical operations, such as activation functions, dropout probabilities, and normalization constants), and \textit{Tokenization \& Metadata} (keys managing the input/output interface and administrative framework integration, such as vocabulary mapping, special token identifiers, and library versions).

    \item \textbf{Step 4.3.3 Identify configuration keys uniquely defining each reuse method:} Finally, we identified keys that appear exclusively in a single reuse method, such that the presence of any of these keys in a model's configuration directly indicates its reuse method. For each category, we counted the frequency of these unique keys in child model configurations and selected the top 10 most frequent ones. 
\end{itemize}

\head{Result}\\
\noindent \textbf{A core set of 68 configuration keys constitutes the platform-level configuration keys shared across all reuse methods, although the specific values of these keys can vary between models.} \Cref{invariant_keys} shows the intersection of configuration keys present in every reuse method analyzed. Despite diverse reuse methods, including fine-tuning, quantization, and distillation, these 68 keys consistently appear across all parent--child model pairs. While specific keys such as \texttt{num\_attention\_heads} are inherent to the Transformer architecture dominant in our dataset, they represent mandatory structural keys required to instantiate model configuration files within the transformers framework. Because these keys are consistently present across all reuse methods, their structural invariance, meaning their presence rather than value invariance, demonstrates that they form the foundational configuration required to instantiate models within the ecosystem. 

Mapping this invariant core serves as an essential preprocessing step to identify which configuration keys can be filtered out as shared architectural noise. By isolating and excluding this common baseline, our lightweight approach can focus its attention entirely on the presence of the remaining discriminative configuration keys. As detailed in our subsequent findings, these reuse-specific configuration keys appear exclusively within individual modification types, providing the distinct structural signals necessary to identify the reuse method without requiring value-level comparisons.

\begin{table*}[t]
\centering
\caption{The 68 ``platform-level" configuration keys present across all analyzed model reuse methods, maintaining their structural presence, though not necessarily identical values, during model reuse.}
\label{invariant_keys}
\vspace{4pt}
\resizebox{\textwidth}{!}{%
\begin{tabular}{p{0.3\textwidth} p{0.65\textwidth}}
\toprule
\textbf{Category} & \textbf{Configuration Keys} \\
\midrule
\textbf{Structural Dimensions} & hidden\_size, num\_hidden\_layers, num\_attention\_heads, num\_key\_value\_heads, num\_heads, intermediate\_size, head\_dim, d\_model, d\_ff, d\_kv, num\_decoder\_layers, num\_layers, max\_position\_embeddings, max\_window\_layers, sliding\_window, use\_sliding\_window, num\_experts\_per\_tok, architectures, torch\_dtype \\
\midrule
\textbf{Architectural Behaviors} & activation, hidden\_act, dense\_act\_fn, hidden\_activation, is\_gated\_act, hidden\_dropout\_prob, attention\_probs\_dropout\_prob, dropout\_rate, attention\_dropout, hidden\_dropout, classifier\_dropout, layer\_norm\_epsilon, layer\_norm\_eps, rms\_norm\_eps, rope\_scaling, rope\_theta, rotary\_emb\_base, rotary\_pct, partial\_rotary\_factor, position\_embedding\_type, initializer\_range, initializer\_factor, use\_cache, gradient\_checkpointing, is\_encoder\_decoder \\
\midrule
\textbf{Tokenization \& Metadata} & vocab\_size, type\_vocab\_size, pad\_token\_id, bos\_token\_id, eos\_token\_id, decoder\_start\_token\_id, tie\_word\_embeddings, auto\_map, transformers\_version, \_name\_or\_path, problem\_type \\
\bottomrule
\end{tabular}%
}
\end{table*}

\noindent \textbf{Beyond the 68 platform-level configuration keys, each reuse method introduces a varying number of specialized configuration keys, ranging from 0 to 354 depending on the reuse method’s complexity.} \Cref{unique_keys} lists these keys for each reuse method. Fine-tuning introduces the largest set of unique keys (354), far exceeding the count of the platform-level core, reflecting its broad scope of tasks, for example, \texttt{n\_topics} corresponds to topic modeling, and \texttt{mm\_connector\_cfg} to multimodal reuse. In contrast, optimization-focused categories introduce more targeted sets: Quantization (77 keys) includes keys such as \texttt{quant\_strategy} and \texttt{is\_bitnet\_config}, indicating compression-specific configurations; Model Merging (13 keys) relies on provenance-tracking keys like \texttt{constituent\_models} and \texttt{merged\_models}; Pruning (71 keys) is characterized by specialized keys such as \texttt{hook\_point\_layer}. PEFT (27 keys) and Distillation (9 keys) have smaller sets of unique keys, while Deduplication introduces no unique keys, confirming it is a purely data-level operation. These exclusive keys provide a reliable basis for identifying a model’s reuse method.

\begin{table*}[t]
\centering
\caption{Distinctive configuration keys that appear exclusively within the specified reuse method, suggesting that the presence of such keys may help identify the reuse method.}
\label{unique_keys}
\vspace{4pt}
\resizebox{\textwidth}{!}{%
\begin{tabular}{l c p{0.65\textwidth}}
\toprule
\textbf{Reuse Method} & \textbf{Unique Keys} & \textbf{Top-10 Distinctive Key Examples} \\
\midrule
\textbf{Finetune} & 354 & n\_topics, qllama\_config, mm\_connector\_cfg, head\_wise\_ranks, llm\_model, llm\_weight, backend\_type, thread\_num, memory, num\_telemetry\_features \\
\midrule
\textbf{Quantization} & 77 & test\_set, quant\_strategy, visual\_tokenizer\_config, tokenizer\_config, prompt\_format\_dict, turbomind, is\_bitnet\_config, training\_config, repo\_type, prompt\_wrapper \\
\midrule
\textbf{Pruning} & 71 & model\_class\_name, hook\_point\_layer, layer\_subtype, hook\_point\_head\_index, use\_cached\_activations, use\_patches\_only, cached\_activations\_path, d\_in, activation\_fn\_str, activation\_fn\_kwargs \\
\midrule
\textbf{PEFT} & 27 & log\_history, additional\_pos\_embed, peft\_model\_id, is\_flat, objective\_type, plm\_name\_or\_path, prefix\_dropout\_prob, use\_layer\_dep, scale\_dropout, best\_metric \\
\midrule
\textbf{Merge} & 13 & wav2vec2model, sources, constituent\_models, architectural\_design, string\_config, f\_dropout, merged\_models, python\_version, deep\_learning\_framework, capabilities \\
\midrule
\textbf{Distillation} & 9 & filter\_disabled, filter\_interval, filter\_nonlinear, filter\_output\_dim, train\_filters, n\_decoder\_layers, similarity\_metric, n\_key\_value\_heads, id1abel \\
\midrule
\textbf{Deduplication} & 0 & \textit{None} \\
\bottomrule
\end{tabular}%
}
\end{table*}

\begin{Summary}
{Summary of RQ$_1$ Findings}{
\begin{itemize}
\item Configuration keys and repository tags exhibit strong inheritance, with child models retaining 98.2\% of parent configuration keys and 88.4\% of parent repository tags. This high retention means these stable keys and tags can be ignored, leaving a much smaller set of newly added keys and tags to consider, whose correlation with target metadata labels suggests their utility for imputing missing metadata.

\item Configuration keys covering internal learning behavior (e.g., \texttt{pos\_att\_type}) and framework compatibility (e.g., \texttt{auto\_map}) frequently differ in terms of their value between parent and child models (e.g., 71.2\% for \texttt{pos\_att\_type}), providing potential reuse signals for automated metadata imputation.

\item Child models extend their schemas during reuse by adding quantization or task-specific keys while removing runtime generation parameters, suggesting that the presence and absence of specific keys, such as quantization\_config.bits, problem\_type, and dense\_act\_fn can differentiate reuse methods and pipeline tags.

\item While a core set of 68 invariant configuration keys is shared across all models, individual reuse methods introduce up to 354 unique keys, providing strong discriminative signals for identifying specific reuse methods.
\end{itemize}}
\end{Summary}

\section{The SemFin Approach}\label{semfin}
Recent initiatives for AIBOMs highlight a critical need for lightweight, automated methods to verify model provenance, a task currently hindered by pervasive missing metadata \citep{rajbahadur2025building}. While a model's learned weights constitute its functional core, deriving lineage from multi-gigabyte binary tensors is computationally prohibitive at ecosystem scale, and relying on unstructured free-text model cards is notoriously unreliable. Instead, configuration files offer an optimal analytical trade-off: they are lightweight, model format-agnostic, and functionally coupled to the model, acting as verifiable intrinsic artifact that developers cannot arbitrarily alter without breaking the model's executability.

Furthermore, in \Cref{RQ1_result} we found that changes to configuration file keys and values have a predictable relationship with how models are reused. \Cref{section_4.1} established that child models inherit 98.2\% of configuration keys and 88.4\% of repository tags, which suggests that model reuse is an incremental and structured process that preserves a stable, learnable semantic fingerprint. Our findings in \Cref{section_4.2} and \Cref{section_4.3} further show that while 324 keys are identified via value changes between parent and child models, we exclude their corresponding values and focus strictly on the presence of these 324 keys to provide specific structural signals. Concurrently, by mapping and filtering out the 68 platform-level configuration keys as shared architectural noise, our approach isolates a restricted feature space of remaining discriminative configuration keys to distinguish between parent and child models.

While resolving all the AIBOM fields remains an extensive open challenge, we propose SemFin as a foundational stepping stone. By aggregating the most discriminative configuration keys and repository tags across reuse methods into a restricted feature space, SemFin explores the feasibility of predicting specific subsets of missing metadata. The choice to predict these five metadata fields (pipeline tag, reuse method, license, model type, and library name) is part of our empirical study design to establish a direct benchmark with existing state of the art literature \citep{horwitz2025we}. This provides a scalable baseline that future generalized AIBOM frameworks can expand upon.

\head{Approach}\\
The design of SemFin, instructed by the results of \Cref{RQ1_result}, involves the following steps:

\begin{figure*}[t]
\centering
\includegraphics[width=\textwidth]{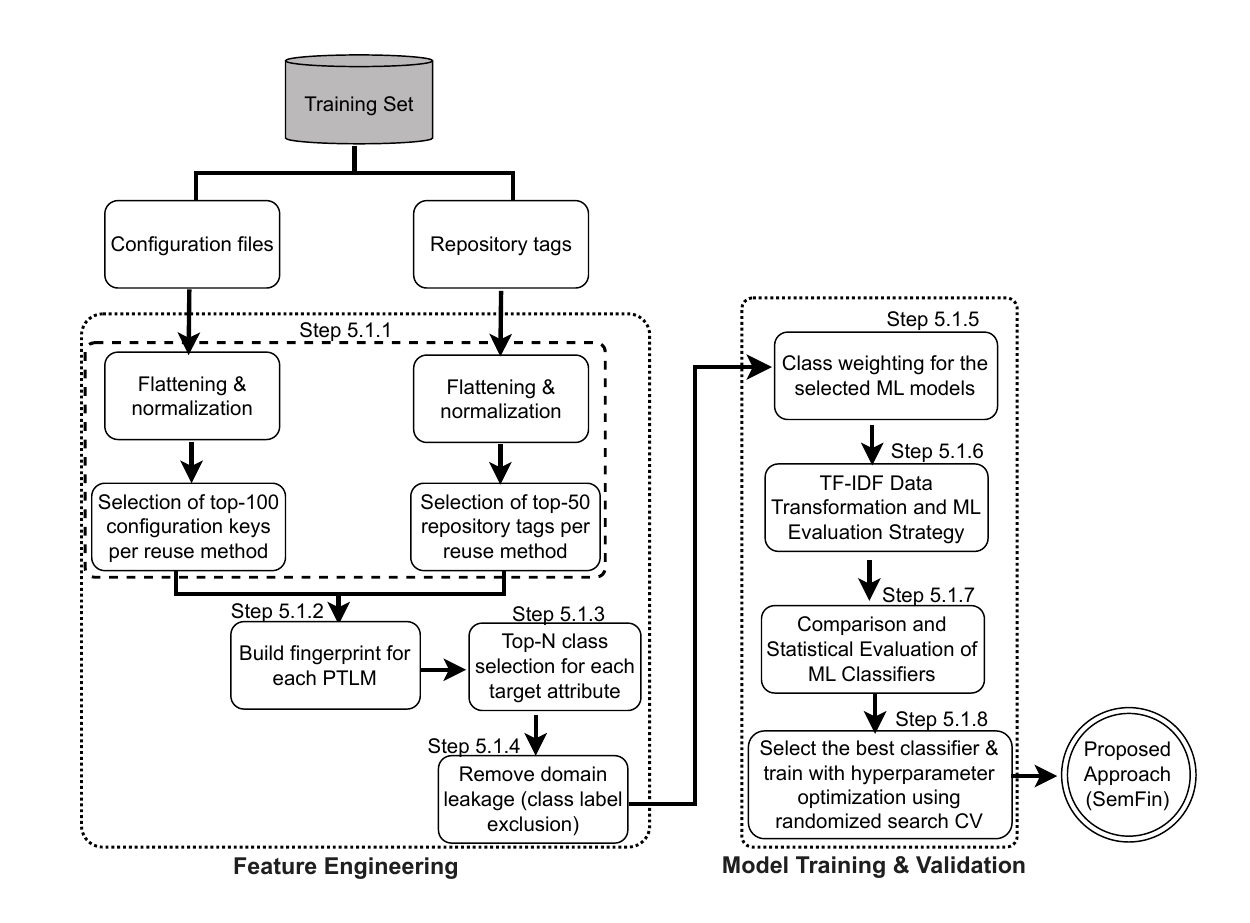}
\caption{Schematic overview of the SemFin approach.}
\label{semfin_technique}
\end{figure*}

\begin{itemize}
    \item \textbf{Step 5.1.1 - Filtering and normalizing configuration keys and HF repository tags:} To construct robust semantic fingerprints from heterogeneous and noisy tag and configuration data, we implemented a multi-stage normalization pipeline to focus on the changes in their presence in configuration keys and repository tags that identify how a model was modified while addressing the issue of missing or inconsistent metadata fields. Because our results in Section 4.1 show that an overwhelming 294,826 unique tags and 1,077 configuration keys are introduced exclusively by child models, relying on these unfiltered configuration keys and repository tags would lead to an explosion in the feature space and extreme sparsity that needs to be reduced.
    \begin{itemize}
        \item \textbf{Feature vocabulary construction from tags:} First, we performed noise reduction by removing administrative metadata, including ISO~639-1 language codes\footnote{https://www.iso.org/iso-639-language-code} (e.g., \textit{en, zh}), administrative prefixes (e.g., \textit{license:, arxiv:, doi:}), and numerical patterns indicative of model size (e.g., 7b, 13m), to prevent overfitting to specific numeric parameter counts. For key--value tags (e.g., \textit{dataset:wikitext}), we retained only the key prefix (e.g., \textit{dataset}), aggregating all such variation into a single structural feature representing the presence of that metadata category. To address the long-tail distribution of user-provided tags and prevent feature sparsity, we applied stratified, reuse-method-aware feature selection. For each reuse method, we identified the top-50 most prevalent repository tags within that specific category, ensuring that frequent signals for every reuse method were captured while discarding idiosyncratic, non-generalizable tags. We then constructed the final feature vocabulary by taking the mathematical union of these lists across all categories. This union automatically collapsed overlapping tags that appeared in multiple reuse methods into a single unique feature, yielding 157 unique tag features rather than the theoretical maximum of 400 (8 reuse methods $\times$ 50 tags). Finally, for each parent--child pair, we retained only the repository tags present in this unified vocabulary, removed duplicate tag strings, and formatted the remaining tags as a single comma-separated string for downstream vectorization.

        \item \textbf{Feature vocabulary construction from configuration keys:} First, we applied normalization and truncation. Following hierarchical flattening, we truncated each configuration key to retain only the first two hierarchy levels (e.g., \textit{model.layers.attention} $\rightarrow$ \textit{model.layers}), reducing dimensionality while preserving high-level architectural structure.

        Second, we performed administrative and generic-key removal by filtering out non-architectural keys, including version control metadata (e.g., commit\_hash), citation identifiers, and numerical model-size patterns (e.g., 7b, 34b). These keys were easily identifiable because they either contained prefixes like ``commit\_", ``arxiv\_", or ``doi\_", matched known framework metadata documented in Hugging Face Transformers, or consisted solely of numerical patterns with ``b" or ``m" suffixes, resulting in removal of 35 unique keys. This matches our findings in \Cref{section_4.2} and \Cref{section_4.3}, where we systematically excluded these same categories during differential analysis and key addition and removal analysis. For repository tags, we did not perform common-tag exclusion because tags lack the equivalent of platform-level configuration keys. Unlike configuration files, which contain mandatory structural keys present across all models, tags are optional user-supplied descriptors with no universal baseline to exclude.
        
        Third, we performed common-key exclusion. While \Cref{section_4.3} identified a broad invariant core of 68 configuration keys based purely on ecosystem presence, translating these exploratory findings into an optimized machine learning feature space requires a stricter operational threshold to avoid eliminating potentially useful sparse signals. Therefore, as a design choice for feature selection, we retained only those configuration keys appearing in at least 50\% of individual model repositories within each individual reuse method category. We then calculated the intersection of these per-method key sets, meaning a configuration key was selected for removal only if it appeared in the majority of model repositories across every reuse method. This strict majority threshold refined the platform-level core down to 13 highly ubiquitous keys. Removing these 13 common keys ensures the feature space strictly captures discriminative variations rather than shared architectural similarities.

        Finally, we performed category-balanced feature selection. Because \Cref{section_4.3} demonstrated that the number of specialized configuration keys varies drastically across reuse methods, ranging from 0 to 354, selecting the most frequent keys on a global, dataset-wide level would inherently bias the model toward reuse methods with larger configuration schemas. To guarantee fair representation and capture the unique signals of every category, we extracted the top 100 most frequent keys from each individual reuse method. This localized extraction ensures that the unique configuration keys characterizing smaller schemas, which would be lost in a global ranking, are successfully preserved within our final feature space. For reuse methods with fewer than 100 total keys, such as Merge with 13 keys or Distillation with 9 keys, we simply took all available keys. By taking the union of these category-specific sets, we constructed our final, balanced vocabulary of 323 unique configuration features.
    \end{itemize}

    \item \textbf{Step 5.1.2 - Merging repository tags and configuration keys as features into fingerprints:} To capitalize on our finding that repository tags and configuration keys provide complementary signals, we generated a feature string for each individual model repository by concatenating the filtered configuration keys and the repository tags into a single unified text string. Fusing these elements together allows SemFin to maintain predictive power even when one source is incomplete or missing. While our findings in \Cref{section_4.2} showed that configuration value changes carry potential reuse signals, we deliberately excluded the corresponding values of those configuration keys from our fingerprints. This decision was informed by our finding that only 5.4\% of unique normalized configuration keys experience any value change between parent and child models. Because these configuration values vary drastically across models, ranging from continuous floats to arbitrarily nested lists, incorporating them would cause an explosion in the feature space and lead to extreme sparsity. By representing models purely through the presence or absence of these configuration keys alongside their repository tags, we ensure that the combined predictive signals are presented to the classifier in a lightweight format.

    \item \textbf{Step 5.1.3 - Selecting the top-N classes:} Some dependent variables (the metadata fields we are trying to impute) exhibit an extreme long-tailed distribution, with many rare classes. For example, library\_name contains 79 classes, pipeline\_tag 15 classes, license 64 classes, and model\_type 990 classes. To reduce sparsity and focus on the most representative categories, we applied a rule-based approach to select the top-N classes for each variable. We chose the threshold of 1,000 occurrences to capture classes that appear frequently enough to provide reliable training signal, and the threshold of 100 to capture moderately frequent classes that still offer predictive value without introducing excessive sparsity. Specifically, we first computed the frequency of each class within each target metadata fields. If ten or more classes had a count of at least 1,000, we retained only those high-volume classes. Otherwise, we kept all classes with a count of at least 100. If no classes met either threshold, meaning all classes were rare, we retained the top 10 most frequent classes as a fallback. In all cases, any class not explicitly retained by the triggered criterion was recoded as an ``other" class. After including the ``other" category, we retained 11 classes for license, 32 classes for model\_type, 11 classes for library\_name, 10 classes for pipeline\_tag, and all 8 classes for reuse\_method. In each dependent variable, the ``other" category did not rank among the top two most prevalent classes, showing that it does not disproportionately influence the analysis.

    \item \textbf{Step 5.1.4 - Removing domain leakage for the empirical evaluation:} To prevent the model from memorizing explicit metadata fields rather than learning underlying structural signals, we removed any features from the semantic fingerprints that directly matched the metadata fields being predicted. This step is essential for any application of SemFin, regardless of which metadata are being imputed, because it ensures the model generalizes to new models where those explicit values are absent.

    We first aggregated all unique classes from the target metadata fields that the user wishes to predict. We converted these classes to lowercase and created a filter list. We then tokenized each semantic fingerprint and automatically removed any token that exactly matched a value on this filter list. For example, if a user is predicting pipeline tags and ``text-generation appears as a the value of a metadata field, our automated process removes the exact word ``text-generation" from the fingerprint, but retains structural variations such as ``text-generation-inference" that do not constitute a direct match. This automated filtering process is applied uniformly across all models in the training set.
    
    Across the metadata fields that we target (pipeline tag, library name, reuse method, license, and model type), the union of all removed metadata values across all target fields totaled 70 unique features, such as bart, finetune, translation, and exl2. By removing these direct value matches, we prevent the model from relying on memorized answers and ensure it learns genuine structural signals that generalize to unseen models.

    \item \textbf{Step 5.1.5 - Machine Learning (ML) model selection and class imbalance:} To robustly predict the five target metadata, we adopted a multi-classifier framework in which a separate model was trained for each dependent variable. 
    \begin{itemize} 
        \item \textbf{Metadata-specific target filtering:} While our previous vectorization steps resulted in complete, fully numerical semantic fingerprints (our input features), our broader training set still contains partially labeled instances where the \textit{target metadata fields} itself is missing. Because our top-N class selection (Step 5.1.3) specifically recoded only rare \textit{known} classes to \texttt{other}, instances entirely lacking metadata remained labeled as missing. Therefore, prior to training each classifier, we applied target-specific filtering. For the specific dependent variable being modeled, we retained only training instances with a valid ground-truth label, explicitly removing instances where the target was \texttt{NaN}, \texttt{None}, \texttt{null}, or \texttt{unknown}. This design ensured that each classifier was supervised exclusively using high-quality observed metadata, while still preserving those models in the dataset to train other classifiers where their respective metadata was present.
        
        \item \textbf{Selected models and class weighting:} To ensure that results are not artifacts of a specific algorithm, we selected four classifiers representing distinct learning families, implemented using the scikit-learn and lightgbm Python libraries. Given that metadata in any application of SemFin may exhibit class imbalance, we evaluated a ``Standard" (unweighted) and a ``Balanced" variant for our tree-based models by assigning class weights inversely proportional to class frequencies. We also employed distance-based weighting for our k-NN classifier. The chosen algorithms and their balancing configurations are:
        \begin{enumerate}
            \item \textbf{Random Forest (RF):} A parallel ensemble of decision trees. We utilized a configuration with 200 estimators and square-root feature sampling ($\sqrt{n\_features}$) based on standard thresholds for convergence and decorrelation \citep{breiman2001random, oshiro2012many}. For the balanced variant, we applied the \texttt{class\_weight='balanced'} parameter within the \texttt{scikit-learn} implementation.
            \item \textbf{LightGBM (LGBM):} A gradient-boosting framework optimized for efficiency. We utilized 200 estimators and a learning rate of 0.05, incorporating L1 and L2 regularization to prevent overfitting to rare tokens in the sparse TF-IDF space \citep{hastie2009elements}. For the balanced variant, we applied the \texttt{is\_unbalance=True} parameter within the native \texttt{lightgbm} API to automatically adjust penalization according to class frequencies.
            \item \textbf{Bagging Ensemble:} We implemented a Bagging classifier using Random Forest as the base estimator (10 RF estimators with bootstrap aggregation) to evaluate ensemble diversity.
            \item \textbf{k-Nearest Neighbors (k-NN):} We included this approach as a non-parametric baseline, evaluating $k=1$ (relying on the single most similar model) and $k=5$ (capturing a majority vote from a small local neighborhood) using Cosine Similarity to capture the semantic orientation of vectors regardless of document length \citep{schutze2008introduction}.
        \end{enumerate}
    \end{itemize}

    \item \textbf{Step 5.1.6 - TF-IDF data transformation and ML evaluation strategy:} To rigorously evaluate model stability within the training partition, we employed Stratified 5-Fold Cross-Validation. A critical component of this pipeline was the prevention of data leakage during feature extraction. Unlike standard pipelines that vectorize the entire dataset prior to splitting, we fitted a new TF-IDF vectorizer exclusively on the training indices of each cross-validation fold. This fitted vectorizer was then used to transform both the training data (for model training) and the validation data (for evaluation within that fold), ensuring that the validation data did not influence the vocabulary or term frequencies. We strictly adhered to this protocol, ensuring that the global vocabulary and term frequencies were derived solely from the training data within each fold. After cross-validation, we fitted a final TF-IDF vectorizer on the complete training set. This final vectorizer is saved alongside the trained model and is applied to any new model when making predictions, ensuring that unseen data is transformed using the same feature space as the training data.
 
    \item \textbf{Step 5.1.7 - Selection of the best machine learning classifiers:} Following the cross-validation procedure, we conducted a comprehensive performance analysis to benchmark the four algorithms (RF, LightGBM, Bagging, and k-NN). To ensure a rigorous comparison, we utilized standardized hyperparameters selected to ensure performance stabilization rather than relying on arbitrary library defaults. Specifically, we fixed the number of estimators at 200 for both tree-based ensembles. This threshold was selected based on evaluating the trade-off between computational cost and error reduction, consistent with findings by \citep{oshiro2012many}, which demonstrate that tree-based ensemble performance typically stabilizes well before this threshold. This ensured that the comparative analysis reflected the models' true predictive capabilities rather than insufficient training. We compiled the results into summary tables reporting two key metrics computed using the \texttt{scikit-learn} library: Micro-Accuracy (MIA) and Macro-Accuracy (MAA). We define MIA as the global aggregation of all correct predictions across all classes, heavily reflecting the model's performance on the dominant majority classes. Conversely, we define MAA as the unweighted mean of the individual accuracies computed independently for each class; this metric treats all classes equally, making it highly sensitive to the model's ability to correctly classify rare, minority categories. 
    
    To rigorously assess whether performance differences between classifiers were statistically significant rather than artifacts of random sampling variation, we applied McNemar's test. McNemar's test is a paired, non-parametric statistical test designed for comparing two classifiers evaluated on the same dataset. It is widely recommended for comparing machine learning models in paired settings and has been frequently used in software engineering research to compare predictive methods. To apply this statistical framework, we conducted this test systematically for each prediction model. For each dependent variable, we ranked the algorithms (Random Forest, LightGBM, Bagging, k-NN) by performance and compared the top-ranked model against the second-best model using their paired prediction vectors. We focused exclusively on instances where the two models disagreed: cases correctly predicted by the top model but misclassified by the runner-up, and vice versa. We tested the null hypothesis that both classifiers have equal error rates using the chi-squared approximation with continuity correction. 
    
    To further quantify the practical magnitude of the difference, we calculated the effect size using Cohen's $g$. While the p-value confirms the existence of a difference, $g$ measures the asymmetry of these disagreements relative to random chance. Following recent applications in machine learning evaluation, we assessed the effect size where values of 0.05, 0.15, and 0.25 correspond to small, medium, and large effects, respectively. Statistical significance was evaluated at $\alpha = 0.05$, with results annotated as $*$ ($p < 0.05$), $**$ ($p < 0.01$), and $***$ ($p < 0.001$). Note that this statistical comparison uses the predictions generated from the standardized hyperparameters established in Step 5.1.5. The separate hyperparameter optimization in Step 5.1.8 is applied only after selecting the winning algorithm, and serves a different purpose: maximizing the performance of the final deployed model rather than comparing algorithms fairly.

    \item \textbf{Step 5.1.8 - Training of Final Model:} To construct our final SemFin approach, we selected the algorithm that achieved the highest best performance from Step 5.1.7. For this best performed model, we performed hyperparameter optimization using Randomized Search Cross-Validation (10 iterations, 5-fold CV) on the training partition to identify the optimal configuration. We then instantiated new classifiers using these optimized parameters and retrained them on the entire training dataset. We also fitted a new TF-IDF vectorizer exclusively on the complete training dataset to ensure the final production vectorizer captures the full vocabulary and term frequencies from all available data, maximizing its representational power for inference on unseen models. These final trained models formally constitute our proposed SemFin tool.  
\end{itemize}

\section{\textbf{RQ$_2$:} \RQb} 
In Section 5, we operationalized the empirical findings from $RQ_1$ to design SemFin, a lightweight, artifact-driven approach for metadata imputation. While we have established that configuration files and repository tags encode rich, co-evolving structural signals, it remains an open question whether fusing these coupled features can practically outperform existing imputation methods. Current baseline techniques, such as the graph-based neighbor averaging and hub-based sibling voting proposed by \citet{horwitz2025we}, rely entirely on inferring metadata from adjacent or sibling models in the reuse graph. Consequently, they frequently fail or abstain when models are isolated or when the surrounding metadata is sparse. In this research question, we empirically validate the effectiveness of SemFin by directly comparing the performance against these propagation-based heuristics. We investigate whether our semantic fingerprints, constructed by fusing the model's configuration keys with its repository tags, can overcome the coverage limitations of the baseline heuristics and improve predictive accuracy, thereby providing a more robust foundation for automated AIBOM generation.

\begin{figure*}[t]
    \centering
    \includegraphics[width=\textwidth]{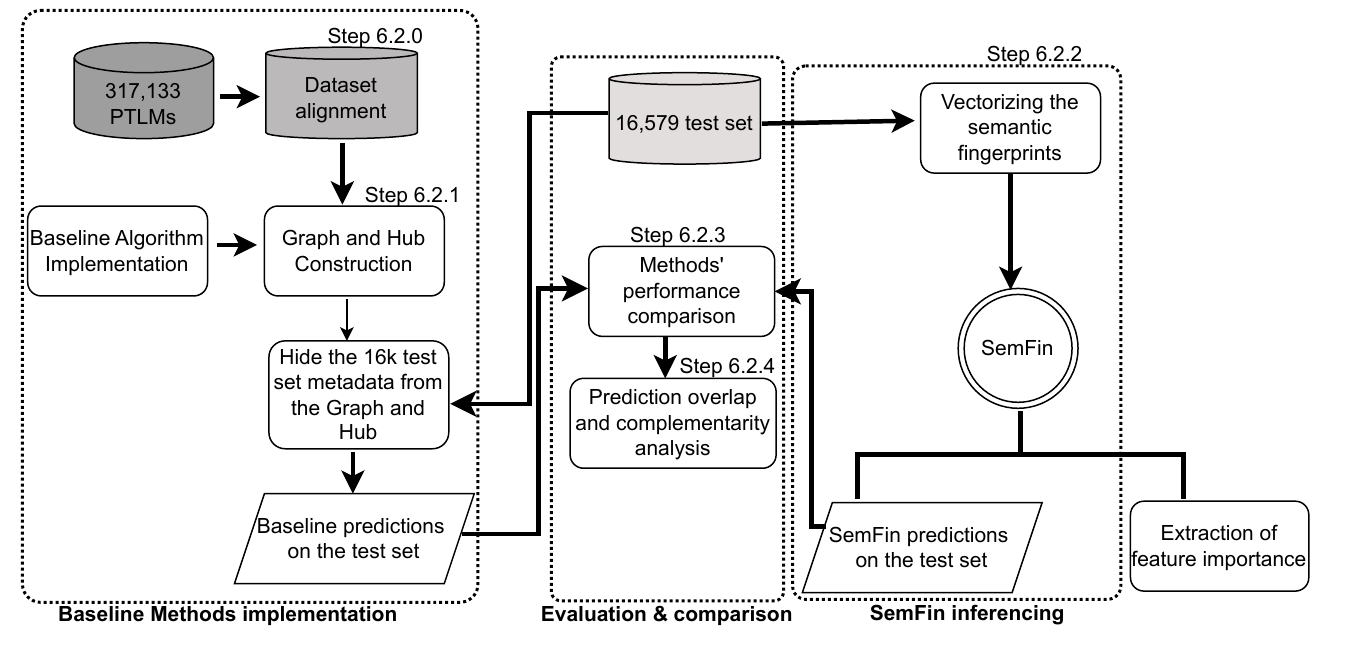}
    \caption{Overview of the experimental workflow for RQ2. The pipeline details the data preparation, the training and optimization of the SemFin models, and the comparative evaluation against Graph and Hub baseline heuristics.}
    \label{rq2_workflow}
\end{figure*}

\noindent\emph{Approach and Results}\\
To address this research question, we implemented a multi-stage experimental framework that spans data preparation, model training, and comparative evaluation against baseline heuristics. We illustrate the complete experimental flow for RQ2 in \Cref{rq2_workflow}. To maintain clarity across these distinct experimental phases, we again structure this section by presenting the specific \textbf{Approach} followed immediately by the corresponding \textbf{Results} for each sub-question.

\subsection{Evaluating model performance and assessing feature importance}\label{ml_comparison}
\head{Approach}\\
To evaluate the performance of the machine learning models used in this study, we report the results of the 5‑fold cross‑validation described in \Cref{semfin}. Having designed the SemFin feature training pipeline, we first present the comparative predictive performance of the candidate classifiers. This evaluation justifies our selection of LightGBM as the optimal classifier for the five metadata in our specific empirical setting, establishing it as the predictive core of SemFin in this study.

To justify the importance of feature fusion, we benchmarked three configurations of the best performing model: (i) trained exclusively on configuration keys, (ii) trained exclusively on repository tags, and (iii) trained on their fusion, all following the same steps in \Cref{semfin}. This ablation study quantifies the contribution of each signal type and validates the decision to combine both sources within SemFin.

To identify the specific tokens (derived from configuration keys and repository tags after fusion and TF‑IDF vectorization) that most strongly contributed to recovering missing model metadata, we analyzed feature importance scores for our selected model. We restricted this analysis exclusively to the top‑performing classifier for two methodological reasons: (i) as demonstrated in our performance evaluation, it yielded the highest predictive accuracy across all dependent variables, validating its role as the core engine for SemFin; and (ii) distance‑based baselines such as k-NN do not produce feature importance scores comparable to the split‑gain metrics provided by tree‑based ensembles.

Consequently, we extracted feature importance directly from the final, optimized LightGBM model. Specifically, we relied on native tree‑based feature importance scores, which estimate each TF‑IDF feature’s contribution to the model’s decisions based on impurity reduction during tree splits \citep{breiman2001random}. Because TF‑IDF assigns higher weights to rare but discriminative tokens \citep{sparck2004statistical,manning2008introduction}, these scores identify the tokens (configuration keys and repository tag components) that most strongly influence predictions. For interpretability, we mapped numerical feature indices back to their original tokens using the fitted TF‑IDF vectorizer, revealing which elements of the semantic fingerprint were most impactful for each task.

We report classifier performance using both macro‑averaged and micro‑averaged metrics. Macro‑averaged scores treat all classes equally, providing insight into performance on rare or underrepresented metadata labels, while micro‑averaged scores weight each individual prediction (instance) equally, reflecting overall predictive accuracy across the dataset. Reporting both averages ensures that our evaluation captures not only the overall predictive strength of SemFin, but also its robustness across classes of varying frequency.

\head{Result}\\
\noindent\textbf{LightGBM consistently achieves the strongest overall performance across all the predicted metadata, establishing it as the most effective SemFin model.} 
\Cref{fingerprint_standard} reports results under the original class distribution, while \Cref{fingerprint_balanced} presents class-weighted results that emphasize minority classes. In the standard setting, LightGBM achieves the highest micro accuracy (78.6\%--98.9\%) and the best macro accuracy across all five tasks. Under class balancing, macro accuracy improves notably, particularly for \textit{reuse\_method} (0.667 $\rightarrow$ 0.823) and \textit{library\_name} (0.785 $\rightarrow$ 0.847). Despite class imbalance challenges, \textit{model\_type} remains highly stable, maintaining $\geq$0.95 performance across both micro and macro metrics.

We further validate these findings using McNemar tests with Bonferroni correction ($\alpha = 0.0025$). For macro accuracy, LightGBM significantly outperforms the second-best models (5-NN or Random Forest) in all comparisons (all $p < 0.001$), with small-to-medium effect sizes ($g = 0.033$--0.294). Medium effects are observed for \textit{model\_type} and \textit{library\_name} ($g \geq 0.262$), indicating consistent gains in disagreement cases, while \textit{pipeline\_tag} and \textit{reuse\_method} show stable moderate effects ($g \approx 0.20$).

In contrast, micro accuracy differences under the balanced setting are not statistically significant after correction, despite 5-NN showing slightly higher values in several tasks. The only consistent exception is \textit{model\_type}, where both models are tied (p = 1.00). While 5-NN is competitive for majority-class prediction, it does not match LightGBM’s statistically superior macro performance, particularly for minority-class recovery.

Following established guidance on evaluation under class imbalance \citep{sokolova2009systematic}, macro accuracy is prioritized because SemFin targets underrepresented metadata by design. Overall, the results confirm that model choice significantly impacts performance, and they empirically justify selecting LightGBM for the remainder of the study. Although alternative models may perform competitively in specific settings, LightGBM provides the most reliable performance for minority-sensitive metadata prediction in our framework.

\begin{table*}[t]
\centering
\caption{Predictive performance of SemFin under the standard class distribution (MIA = Micro Accuracy, MAA = Macro Accuracy).}
\label{fingerprint_standard}
\resizebox{\textwidth}{!}{%
\begin{tabular}{lcccccccccc}
\toprule
\textbf{Metadata} &
\textbf{RF MIA} & \textbf{RF MAA} &
\textbf{LGB MIA} & \textbf{LGB MAA} &
\textbf{Bag MIA} & \textbf{Bag MAA} &
\textbf{1-NN MIA} & \textbf{1-NN MAA} &
\textbf{5-NN MIA} & \textbf{5-NN MAA} \\
\midrule
Pipeline Tag   & 0.901 & 0.651 & \textbf{0.934} & \textbf{0.779} & 0.847 & 0.506 & 0.879 & 0.709 & 0.918 & 0.753 \\
License         & 0.745 & 0.400 & \textbf{0.786} & \textbf{0.573} & 0.719 & 0.306 & 0.698 & 0.521 & 0.770 & 0.560 \\
Reuse Method & 0.891 & 0.524 & \textbf{0.914} & \textbf{0.667} & 0.878 & 0.412 & 0.870 & 0.560 & 0.896 & 0.628 \\
Model Type     & 0.968 & 0.931 & \textbf{0.978} & \textbf{0.955} & 0.950 & 0.891 & 0.959 & 0.928 & 0.970 & 0.947 \\
Library Name   & 0.981 & 0.442 & \textbf{0.989} & \textbf{0.785} & 0.968 & 0.262 & 0.974 & 0.610 & 0.985 & 0.591 \\
\bottomrule
\end{tabular}}
\end{table*}

\begin{table*}[t]
\centering
\caption{Predictive performance of SemFin after class balancing to emphasize minority classes (MIA = Micro Accuracy, MAA = Macro Accuracy).}
\label{fingerprint_balanced}
\resizebox{\textwidth}{!}{%
\begin{tabular}{lcccccccccc}
\toprule
\textbf{Metadata} &
\textbf{RF MIA} & \textbf{RF MAA} &
\textbf{LGB MIA} & \textbf{LGB MAA} &
\textbf{Bag MIA} & \textbf{Bag MAA} &
\textbf{1-NN MIA} & \textbf{1-NN MAA} &
\textbf{5-NN MIA} & \textbf{5-NN MAA} \\
\midrule
Pipeline Tag   & 0.907 & 0.826 & 0.905 & \textbf{0.838} & 0.884 & 0.788 & 0.879 & 0.709 & \textbf{0.923} & 0.770 \\
License         & 0.684 & 0.708 & 0.669 & \textbf{0.713} & 0.586 & 0.656 & 0.698 & 0.521 & \textbf{0.781} & 0.584 \\
Reuse Method & 0.779 & 0.807 & 0.773 & \textbf{0.823} & 0.726 & 0.748 & 0.870 & 0.560 & \textbf{0.900} & 0.644 \\
Model Type     & 0.969 & 0.967 & \textbf{0.973} & \textbf{0.972} & 0.945 & 0.949 & 0.959 & 0.928 & \textbf{0.973} & 0.951 \\
Library Name   & 0.973 & 0.826 & 0.967 & \textbf{0.847} & 0.933 & 0.805 & 0.974 & 0.610 & \textbf{0.987} & 0.646 \\
\bottomrule
\end{tabular}}
\end{table*}

\begin{table*}[t]
\centering
\caption{Statistical Comparison of Classifiers Using McNemar's Test (Standard class distribution.)}
\label{mcnemar_unbalanced}
\small
\begin{tabular}{@{}lccccc|ccccc@{}}
\toprule
& \multicolumn{5}{c}{\textbf{Micro Accuracy}} & \multicolumn{5}{c}{\textbf{Macro Accuracy}} \\
\cmidrule(lr){2-6} \cmidrule(l){7-11}
\textbf{Metadata} & \textbf{Best} & \textbf{2nd} & $\Delta$ & \textbf{p-val} & \textbf{\textit{g}} & 
\textbf{Best} & \textbf{2nd} & $\Delta$ & \textbf{p-val} & \textbf{\textit{g}} \\
\midrule
Pipeline Tag & LGB & 5-NN & 0.016 & $<$0.001*** & 0.209 & LGB & 5-NN & 0.026 & $<$0.001*** & 0.204 \\
License & LGB & 5-NN & 0.016 & $<$0.001*** & 0.080 & LGB & 5-NN & 0.014 & $<$0.001*** & 0.033 \\
Reuse Method & LGB & 5-NN & 0.018 & $<$0.001*** & 0.202 & LGB & 5-NN & 0.039 & $<$0.001*** & 0.088 \\
Model Type & LGB & 5-NN & 0.008 & $<$0.001*** & 0.265 & LGB & 5-NN & 0.008 & $<$0.001*** & 0.294 \\
Library Name & LGB & 5-NN & 0.005 & $<$0.001*** & 0.262 & LGB & 1-NN & 0.175 & $<$0.001*** & 0.266 \\
\bottomrule
\end{tabular}
\vspace{0.2cm} 
\parbox{\linewidth}{%
\footnotesize \textit{Note: LGB = LightGBM. Significance codes: *** $p<0.001$ after Bonferroni correction ($\alpha = 0.05/20 = 0.0025$). $\Delta$: Difference between best and 2nd best accuracy. $g$: Cohen's g Effect Size. All differences are statistically significant.}}
\end{table*}

\begin{table*}[t]
\centering
\caption{Statistical Comparison of Classifiers Using McNemar's Test (Balanced class distribution with class weighting.)}
\label{mcnemar_balanced}
\small
\begin{tabular}{@{}lccccc|ccccc@{}}
\toprule
& \multicolumn{5}{c}{\textbf{Micro Accuracy}} & \multicolumn{5}{c}{\textbf{Macro Accuracy}} \\
\cmidrule(lr){2-6} \cmidrule(l){7-11}
\textbf{Metadata} & \textbf{Best} & \textbf{2nd} & $\Delta$ & \textbf{p-val} & \textbf{\textit{g}} & 
\textbf{Best} & \textbf{2nd} & $\Delta$ & \textbf{p-val} & \textbf{\textit{g}} \\
\midrule
Pipeline Tag & 5-NN & RF & 0.016 & $<$0.001*** & 0.128 & LGB & RF & 0.012 & $<$0.001*** & 0.122 \\
License & 5-NN & 1-NN & 0.083 & $<$0.001*** & 0.229 & LGB & RF & 0.004 & $<$0.001*** & 0.044 \\
Reuse Method & 5-NN & 1-NN & 0.030 & $<$0.001*** & 0.181 & LGB & RF & 0.016 & $<$0.001*** & 0.183 \\
Model Type & 5-NN & LGB & 0.000 & 1.000 & 0.012 & LGB & RF & 0.006 & $<$0.001*** & 0.024 \\
Library Name & 5-NN & 1-NN & 0.012 & $<$0.001*** & 0.292 & LGB & RF & 0.022 & $<$0.001*** & 0.136 \\
\bottomrule
\end{tabular}
\vspace{0.2cm} 
\parbox{\linewidth}{%
\footnotesize \textit{Note: LGB = LightGBM, RF = Random Forest. Significance codes: *** $p<0.001$ after Bonferroni correction ($\alpha = 0.05/20 = 0.0025$). $\Delta$: Difference between best and 2nd best accuracy. $g$: Cohen's g Effect Size.}}
\end{table*}

\noindent\textbf{SemFin benefits from the combination of two complementary aspects of Hugging Face metadata, as neither configuration files nor repository tags alone are sufficient to resolve missing metadata.} As shown in \Cref{tab:lgb_comparison}, the fused representation (Tags + Config) consistently achieves the highest micro accuracy (MIA) across all attributes. For example, for \textit{pipeline\_tag}, MIA increases from 0.731 (Tags) and 0.836 (Config) to 0.934 (Fusion), while for \textit{license}, it increases from 0.634 and 0.738 to 0.786. Similar trends are observed for \textit{reuse\_category} (0.884 / 0.839 vs.\ 0.914), \textit{model\_type} (0.465 / 0.969 vs.\ 0.978), and \textit{library\_name} (0.982 / 0.961 vs.\ 0.989).

The comparison between individual sources shows that configuration-only models achieve higher MIA than tag-only models for \textit{pipeline\_tag} (0.836 vs.\ 0.731) and \textit{model\_type} (0.969 vs.\ 0.465), while tag-only models achieve higher MIA for \textit{reuse\_category} (0.884 vs.\ 0.839) and \textit{library\_name} (0.982 vs.\ 0.961). For \textit{license}, configuration-only also yields higher MIA (0.738 vs.\ 0.634). 

A similar pattern is observed for macro accuracy (MAA). For instance, in the balanced setting for \textit{library\_name}, MAA increases from 0.623 (Tags) and 0.657 (Config) to 0.847 (Fusion), while for \textit{pipeline\_tag}, it increases from 0.564 and 0.613 to 0.838. These results indicate that repository tags and configuration files capture different but complementary signals, and their combination leads to consistently higher and more balanced performance across both MIA and MAA.

\begin{table*}[t]
\centering
\caption{LightGBM Performance: Tags Only vs. Config Only vs. Tags+Config}
\label{tab:lgb_comparison}
\resizebox{\textwidth}{!}{%
\begin{tabular}{l|cccc|cccc|cccc}
\toprule
 & \multicolumn{4}{c|}{\textbf{Tags Only}} & \multicolumn{4}{c|}{\textbf{Config Only}} & \multicolumn{4}{c}{\textbf{Tags + Config}} \\
\cmidrule(lr){2-5} \cmidrule(lr){6-9} \cmidrule(lr){10-13}
 & \multicolumn{2}{c}{Standard} & \multicolumn{2}{c|}{Balanced} & \multicolumn{2}{c}{Standard} & \multicolumn{2}{c|}{Balanced} & \multicolumn{2}{c}{Standard} & \multicolumn{2}{c}{Balanced} \\
\cmidrule(lr){2-3} \cmidrule(lr){4-5} \cmidrule(lr){6-7} \cmidrule(lr){8-9} \cmidrule(lr){10-11} \cmidrule(lr){12-13}
\textbf{Attribute} & \textbf{MIA} & \textbf{MAA} & \textbf{MIA} & \textbf{MAA} & \textbf{MIA} & \textbf{MAA} & \textbf{MIA} & \textbf{MAA} & \textbf{MIA} & \textbf{MAA} & \textbf{MIA} & \textbf{MAA} \\
\midrule
Pipeline Tag 
& 0.731 & 0.466 & 0.583 & 0.564 
& 0.836 & 0.481 & 0.786 & 0.613 
& \textbf{0.934} & 0.779 & 0.905 & \textbf{0.838} \\

License 
& 0.634 & 0.285 & 0.354 & 0.467 
& 0.738 & 0.476 & 0.569 & 0.635 
& \textbf{0.786} & 0.573 & 0.669 & \textbf{0.713} \\

Reuse Method 
& 0.884 & 0.459 & 0.587 & 0.691 
& 0.839 & 0.428 & 0.634 & 0.664 
& \textbf{0.914} & 0.667 & 0.773 & \textbf{0.823} \\

Model Type 
& 0.465 & 0.209 & 0.218 & 0.336 
& 0.969 & 0.943 & 0.946 & 0.949 
& \textbf{0.978} & 0.955 & 0.973 & \textbf{0.972} \\

Library Name 
& 0.982 & 0.469 & 0.824 & 0.623 
& 0.961 & 0.315 & 0.450 & 0.657 
& \textbf{0.989} & 0.785 & 0.967 & \textbf{0.847} \\

\bottomrule
\end{tabular}%
}
\end{table*}

\noindent\textbf{Repository tags are the most influential features across all prediction tasks, while configuration keys such as \textit{problem\_type}, \textit{tokenizer\_class}, and \textit{eos\_token\_id} drive task specific predictions.}
\Cref{feature_importance} presents the union of the top ten most important input features for predicting each of the five metadata targets: pipeline tag, license, reuse method, model type, and library name. The importance scores highlight the relative contribution of repository-level tags and configuration-file keys, enabling comparison between ecosystem-level signals and architecture-level signals.

Repository tag features such as \textit{autotrain\_compatible}, \textit{safetensors}, and \textit{endpoints\_compatible} consistently rank among the most influential predictors across all targets, each contributing roughly 10\%--14\% of total importance. In particular, \textit{safetensors} appears across all five tasks, indicating strong and stable co-occurrence patterns with configuration attributes such as \textit{use\_cache}, and reinforcing its role as a general indicator of model provenance.

In contrast, configuration-file features provide target-specific predictive signals. For \textit{pipeline\_tag}, features such as \textit{dataset} and \textit{problem\_type} are highly influential, reflecting task-level semantics. For \textit{library\_name}, features including \textit{eos\_token\_id} and \textit{attention\_probs\_dropout\_prob} capture implementation-level differences across frameworks. For \textit{model\_type}, \textit{tokenizer\_class} emerges as a dominant predictor, reflecting architecture-level distinctions. Overall, these patterns show that the fusion of repository tags and configuration keys provides complementary signals: repository tags capture ecosystem‑level provenance (e.g., safetensors, autotrain\_compatible), while configuration keys encode architectural and implementation details (e.g., problem\_type, tokenizer\_class, attention\_dropout). Together, this heterogeneous signal combination enables SemFin to recover missing metadata across diverse categories.

\begin{table*}[t]
\centering
\caption{Union of Top-10 Most Influential Features for Recovering Missing Metadata. Feature importance scores (Importance \%) represent the contribution of each feature across the five optimized models. The Type column indicates the observed source(s) of each feature in the dataset: Hugging Face repository metadata (Repo Tag), model configuration files (Config Key), or both sources (Repo + Config). Bold indicates the highest value in each column.}
\label{feature_importance}
\vspace{2mm}
\resizebox{\textwidth}{!}{%
\begin{tabular}{@{}l l r r r r r@{}}
\toprule
\textbf{Feature} & \textbf{Type} & \textbf{Pipeline Tag} & \textbf{License} & \textbf{Reuse Method} & \textbf{Model Type} & \textbf{Library Name} \\
\midrule
autotrain\_compatible & Repo Tag & \textbf{3.97} & 4.05 & \textbf{5.39} & 4.28 & 2.91 \\
safetensors & Repo Tag & 3.04 & \textbf{4.49} & 4.63 & \textbf{4.48} & \textbf{5.37} \\
endpoints\_compatible & Repo Tag & 3.34 & 3.82 & 4.10 & 3.95 & 3.32 \\
dataset & Both (Repo + Config) & 3.27 & --- & --- & --- & --- \\
problem\_type & Both (Repo + Config) & 2.83 & --- & --- & --- & --- \\
generated\_from\_trainer & Config Key & 2.72 & 2.75 & 2.08 & --- & --- \\
output\_past & Config Key & 2.70 & --- & --- & 2.12 & --- \\
tensorboard & Config Key & 2.63 & --- & --- & 1.93 & --- \\
gradient\_checkpointing & Config Key & 2.28 & --- & --- & 2.45 & --- \\
classifier\_dropout & Config Key & 2.26 & --- & --- & 3.02 & 2.64 \\
attention\_bias & Config Key & --- & 3.37 & 2.37 & --- & --- \\
conversational & Repo Tag & --- & 3.21 & 2.90 & --- & 2.62 \\
attention\_dropout & Config Key & --- & 2.99 & 3.12 & --- & 2.41 \\
pad\_token\_id & Config Key & --- & 2.85 & 2.66 & 2.06 & 3.85 \\
bos\_token\_id & Config Key & --- & 2.65 & 2.28 & --- & 3.26 \\
text-generation-inference & Repo Tag & --- & 2.46 & 2.46 & 2.96 & --- \\
tokenizer\_class & Config Key & --- & --- & --- & 1.94 & --- \\
attention\_probs\_dropout\_prob & Config Key & --- & --- & --- & --- & 2.89 \\
eos\_token\_id & Config Key & --- & --- & --- & --- & 2.68 \\
\bottomrule
\end{tabular}%
}
\begin{flushleft}
\footnotesize \textit{Note}: Values are importance percentages. The table includes all features that ranked in the top-10 for at least one predicting metadata target. Dashes (---) indicate the feature did not rank in the top-10 for that specific task. Repo Tag refers to Hugging Face repository metadata, Config Key refers to model configuration file parameters, and Both (Repo + Config) indicates features observed in both sources within the dataset extraction pipeline.
\end{flushleft}
\end{table*}

\subsection{Comparison between the SemFin approach and the Graph Avg and Hub Avg baseline approaches}
In Section 5, we detailed the implementation of the SemFin approach, leveraging the reuse signals (configuration keys and tags) identified in RQ1 to construct robust model fingerprints. We now turn to empirically validating the effectiveness of this proposed approach by comparing it directly against the Graph Avg and Hub Avg baselines.

\head{Approach}\\
To assess the predictive performance of SemFin and the compared baseline methods on this aligned dataset, we follow these steps: 

\begin{itemize}

    \item \textbf{Step 6.2.0 — Dataset alignment and experimental setup:} To ensure consistency with prior work and enable direct comparison with the ``Graph" and ``Hub" methods proposed by \citet{horwitz2025we}, we aligned our dataset with their benchmark. Specifically, we restricted our analysis to models present in both our mined collection and their public dataset\footnote{\url{https://huggingface.co/datasets/Eliahu/ModelAtlasData}}. This filtering yielded 315,106 PTLMs (99.36\% of the original 317,133 models).

    We focus on five metadata fields (\texttt{pipeline\_tag}, \texttt{license}, \texttt{reuse\_method}, \texttt{model\_type}, and \texttt{library\_name}) to establish a strict one-to-one comparison with prior work. From the aligned dataset, we first identified a high-quality subset of 82,892 PTLMs with complete ground-truth annotations across all five target metadata fields. From this subset, we constructed a stratified test set of 16,579 models (20\%), using reuse method as the stratification variable. This ensures that rare categories (e.g., Deduplication with 529 models) are preserved in the evaluation set, with a maximum class proportion deviation of 0.46\%. This test set is strictly held out and used exclusively for comparative evaluation against Graph Avg and Hub Avg in this section.
    
    The remaining 298,527 models (i.e., the full dataset minus the held-out test set) constitute the dataset used to train SemFin. This dataset is partially labeled and is processed using the pipeline described in Section~\ref{semfin}. During training, we apply target-specific filtering (Step 5.1.5) so that each classifier is trained only on instances with available ground-truth labels for the corresponding metadata.
    
    Model training and validation are performed using Stratified 5-Fold Cross-Validation as described in Step 5.1.6. Within each fold, feature extraction (TF-IDF) and model training are conducted strictly on the training split, with evaluation performed on the validation split, ensuring no data leakage. This design ensures that SemFin is trained and internally validated on a large, partially labeled corpus, while final performance is evaluated on a fully labeled, unseen test set.

    \item \textbf{Step 6.2.1 — Re-implementing the Graph Avg and Hub Avg:} We re-implemented the Graph Avg and Hub Avg imputation procedures following the methodology described by \citep{horwitz2025we} for predictive evaluation.
    \begin{itemize}
        \item \textbf{Graph Construction:} We constructed a global ``Model Atlas'' where each unique model in our dataset of 315,106 PTLMs is represented as a \textbf{node}. An \textbf{edge} is created whenever a model (the child) specifies another model (the parent) in its \texttt{initialization model} metadata. When a model listed multiple parents, we treated each relationship independently, yielding parallel lineage branches. While lineage is directional, we implemented an \textbf{undirected graph topology} for the evaluation, allowing search algorithms to traverse the network ``inward and outward'' to identify ancestors, descendants, and siblings.
        
        \item \textbf{Graph Avg Implementation:} Using the constructed graph, we re-implemented Graph Avg as a graph-based $k$-nearest-neighbor ($k$-NN) imputer. For a target test model, the algorithm performs a breadth-first search to identify the $k=5$ nearest labeled training nodes within a maximum five-hop radius. For example, in a reuse chain $A \rightarrow B \rightarrow C \rightarrow D$, when $D$ is in the test set, Graph Avg aggregates labels from $C$, $B$, and $A$ (when available) and predicts via a \textbf{majority vote} over the labels of these neighbors. If no labeled neighbors exist within the cutoff, the method abstains.

        \item \textbf{Hub Avg Implementation:} This method focuses on ``hubs'' (parent models) in the constructed graph to perform sibling-based voting. For a test model $D$ derived from parent $A$, Hub Avg identifies all other children of $A$ (siblings) that are not in the test set. The prediction is determined by the most frequent label among these siblings. If no siblings exist (i.e., $A$ has no other labeled children), Hub Avg abstains from making a prediction for that test instance. In cases of multi-parent models, sibling groups are constructed independently for each parent and pooled before voting. Ties are broken deterministically using the alphabetical ordering of metadata values.
    \end{itemize}

    \textbf{Experimental Protocol and Label Masking:} To ensure the baselines operated under optimal structural conditions, we allowed them to navigate the full 315,106-node graph topology when searching for labeled training neighbors. Crucially, we maintained full connectivity but masked the metadata labels of the 16,579 models in our held-out test set. By retaining test nodes as ``bridges'' without their labels, we granted the baselines maximum structural paths to reach labeled training neighbors. Performance, however, was calculated exclusively on these 16,579 masked test nodes. Furthermore, because our filtering (Step 4.1.1) removed unresolvable parent references, these heuristics benefited from a cleaner lineage graph than exists in the raw ecosystem, explicitly favoring the baselines.

    \textbf{Evaluation and Significance Testing:} We evaluated SemFin using 5-fold stratified cross-validation on the training set, then tested the final model on a held-out test set. SemFin always returns a prediction for every test sample because it operates as a supervised classifier over the fixed feature space. For the Graph Avg and Hub Avg baseline approaches, these methods sometimes cannot make a prediction. This happens when a model is disconnected from the graph for $k$-NN or has no valid siblings for hub averaging. For these baseline methods, we report two accuracy values. The first is \textit{penalized accuracy}, where we treat an abstention as an incorrect prediction and divide by the total number of test samples. The second is \textit{non-penalized accuracy}, where we calculate accuracy only over the subset of test samples that received a prediction. We report both values to provide a complete picture of baseline performance.

    \item \textbf{Step 6.2.2 — Evaluating the performance of SemFin on the test set:}
    \begin{itemize}
        \item \textbf{Vectorizing the semantic fingerprints:} To convert the semantic fingerprints of each model into a numerical representation suitable for machine learning, we vectorized the training set using the Term Frequency–Inverse Document Frequency (TF–IDF) scheme from the scikit-learn library \citep{pedregosa2011scikit}. 
        
        We configured the TF–IDF vectorizer to utilize the complete effective vocabulary extracted from the training data, resulting in a total of 408 unique features. Consequently, the full set of available fingerprint features was retained without truncation. The vectorizer was fitted exclusively on the training set to prevent information leakage into the testing data.
        
        This process produced sparse TF–IDF vectors that capture the complete semantic information available in the fingerprints while ensuring a fair and reproducible evaluation setup.

        \noindent\textbf{Final SemFin Model Configuration.} Following the training procedure described in Section~\ref{semfin}, we first report the optimized hyperparameter configurations of the final SemFin models in \Cref{final_hyperparams}. These parameters were obtained using Randomized Search with 5-fold cross-validation on the training dataset defined in Step 6.2.0. Using these configurations, the final models were retrained on the full training dataset (298,527 models) and used for inference on the held-out test set.

        \begin{table*}[t]
            \centering
            \caption{Final optimized hyperparameters for the selected SemFin models. Parameters were identified via Randomized Search (5-fold cross-validation) and used to train the final models on the full training dataset.}
            \label{final_hyperparams}
            \resizebox{0.9\textwidth}{!}{
            \begin{tabular}{lcccccc}
            \toprule
            \textbf{Target Metadata} & \textbf{Learning Rate} & \textbf{Max Depth} & \textbf{Estimators} & \textbf{Num Leaves} & \textbf{Subsample} & \textbf{Colsample} \\
            \midrule
            Pipeline Tag   & 0.021 & 30 & 393 & 21  & 0.673 & 0.860 \\
            License        & 0.030 & 20 & 187 & 119 & 0.657 & 0.778 \\
            Reuse Method   & 0.071 & 10 & 352 & 108 & 0.716 & 0.673 \\
            Model Type     & 0.021 & 30 & 393 & 21  & 0.673 & 0.860 \\
            Library Name   & 0.071 & 10 & 352 & 108 & 0.716 & 0.673 \\
            \bottomrule
            \end{tabular}
            }
        \end{table*}

        \item \textbf{Generating predictions on the test set:} To assess generalization, we evaluated the final trained models (LightGBM variants) on a held-out test set of 16,579 PTLMs not used during training (see Step 6.2.0). For each of the five dependent variables (i.e., \texttt{pipeline\_tag}, \texttt{license}, \texttt{reuse\_method}, \texttt{model\_type}, and \texttt{library\_name}), we used the final production LightGBM classifier. As detailed in Step 5.1.8, this classifier was instantiated with the optimal hyperparameters identified during 5-fold cross-validation and retrained on the complete training dataset, along with its fitted TF–IDF vectorizer. The test models' semantic fingerprints were transformed using the same feature space as the training set. 
     
        Each classifier generated predictions for its respective metadata. We assessed performance using Micro-Accuracy, Macro-Accuracy, Micro-F1, and Macro-F1. Micro metrics emphasize frequent classes, while macro metrics reflect performance on minority categories. All metrics were computed on the full test set and summarized in a table. The Macro-F1 and micro-F1 results are available in our replication package \citep{SemFin}.
    \end{itemize}

    \item \textbf{Step 6.2.3 — Comparison between SemFin, Hub Avg, and Graph Avg, including statistical significance testing:} We compared the performance of SemFin against the Hub Average and Graph Average baselines for each of the five target metadata: \texttt{pipeline\_tag}, \texttt{license}, \texttt{reuse\_method}, \texttt{model\_type}, and \texttt{library\_name}. For each metadata, we report the accuracy obtained by Graph and Hub averages both with and without penalty adjustments, alongside the Micro-Accuracy achieved by SemFin.
    
    To quantify the improvement of SemFin over the baselines, we calculated the difference between SemFin's accuracy and the corresponding Graph and Hub accuracies. Formally, for each metadata $a$ and baseline $b$:
    
    \[
    \text{Improvement}_{a,b} = \text{Accuracy}_{\text{SemFin}, a} - \text{Accuracy}_{b, a}
    \]
    
    Positive values indicate that SemFin outperforms the baseline, while negative values indicate the opposite. This calculation was performed for all metadata and both baseline settings (with and without penalty).
    
    To determine whether these performance gains are statistically significant, we applied the same paired statistical methodology defined in Step 5.1.7. For each dependent variable, we constructed contingency tables pairing the predictions of our optimal SemFin model against each baseline. We then calculated McNemar's test statistic ($\chi^2$) to assess significance and Cohen's $g$ to quantify the effect size. By focusing on the discordant pairs ($n_{10}$ and $n_{01}$), this comparison isolates the specific instances where SemFin's artifact-driven approach provides a predictive advantage over the lineage-based propagation heuristics. We report significance at the $p < 0.05$, $p < 0.01$, and $p < 0.001$ levels, with results summarized in tables showing consistent gains across all metadata and settings.

    \item \textbf{Step 6.2.4 — Prediction overlap and complementarity analysis:} 
    To evaluate the unique predictive advantage and complementarity of SemFin relative to Graph Avg and Hub Avg, we analyzed prediction correctness at the instance level. For each test instance, we first determined whether each method's prediction was correct. Based on these correctness indicators, we computed seven categories, defined as follows:
    
    \begin{itemize}
        \item \textbf{``Only SemFin correct'':} Instances where SemFin predicted correctly while \emph{both} Graph Avg and Hub Avg were incorrect (strict three-way comparison).
    
        \item \textbf{``Only Graph correct'':} Instances where Graph Avg predicted correctly while both SemFin and Hub Avg were incorrect (strict three-way comparison).
    
        \item \textbf{``Only Hub correct'':} Instances where Hub Avg predicted correctly while both SemFin and Graph Avg were incorrect (strict three-way comparison).
    
        \item \textbf{``SemFin correct (Graph wrong)'':} Instances where SemFin predicted correctly and Graph Avg was incorrect, regardless of Hub Avg's performance (binary comparison; Hub Avg ignored).
    
        \item \textbf{``SemFin correct (Hub wrong)'':} Instances where SemFin predicted correctly and Hub Avg was incorrect, regardless of Graph Avg's performance (binary comparison; Graph Avg ignored).
    
        \item \textbf{``Graph correct (SemFin wrong)'':} Instances where Graph Avg predicted correctly and SemFin was incorrect, regardless of Hub Avg's performance (binary comparison; Hub Avg ignored).
    
        \item \textbf{``Hub correct (SemFin wrong)'':} Instances where Hub Avg predicted correctly and SemFin was incorrect, regardless of Graph Avg's performance (binary comparison; Graph Avg ignored).
    \end{itemize}
    
    The first three categories correspond to strict three-way cases in which exactly one method is correct. The remaining four categories are pairwise comparisons between SemFin and one baseline, where the third method is intentionally ignored. Consequently, these seven categories are \emph{not} mutually exclusive and do not form a complete partition of the dataset.
    
    This analysis allows us to quantify (i) how often SemFin uniquely succeeds where both Graph Avg and Hub Avg fail, and (ii) how SemFin compares pairwise against each baseline independently. The results are presented as a matrix reporting the percentage of test instances falling into each category across all five evaluated attributes.
\end{itemize}

\head{Result}\\
\noindent\textbf{SemFin provides higher average accuracy (0.84) across all prediction tasks compared to lineage-based heuristics (0.78 for Graph and 0.81 for Hub methods under no-penalty setting).}
\Cref{baseline_comparison} compares SemFin against Graph Avg and Hub Avg under penalized (P) and non-penalized (NP) settings. The performance values reported here differ slightly from the cross-validation metrics in \Cref{fingerprint_standard} because they reflect results on a strictly unseen test population.

SemFin achieves its most significant gains on \texttt{Pipeline Tag} metadata with a task-specific accuracy of 0.913, exceeding the strongest baseline (Hub NP) by +0.092 and the penalized variant by +0.228. The largest improvement is observed for \textit{Reuse Method}, where SemFin surpasses Graph (P) by +0.314 and Hub (NP) by +0.153. These results strongly align with our findings in \Cref{section_4.3}, where we identified discriminative configuration keys (e.g., \texttt{attention\_dropout} and \texttt{classifier\_dropout}) that uniquely define different model reuse methods. The effectiveness of the semantic fingerprint stems from its ability to leverage these unique technical identifiers, whereas graph-based methods rely solely on neighbor consistency.

For metadata like \textit{Model Type} and \textit{Library Name}), the non-penalized baselines achieve slightly higher raw accuracy (up to 0.982) than SemFin. This marginal difference ($\approx -0.005$) occurs because these specific metadata fields are rarely changed during model reuse and are inherited directly along the lineage graph, favoring direct lookups when neighbors are available.

\begin{table*}[t]
\centering
\caption{Comparative Performance of SemFin vs. Baselines (Graph Avg \& Hub Avg). (P) denotes variants with Penalty, and (NP) denotes No Penalty. SemFin achieves higher predictive accuracy than lineage-based heuristics for most predicted metadata, particularly for \textit{Reuse Method} and \textit{Pipeline Tag} metadata, though it is slightly worse by Graph(NP) and Hub(NP) on \textit{Model Type}.}
\label{baseline_comparison}
\resizebox{\textwidth}{!}{%
\begin{tabular}{l ccccc | ccccc}
\toprule
 & \multicolumn{5}{c}{\textbf{Performance Score}} & \multicolumn{4}{c}{\textbf{Improvement ($\Delta$) with SemFin}} \\
\cmidrule(lr){2-6} \cmidrule(lr){7-10}
\textbf{Metadata} & \textbf{Graph(P)} & \textbf{Graph(NP)} & \textbf{Hub(P)} & \textbf{Hub(NP)} & \textbf{SemFin} & \textbf{vs Graph(P)} & \textbf{vs Graph(NP)} & \textbf{vs Hub(P)} & \textbf{vs Hub(NP)} \\
\midrule
Pipeline Tag & 0.697 & 0.753 & 0.684 & 0.821 & \textbf{0.913} & +0.216 & +0.160 & +0.228 & +0.092 \\
License & 0.626 & 0.676 & 0.544 & 0.653 & \textbf{0.742} & +0.116 & +0.066 & +0.198 & +0.089 \\
Reuse Method & 0.518 & 0.560 & 0.567 & 0.680 & \textbf{0.832} & +0.314 & +0.272 & +0.266 & +0.153 \\
Model Type & 0.909 & \textbf{0.982} & 0.819 & \textbf{0.982} & 0.977 & +0.068 & -0.006 & +0.158 & -0.005 \\
Library Name & 0.874 & 0.944 & 0.792 & \textbf{0.950} & 0.945 & +0.071 & +0.000 & +0.153 & -0.005 \\
\bottomrule
\end{tabular}%
}
\end{table*}

To verify that the observed gains are not attributable to random variation, we apply McNemar’s test, a non-parametric test for paired nominal outcomes (correct vs. incorrect predictions). Across all predicted metadata, SemFin significantly outperforms both Graph(P) and Hub(P) baselines at the highest confidence level ($p < 0.001$). The improvements are particularly pronounced for metadata that cannot be reliably inferred through hub- or graph-level label propagation alone, such as \textit{reuse\_method}, where SemFin improves accuracy by +0.314 over Graph(P) and +0.266 over Hub(P). Even for highly standardized metadata with many classes, such as \textit{model\_type}, where baseline performance is already strong, SemFin still delivers statistically significant gains (+0.068 vs.\ Graph(P) and +0.158 vs.\ Hub(P)). These results demonstrate that configuration-based semantic fingerprints provide a robust and statistically distinguishable advantage over graph- and hub-based majority-vote label propagation methods. To further quantify the magnitude of this advantage, we analyzed the effect size using Cohen's $g$. The results indicate a \emph{large effect size} ($g > 0.25$) for nearly all metadata comparisons, confirming that the performance gap is substantial. For instance, \textit{model\_type} and \textit{library\_name} exhibit extremely strong effects (up to $g=0.428$ vs.\ Hub). The only exception is the comparison of \textit{license} against Graph(P), which shows a \emph{medium effect size} ($g=0.150$); however, even here, SemFin maintains a practical lead.

\begin{table*}[t]
\centering
\caption{Statistical Significance and Effect Size of Performance Improvements (SemFin vs. Penalized Baselines). McNemar's test confirms significance ($p < 0.001$), while Cohen's $g$ indicates the magnitude of the effect ($0.15=$ Medium, $0.25=$ Large).}
\label{significance_test}

\resizebox{\textwidth}{!}{%
\begin{tabular}{l ccc | cc cc cc}
\toprule
 & \multicolumn{3}{c}{\textbf{Performance Score}} & \multicolumn{6}{c}{\textbf{Improvement \& Significance}} \\
\cmidrule(lr){2-4} \cmidrule(lr){5-10}
\textbf{Metadata} 
& \textbf{Graph(P)} 
& \textbf{Hub(P)} 
& \textbf{SemFin} 
& \textbf{$\Delta$ vs Graph(P)} 
& \textbf{Sig.} 
& \textbf{g(G)} 
& \textbf{$\Delta$ vs Hub(P)} 
& \textbf{Sig.} 
& \textbf{g(H)} \\
\midrule
Pipeline Tag & 0.697 & 0.684 & \textbf{0.913} & +0.216 & *** & +0.370 & +0.228 & *** & +0.360 \\
License & 0.626 & 0.544 & \textbf{0.742} & +0.116 & *** & +0.150 & +0.198 & *** & +0.260 \\
Reuse Method & 0.518 & 0.567 & \textbf{0.832} & +0.314 & *** & +0.337 & +0.266 & *** & +0.321 \\
Model Type & 0.909 & 0.819 & \textbf{0.977} & +0.068 & *** & +0.350 & +0.158 & *** & +0.428 \\
Library Name & 0.874 & 0.792 & \textbf{0.945} & +0.071 & *** & +0.347 & +0.153 & *** & +0.427 \\
\bottomrule
\end{tabular}%
}

\vspace{5mm}

\resizebox{\textwidth}{!}{%
\begin{tabular}{l ccc | cc cc cc}
\toprule
 & \multicolumn{3}{c}{\textbf{Performance Score}} & \multicolumn{6}{c}{\textbf{Improvement \& Significance}} \\
\cmidrule(lr){2-4} \cmidrule(lr){5-10}
\textbf{Metadata} 
& \textbf{Graph(NP)} 
& \textbf{Hub(NP)} 
& \textbf{SemFin} 
& \textbf{$\Delta$ vs Graph(NP)} 
& \textbf{Sig.} 
& \textbf{g(GNP)} 
& \textbf{$\Delta$ vs Hub(NP)} 
& \textbf{Sig.} 
& \textbf{g(HNP)} \\
\midrule
Pipeline Tag & 0.753 & 0.821 & \textbf{0.913} & +0.160 & *** & +0.307 & +0.092 & *** & +0.174 \\
License & 0.676 & 0.653 & \textbf{0.742} & +0.066 & *** & +0.054 & +0.089 & *** & +0.044 \\
Reuse Method & 0.560 & 0.680 & \textbf{0.832} & +0.272 & *** & +0.273 & +0.153 & *** & +0.147 \\
Model Type & \textbf{0.982} & \textbf{0.982} & 0.977 & -0.006 & *** & -0.101 & -0.005 & *** & -0.122 \\
Library Name & 0.944 & \textbf{0.950} & 0.945 & +0.000 & * & -0.044 & -0.005 & *** & -0.164 \\
\bottomrule
\multicolumn{10}{l}{\footnotesize \textit{Significance Levels: *** $p < 0.001$, ** $p < 0.01$, * $p < 0.05$}}
\end{tabular}%
}
\end{table*}

Finally, while our re-implementation followed the methodology of \citep{horwitz2025we}, the dataset filtering described in Step 3.1.1 ensured that the resulting lineage graph was more coherent than the raw ecosystem, leading to baseline performance increases across all attributes compared to their original study. For instance, the license prediction accuracy of this re-implemented baseline (0.67) is significantly higher than the 0.49 reported in the original paper. Despite this higher-performing baseline environment, SemFin provides more robust results across the full test population.

\begin{figure*}[t]
\centering
\includegraphics[width=\textwidth]{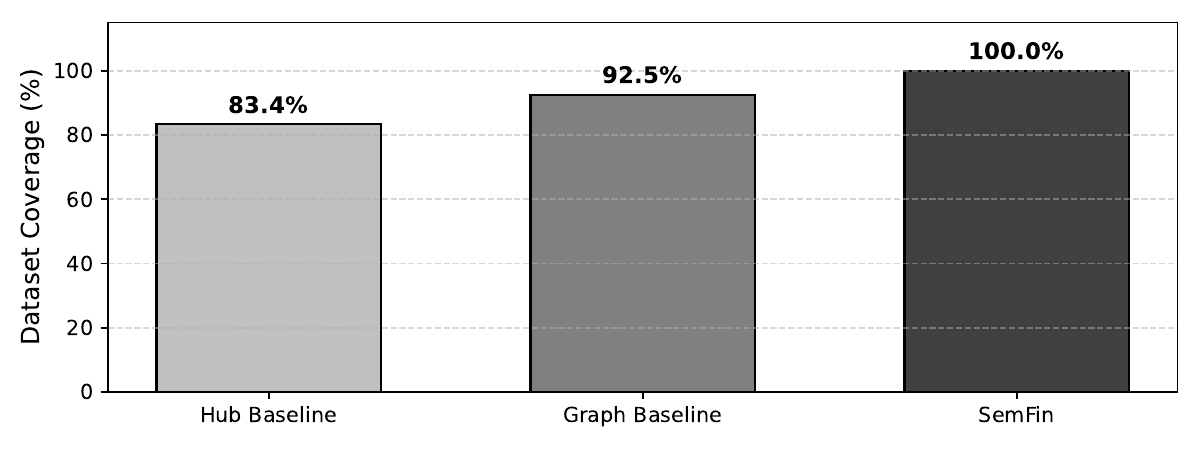}
\caption{Coverage Comparison between SemFin (Our Method) and Baseline Averages. SemFin achieves dataset coverage (100\%) because it relies on internal configuration files rather than external metadata or graph connectivity.}
\label{coverage_comparison}
\end{figure*}

\noindent\textbf{SemFin always provides a prediction for any model in our test set (100\% coverage), whereas baseline methods must abstain when metadata is missing or the model is disconnected.} \Cref{coverage_comparison} compares the prediction coverage of SemFin with Graph Avg and Hub Avg. In this context, coverage refers to the percentage of test instances for which a method can generate a prediction, regardless of correctness. SemFin attains 100.0\% coverage on this population because it operates as a supervised classifier over a fixed feature space based on the presence or absence of configuration keys and repository tags, allowing it to always compute an output vector. This coverage applies specifically to models within the transformers framework that contain configuration files. Our data filtering in Step 3.1.1 explicitly restricted our scope to these configuration-bearing models to ensure a fair baseline comparison, meaning SemFin targets models that actively provide configuration files within the ecosystem.

In contrast, baseline methods suffer from data sparsity constraints that force them to abstain. Hub Avg provides predictions for only 83.4\% of models due to missing or incomplete hub metadata, such as missing \texttt{model\_type} or \texttt{library\_name} fields on the model card. Similarly, Graph Avg reaches 92.5\% coverage but is inherently constrained by topological connectivity. Since it relies on aggregating votes from adjacent neighbors, it fails to generate predictions for isolated models that are disconnected from the reuse graph. The resulting 16.6\% applicability gap between SemFin and Hub Avg demonstrates the advantage of relying on structural configuration features rather than extrinsic platform metadata that may be optional or incomplete.

SemFin's 100\% coverage is not an empirical surprise but a direct consequence of its design: as a supervised classifier operating on configuration keys and repository tags, it produces a prediction for every model that has these artifacts.

\noindent\textbf{SemFin not only statistically outperforms the baseline methods, but also addresses predictive gaps where Graph Avg and Hub Avg fail.} Following the methodology defined in Step 6.2.4, the results of the prediction overlap analysis are presented in \Cref{overlap_matrix}. The analysis confirms that SemFin successfully identifies model metadata in cases where Graph Avg and Hub Avg provide no valid signal. For \textit{Reuse Method}, SemFin uniquely succeeds where both Graph Avg and Hub Avg fail in 27.4\% of all test instances. This high percentage in the ``Only SemFin correct'' category indicates that the internal configuration keys capture a primary signal that remains accessible even when models are isolated or disconnected from the reuse graph. While Graph Avg demonstrates its highest unique contribution in the \textit{License} category at 5.7\%, its unique success across other attributes is significantly lower, representing only 1.7\% for \textit{Reuse Method}.

When comparing SemFin pairwise against the baselines, the results show a consistent trend of more robust error recovery. For \textit{Pipeline Tag}, the category SemFin correct (Graph wrong) accounts for 25.4\% of instances, while the inverse Graph correct (SemFin wrong) accounts for only 3.8\%. This 21.6 percentage point gap demonstrates that the fusion of HF repository tags and configuration keys is significantly more robust than the neighbor-consistency logic used by Graph Avg. Even for metadata with consistent labels across models, such as \textit{Model Type}, SemFin independently recovers 8.1\% of the test set where lineage chains for Graph Avg and Hub Avg are broken or inconsistent. These findings verify that the semantic fingerprint effectively covers the structural gaps where Graph Avg and Hub Avg lack sufficient connectivity for traditional label propagation. At the same time, the complementary strengths of Graph Avg and Hub Avg in certain categories (e.g., License) suggest that a hybrid approach combining semantic fingerprints with graph-based propagation could further improve coverage and accuracy, an avenue we leave for future work.

\begin{table*}[t]
\centering
\caption{Prediction Overlap and Complementarity Matrix. Values represent the percentage of the total test set ($N=16,579$) falling into each category defined in Step 6.2.4. \textbf{Bold} indicates the highest value per comparison group.}
\label{overlap_matrix}
\resizebox{\textwidth}{!}{%
\begin{tabular}{l ccc cccc}
\toprule
 & \multicolumn{3}{c}{\textbf{Three-Way Comparison (Strict)}} & \multicolumn{4}{c}{\textbf{Pairwise Binary Comparison}} \\
\cmidrule(lr){2-4} \cmidrule(lr){5-8}
\textbf{Metadata} & \textbf{Only SemFin} & \textbf{Only Graph} & \textbf{Only Hub} & \textbf{SemFin correct} & \textbf{SemFin correct} & \textbf{Graph correct} & \textbf{Hub correct} \\
 & \textbf{correct} & \textbf{correct} & \textbf{correct} & \textbf{(Graph wrong)} & \textbf{(Hub wrong)} & \textbf{(SemFin wrong)} & \textbf{(SemFin wrong)} \\
\midrule
Pipeline Tag & \textbf{17.3\%} & 0.7\% & 1.4\% & 25.4\% & \textbf{27.3\%} & 3.8\% & 4.5\% \\
License      & \textbf{18.2\%} & 5.7\% & 1.3\% & 25.2\% & \textbf{28.9\%} & 13.6\% & 9.1\% \\
Reuse Method & \textbf{27.4\%} & 1.7\% & 1.5\% & \textbf{39.0\%} & 34.0\% & 7.6\% & 7.4\% \\
Model Type   & \textbf{8.1\%}  & 0.2\% & 0.1\% & 8.3\%  & \textbf{17.1\%} & 1.5\% & 1.3\% \\
Library Name & \textbf{8.0\%}  & 0.3\% & 0.0\% & 8.7\%  & \textbf{16.6\%} & 1.6\% & 1.3\% \\
\bottomrule
\end{tabular}%
}
\end{table*}

\begin{Summary}
{Summary of RQ$_2$ Findings}{
\begin{itemize}
    \item As the core component of SemFin, LightGBM achieved the highest micro-accuracy (up to 98.9\%), significantly outperforming ($p < 0.001$) Random Forest and $k$-NN baselines across all metadata fields.
    \item Framework integration metadata (e.g., \texttt{autotrain\_compatible}) and architectural behaviors dominate prediction, serving as reliable intrinsic proxies for model provenance.
    \item SemFin yields statistically significant improvements ($p < 0.001$) over baselines across all five metadata fields, improving accuracy by +31.4\% for \texttt{reuse\_method}, +22.8\% for \texttt{pipeline\_tag}, +19.8\% for \texttt{license}, +15.8\% for \texttt{model\_type}, and +15.3\% for \texttt{library\_name}.
    \item SemFin guarantees 100\% coverage, solving the cold-start problem for isolated models where baselines abstain (up to 16.6\% of cases). Complementarity analysis confirms that SemFin uniquely succeeds where both Graph Avg and Hub Avg fail in 8.1-27.4\% of instances across metadata types.
\end{itemize}}
\end{Summary}

\section{\textbf{RQ$_3$:} \RQc}
Having established in RQ$_2$ that SemFin can accurately recover missing metadata fields, we now apply this approach to the 167,089 (52.7\%) PTLMs that lack an explicit reuse method and the 188,976 models that lack license metadata. Relying solely on the small subset of models with declared metadata creates lineage mirages: false structural patterns regarding the nature of reuse and licensing that emerge from missing information rather than genuine model change.

\begin{figure}[t]
\centering
\includegraphics[width=\linewidth]{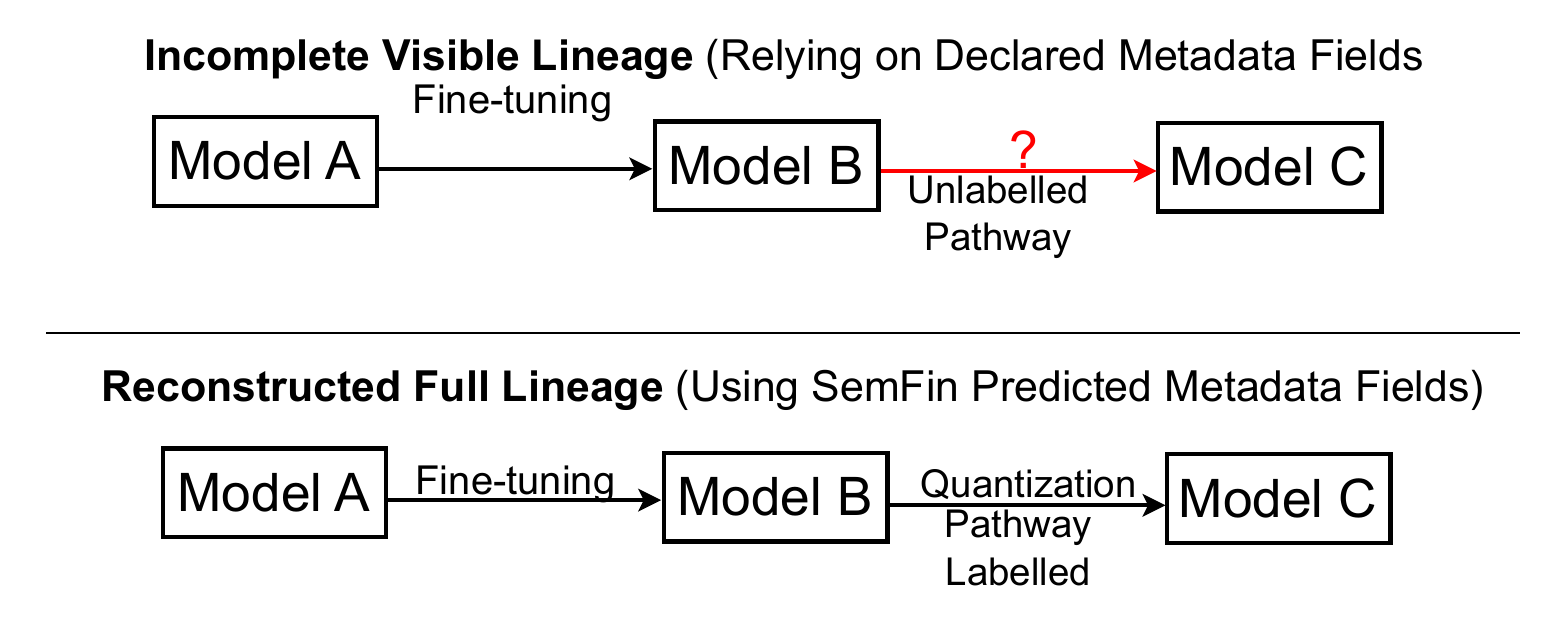}
\caption{The mechanism of lineage mirages caused by missing metadata. The top diagram shows the incomplete visible graph, where the parent--child dependency edges between Model A, Model B, and Model C are fully visible, but the second edge lacks a functional annotation. The bottom diagram shows the reconstructed graph, where the underlying topological edges remain unchanged, but SemFin successfully recovers the missing quantization label to reveal the true ``lineage chain".}
\label{lineage_mirage_fig}
\end{figure}

As illustrated in \Cref{lineage_mirage_fig}, while the underlying parent--child dependency edges are stored for all models in HF, the missing reuse method labels of 52.7\% completely distort the observable reuse lineages. For example, consider a true lineage chain: Model A $\rightarrow$ Fine-tuning $\rightarrow$ Model B $\rightarrow$ Quantization $\rightarrow$ Model C. A practitioner auditing this lineage using only declared metadata sees the full structural path but lacks the annotations to understand the engineering actions taken along the edges. If the fine-tuning annotation is missing, the relationship between Model A and Model B becomes functionally blank. As a result, while the structural connection from Model B to Model C remains visible, the true multi-step evolutionary history leading to Model C is obscured, leaving reviewers unable to trace the full lineage of modifications. These mirages actively misrepresent the functional characteristics, reuse distributions, and development history across model lineages.

\begin{figure*}[t]
\centering
\includegraphics[width=\textwidth]{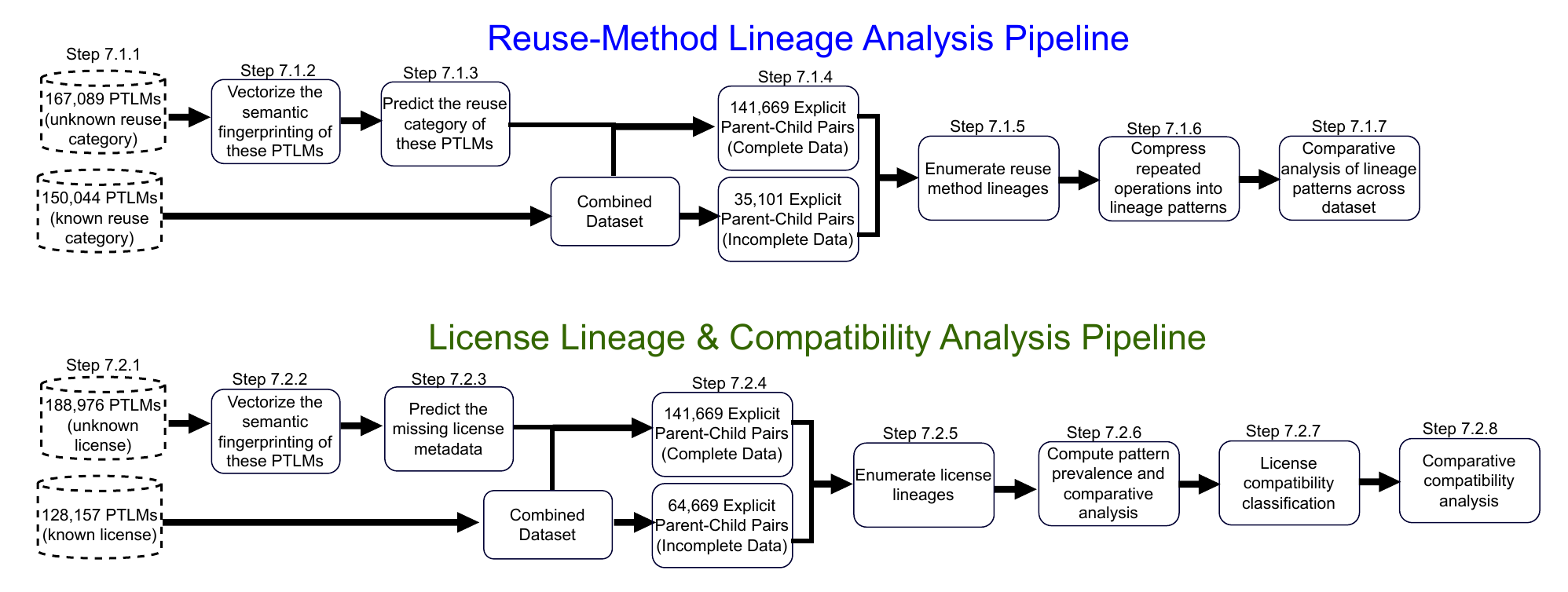}
\caption{Overview of the RQ3 analysis workflow. The top pipeline reconstructs missing reuse-method metadata using SemFin and compares reuse-method lineage patterns between incomplete and complete datasets. The bottom pipeline predicts missing license metadata, extracts license lineage patterns, and evaluates license compatibility across incomplete and complete datasets.}
\label{rq3_workflow}
\end{figure*}

In this RQ, we move beyond performance evaluation to apply a dataset completed with SemFin, reconstructing the reuse method and license lineage patterns for unlabeled models. By contrasting the incomplete visible lineage with this reconstructed lineage, we analyze the impact of missing data on ecosystem analysis and reveal how it obscures the true distribution of reuse methods and license compatibility across the lineage graph, as illustrated in \Cref{rq3_workflow}. We organize this section by presenting the specific \emph{Approach} and corresponding \emph{Results} for each sub-question.

\subsection{Reuse method lineage patterns analysis}
Building on the high macro accuracy of SemFin in RQ$_2$, we now apply SemFin to the 52.7\% of the models lacking explicit reuse metadata. This analysis focuses on how models are transformed through sequences of reuse operations (e.g., Finetune, Merge, Quantization). Missing metadata not only removes lineage links but also distorts the observed order and composition of reuse methods. By comparing pattern distributions between incomplete and complete datasets, we quantify biases introduced by missing metadata and reveal the true diversity of model development pathways. For example, what are really the most common reuse methods? Does quantization typically happen at the end of a model lineage? \\

\head{Approach}\\
To understand the implications of incomplete metadata, we follow the procedure below. All analyses were implemented in Python, using \texttt{LightGBM} for classification and \texttt{NetworkX} for graph processing.

\begin{itemize}

\item \textbf{Step 7.1.1 Retrieving PTLMs with unknown reuse method:} We first separate the dataset based on whether the reuse method is explicitly declared, as described in \Cref{reuse_extraction}. This step yields 167,089 PTLMs without an identified reuse method. This missing metadata field corresponds directly to models that specify a parent model but omit the reuse method via which they were derived from their parent model(s). Consequently, Hugging Face's native model tree in fact does not render these entries, creating broken lineage links that SemFin repairs.

\item \textbf{Step 7.1.2 Vectorizing the semantic fingerprints of these 167,089 PTLMs:} To predict the reuse method for PTLMs with unknown reuse metadata using the trained model, we vectorize their semantic fingerprints using the vectorization class defined in Step~5.1.6.

\item \textbf{Step 7.1.3 Predicting the PTLMs' missing reuse method metadata field:}
We use the LightGBM classifier, identified as the top-performing model in RQ$_2$, to infer unknown reuse method metadata fields of the 167,089 PTLMs. Given the model's robust predictive performance on the held-out test set (Micro-Accuracy = 0.826 and Macro-Accuracy = 0.823 in RQ2), we treat the PTLM's predictions as high-confidence proxies for the ground truth. For each PTLM, we feed the vectorized semantic fingerprint into the classifier to generate a predicted reuse method metadata field, and merge these predictions back into the dataset to create a complete lineage graph. While this imputation allows for a complete sequence of reuse methods in the lineage chain, we explicitly acknowledge that the prediction introduces potential noise corresponding to the classifier's observed error rate ($\approx$17\%, derived from $1-$Micro-Accuracy). However, the impact of this initial error is heavily mitigated by the filtering protocol executed in the subsequent step, which isolates the active corpus and discards structural anomalies before lineage construction.

\item \textbf{Step 7.1.4 Filtering and resolving parent-child relationships in both complete and incomplete dataset:}
To ensure consistent lineage analysis across both dataset variants, we apply a uniform filtering criterion that retains only parent-child pairs where both the parent and child models are present as verifiable entries in our dataset collection. This prevents "dangling" references to generic parent names (e.g., bert-base-uncased) or local filesystem paths for which repository-level metadata or configuration artifacts were not retrieved. After filtering, the complete dataset resulted in 141,669 parent-child pairs, while the incomplete dataset yielded 35,101 parent-child pairs.

For models that specify multiple parents (in case of merge models), we select a single primary parent based on the highest download count among parents that are not the model itself, using download popularity as a proxy for the most canonical lineage path. We explicitly filter out self-references (where a model lists itself as a parent) before selecting the primary parent. Using a single parent is important because retaining all parents would cause the same merge model to appear multiple times in the dataset, once per declared parent, artificially inflating lineage path counts and biasing comparative frequency analysis. Apart from preventing path explosion and ensuring that each model contributes proportionally to lineage statistics, selecting one canonical parent also enables a fair comparison between merge and other reuse methods across both the incomplete and complete datasets. Ties in download count are broken alphabetically by model name to ensure deterministic and reproducible results. While our empirical analysis reveals no cases where a merge model lists only self-references, for theoretical completeness, any model where all declared parents are self-references is treated as having no valid parent. 

\item \textbf{Step 7.1.5 Enumerating reuse-method lineages}
We enumerate all root-to-leaf lineages using \Cref{lineage_enumeration} in order to represent each dataset as a directed graph in which nodes correspond to models and edges represent reuse pathways annotated with reuse methods. The algorithm first identifies root nodes as models that appear only as parents (never as children). Starting from each root, it traverses the reuse graph to extract all raw root-to-leaf paths, where each lineage captures the ordered sequence of reuse methods along a model's development history.

\begin{algorithm}[t]
\caption{Lineage Pattern Enumeration from Model Reuse Graph}
\label{lineage_enumeration}
\begin{algorithmic}[1]

\Require Directed reuse graph $G=(V,E)$
\Ensure Lineage patterns with prevalence statistics

\State $Roots \gets$ models that appear only as parents 
\State Initialize $PatternCounter$ 
\State Initialize $PatternModels$ \Comment{Stores sets of unique models per pattern} 

\Statex
\Statex \Comment{=== MAIN ENUMERATION ===} 

\ForAll{$root \in Roots$} 
    \ForAll{lineage $L$ obtained from a root-to-leaf traversal of $G$} 

        \State $ReuseMethods \gets$ sequence of reuse methods in $L$ 

        \If{$|ReuseMethods| \leq 1$} 
            \State \textbf{continue} 
            \Comment{Skip trivial lineages} 
        \EndIf 

        \State $Pattern \gets$ collapse consecutive duplicate reuse methods 
        \Comment{Create lineage pattern} 

        \State $PatternCounter[Pattern] \gets PatternCounter[Pattern] + 1$ 
        \State $PatternModels[Pattern] \gets PatternModels[Pattern] \cup \{m \mid m \in L\}$ \Comment{Collect unique models}

    \EndFor 
\EndFor 

\Statex
\Statex \Comment{=== POST-PROCESSING ===} 

\ForAll{$Pattern \in PatternModels$} 
    \State Compute prevalence from $PatternCounter[Pattern]$ and model participation from $|PatternModels[Pattern]|$
\EndFor 

\State \Return lineage patterns and prevalence measures 

\end{algorithmic}
\end{algorithm}

\item \textbf{Step 7.1.6 Compressing repeated operations into lineage patterns:}
Raw lineages often contain consecutive repetitions of the same reuse method, which inflate path lengths without providing additional semantic information about meaningful transitions between distinct transformation types. For example, a practitioner may upload a model as \textit{Finetune}, then later fine-tune the same model again, producing the lineage \textit{Finetune} $\rightarrow$ \textit{Finetune}.
To focus on meaningful transitions between reuse methods, we compress each lineage by merging consecutive duplicate reuse methods into a single step. This compression is performed by the function \textsc{RemoveConsecutiveDuplicates} in \Cref{lineage_enumeration}. For instance, the lineage \textit{Finetune} $\rightarrow$ \textit{Finetune} $\rightarrow$ \textit{Quantization} becomes the lineage pattern \textit{Finetune} $\rightarrow$ \textit{Quantization}. This ensures that lineage patterns reflect true changes in model transformation strategy rather than repeated identical operations. We then compute the prevalence of each compressed lineage pattern and the number of unique models participating in lineages exhibiting that pattern.

\item \textbf{Step 7.1.7 Comparative analysis of lineage patterns across datasets:}
For each unique lineage pattern, we compute its prevalence (number of occurrences across all root-to-leaf lineages) and the number of unique models associated with that pattern. We report both absolute counts and normalized percentages to distinguish widely recurring patterns from those affecting diverse model populations. We apply the same pipeline to both the incomplete dataset (using only explicitly declared reuse methods) and the complete dataset (augmented with SemFin predictions). We then compare the resulting patterns to identify patterns shared across datasets, patterns introduced by imputation, and patterns that appear to disappear after imputation.
\end{itemize}

\head{Result}\\

\noindent\textbf{The complete dataset reveals 131,788 lineage pattern occurrences across 141,654 unique models, exposing a substantially richer set of 155 unique lineage patterns than what is observable from practitioner-declared metadata alone.} This delta between the total number of unique models and valid lineage occurrences is primarily due to the exclusion of trivial lineages (paths with $\le 1$ reuse methods), isolated models, and root models serving strictly as network initializations. By applying SemFin to impute missing fields, the observable data expands substantially from 31,795 occurrences and 35,092 unique models in the incomplete dataset, which contains only 76 unique lineage patterns. This structural recovery uncovers 86 unique lineage patterns that are present in the complete dataset but entirely absent from the incomplete metadata. At the same time, 7 lineage patterns identified in the incomplete dataset do not appear in the complete dataset. Rather than indicating metadata revisions, these patterns disappeared because filling the missing context resolved structural fragments, extending shorter sequences into more accurate, multi-step lineage trajectories. Additionally, 69 lineage patterns are common to both datasets.

These findings suggest that practitioner-declared metadata alone significantly underrepresents the true diversity of model reuse behaviors. This highlights the importance of automated metadata imputation for accurately characterizing how models are reused in practice.

\noindent\textbf{An analysis of the most prevalent reuse method lineage patterns across both datasets reveals that \textit{Finetune} becomes an even more dominant pattern, while \textit{Quantization} remains the second most prevalent despite a substantial decline in relative share after metadata imputation (\Cref{lineage_patterns}).} In the incomplete dataset, \textit{Finetune} accounts for 39.12\% of occurrences across 13,668 unique models, increasing to 56.62\% across 79,062 unique models in the complete dataset. This indicates that fine-tuning activities are heavily under-reported by practitioners during model upload. Conversely, although \textit{Quantization} decreases in relative share from 28.91\% to 14.55\%, its absolute number of unique models increases from 9,303 to 19,769.

Crucially, the imputation of missing metadata fundamentally reshapes our understanding of the ecosystem's structural complexity by demonstrating that post-imputation lineage patterns are significantly simpler, shifting heavily toward single-step trajectories. While the standalone \textit{Merge} pattern accounts for 16.76\% of occurrences in the incomplete dataset (6,150 models), its relative share drops sharply to 4.38\% (6,443 models) in the complete dataset. This drop occurs because a vast array of previously unlabelled single-step reuse methods (such as \textit{Finetune} or \textit{Peft}) were successfully attributed, causing the relative prominence of merge operations to decline. 

Consequently, single-step patterns entirely dominate the top positions of the complete dataset, expanding from just the top 3 ranks in the incomplete dataset to occupying the top 6 positions. Specifically, single-step sequences such as \textit{Peft} (6.87\% across 9,467 models), \textit{Distillation} (3.61\% across 5,023 models), and \textit{Pruning} (2.53\% across 3,493 models) become highly prominent after imputation. They actively displace the multi-step sequences that artificially appeared more dominant due to data sparsity, such as \textit{Finetune $\rightarrow$ Merge} (declining from 2.66\% to 1.13\%) and \textit{Finetune $\rightarrow$ Quantization} (declining from 3.01\% to 2.17\%).

Even when multi-step trajectories emerge in the complete dataset, they remain structurally simple, rarely expanding beyond two distinct transformation steps. For instance, alternative multi-step pathways such as \textit{Quantization $\rightarrow$ Finetune} account for 1.33\% across 2,016 unique models, while other specialized multi-step trajectories including \textit{Distillation $\rightarrow$ Finetune} (0.47\%), \textit{Pruning $\rightarrow$ Finetune} (1.18\%), \textit{Peft $\rightarrow$ Quantization} (0.91\%), and \textit{Peft $\rightarrow$ Finetune} (0.69\%) maintain a minor but observable presence. Several complex multi-step sequences present only in the incomplete dataset (e.g., \textit{Finetune $\rightarrow$ Merge $\rightarrow$ Finetune $\rightarrow$ Merge}) entirely disappear after completion due to path compression or upstream structural corrections. 

Overall, these findings demonstrate that practitioners seldom vary their reuse strategies within a given model lineage. In the vast majority of cases, they apply a single-step pattern, with complex multi-step trajectories restricted to a small minority of lineages that rarely exceed two steps.

\begin{table}[t]
\centering
\caption{Top-15 lineage patterns across incomplete and complete datasets. Dashes (--) indicate that the pattern was not observed in the corresponding dataset.}
\label{lineage_patterns}
\begin{tabular}{l c c c c}
\hline
\textbf{Lineage Pattern} & \textbf{\begin{tabular}[c]{@{}c@{}}Incomplete\\ Prevalence (\%)\end{tabular}} & \textbf{\begin{tabular}[c]{@{}c@{}}\#Incomplete Unique\\ Models\end{tabular}} & \textbf{\begin{tabular}[c]{@{}c@{}}Complete\\ Prevalence (\%)\end{tabular}} & \textbf{\begin{tabular}[c]{@{}c@{}}\#Complete Unique\\ Models\end{tabular}} \\ \hline
Finetune & 39.12 & 13,668 & 56.62 & 79,062 \\
Quantization & 28.91 & 9,303 & 14.55 & 19,769 \\
Peft & 1.48 & 483 & 6.87 & 9,467 \\
Merge & 16.76 & 6,150 & 4.38 & 6,443 \\
Distillation & 1.66 & 547 & 3.61 & 5,023 \\
Pruning & 0.25 & 91 & 2.53 & 3,493 \\
Finetune $\rightarrow$ Quantization & 3.01 & 1,313 & 2.17 & 3,710 \\
Quantization $\rightarrow$ Finetune & 0.49 & 236 & 1.33 & 2,016 \\
Finetune $\rightarrow$ Merge & 2.66 & 1,251 & 1.13 & 2,177 \\
Finetune $\rightarrow$ Peft & -- & -- & 0.57 & 936 \\
Deduplication & 1.33 & 424 & 0.45 & 597 \\
Distillation $\rightarrow$ Finetune & -- & -- & 0.47 & 712 \\
Pruning $\rightarrow$ Finetune & -- & -- & 1.18 & 2,250 \\
Peft $\rightarrow$ Quantization & -- & -- & 0.91 & 1,469 \\
Peft $\rightarrow$ Finetune & -- & -- & 0.69 & 1,155 \\
Finetune $\rightarrow$ Merge $\rightarrow$ Finetune $\rightarrow$ Merge & 0.77 & 292 & -- & -- \\
Merge $\rightarrow$ Quantization & 0.84 & 460 & -- & -- \\
Merge $\rightarrow$ Finetune $\rightarrow$ Merge & 0.55 & 259 & -- & -- \\
Merge $\rightarrow$ Finetune & 0.42 & 271 & -- & -- \\
Quantization $\rightarrow$ Finetune $\rightarrow$ Merge & 0.15 & 116 & -- & -- \\ \hline
\end{tabular}
\end{table}

\noindent\textbf{Closer inspection of the seven lineage patterns that initially appeared to vanish reveals that missing metadata can substantially distort reuse method lineage patterns, as lineages cannot be accurately reconstructed without complete metadata.} To understand these disappearing patterns, we manually extracted the unique multi-step trajectories from the complete dataset and cross-referenced them to locate where the 7 missing patterns from the incomplete dataset were absorbed. By tracking these structural transitions at the pattern level, we evaluated whether the original sequences were reclassified, compressed, or structurally expanded during metadata completion. Relying on incomplete metadata distorts the true chain of modifications, leading to misleading representations of model development history.

This comparative tracking shows that all 7 apparently missing lineage patterns were structurally extended in the complete dataset due to previously hidden upstream context. Table~\ref{updated_lineage_patterns} explicitly maps these seven patterns, showing how the incomplete sequences omit foundational upstream reuse method labels. Once missing labels are imputed, the complete patterns reveal that steps such as \textit{Merge} or an earlier \textit{Finetune} actually initiated the lineage trajectories, resolving the structural distortion caused by missing data.

These structural transitions collectively demonstrate that incomplete metadata systematically obscures the true starting reuse method, depth, and evolutionary complexity of reuse method lineages. This underscores the critical importance of metadata completion for accurately characterizing how models are reused in practice.

\begin{table}[t]
\centering
\caption{Reuse method lineage patterns that appeared to vanish in the incomplete dataset but were structurally extended in the complete dataset after metadata imputation.}
\label{updated_lineage_patterns}
\vspace{2mm}
\small
\begin{tabularx}{\textwidth}{X c X}
\toprule
\textbf{Incomplete Pattern} & \textbf{Count} & \textbf{Found Examples} \\
\midrule
Finetune $\rightarrow$ Merge $\rightarrow$ Finetune $\rightarrow$ Merge $\rightarrow$ Finetune $\rightarrow$ Merge & 11 & Merge $\rightarrow$ Finetune $\rightarrow$ Merge $\rightarrow$ Finetune $\rightarrow$ Merge $\rightarrow$ Finetune $\rightarrow$ Merge \\
\addlinespace
Finetune $\rightarrow$ Merge $\rightarrow$ Finetune $\rightarrow$ Merge $\rightarrow$ Finetune $\rightarrow$ Quantization & 4 & Merge $\rightarrow$ Finetune $\rightarrow$ Merge $\rightarrow$ Finetune $\rightarrow$ Merge $\rightarrow$ Finetune $\rightarrow$ Merge $\rightarrow$ Quantization \\
\addlinespace
Finetune $\rightarrow$ Merge $\rightarrow$ Finetune $\rightarrow$ Merge $\rightarrow$ Distillation & 3 & Merge $\rightarrow$ Finetune $\rightarrow$ Merge $\rightarrow$ Finetune $\rightarrow$ Merge $\rightarrow$ Finetune $\rightarrow$ Merge $\rightarrow$ Distillation $\rightarrow$ Merge \\
\addlinespace
Finetune $\rightarrow$ Merge $\rightarrow$ Finetune $\rightarrow$ Merge $\rightarrow$ Finetune $\rightarrow$ Merge $\rightarrow$ Finetune $\rightarrow$ Quantization & 2 & Merge $\rightarrow$ Finetune $\rightarrow$ Merge $\rightarrow$ Finetune $\rightarrow$ Merge $\rightarrow$ Finetune $\rightarrow$ Merge $\rightarrow$ Finetune $\rightarrow$ Quantization \\
\addlinespace
Finetune $\rightarrow$ Merge $\rightarrow$ Finetune $\rightarrow$ Merge $\rightarrow$ Distillation $\rightarrow$ Merge & 1 & Merge $\rightarrow$ Finetune $\rightarrow$ Merge $\rightarrow$ Finetune $\rightarrow$ Merge $\rightarrow$ Finetune $\rightarrow$ Merge $\rightarrow$ Distillation $\rightarrow$ Merge \\
\addlinespace
Finetune $\rightarrow$ Merge $\rightarrow$ Finetune $\rightarrow$ Merge $\rightarrow$ Distillation $\rightarrow$ Finetune & 1 & Merge $\rightarrow$ Finetune $\rightarrow$ Merge $\rightarrow$ Finetune $\rightarrow$ Merge $\rightarrow$ Finetune $\rightarrow$ Merge $\rightarrow$ Distillation $\rightarrow$ Finetune \\
\addlinespace
Merge $\rightarrow$ Pruning & 1 & Merge $\rightarrow$ Finetune $\rightarrow$ Merge $\rightarrow$ Pruning \\
\bottomrule
\end{tabularx}
\end{table}

\subsection{License lineage compatibility analysis}
Building on the predictive performance of SemFin established in RQ$_2$, we apply SemFin to infer missing license metadata for models lacking explicit license declarations. This analysis examines how licenses evolve across lineages and whether observed licenses in a lineage remain legally compatible. Missing license metadata not only obscures portions of lineage structures but may also conceal incompatible licenses within a lineage, such as non-commercial followed by commercial licenses or violations of share-alike requirements. By comparing license lineage patterns derived from incomplete and complete metadata, we assess how metadata imputation affects the visibility of licensing conflicts and identify the most common incompatible lineage patterns that emerge once missing metadata is recovered.

\head{Approach}

To evaluate license compatibility across lineage chains under both incomplete and complete metadata conditions, we follow the procedure below. 
\begin{itemize}
\item \textbf{Step 7.2.1 Separating known and unknown license declarations:}
We partition the dataset based on the availability of license metadata. Models with explicitly declared licenses are retained as known instances, while those with missing or unknown license values are designated for license prediction. We identified 128,157 models with known license metadata, while 188,976 models did not include license information.

\item \textbf{Step 7.1.2 Vectorizing the semantic fingerprints of these 188,976 PTLMs:} To predict the license for PTLMs with unknown license metadata using the trained model, we vectorize their semantic fingerprints using the vectorization procedure defined in Step~5.1.6.

\item \textbf{Step 7.2.3 Predicting missing license metadata:}
For models lacking license declarations, we apply the pre-trained SemFin classifier from Step 5.1.8 to predict license metadata. The predicted license labels are then combined with explicitly declared licenses to produce a metadata-complete dataset. As with any prediction-based approach, this imputation may introduce classification errors; therefore, the implications of prediction uncertainty are discussed in the threats to validity section.

\item \textbf{Step 7.2.4 Filtering and resolving parent-child relationships for license analysis:}
We apply the same filtering and parent selection procedure described in Step 7.1.4 to ensure consistency. This produces two comparable datasets: (i) an incomplete dataset containing only explicitly declared licenses (64,669 parent-child pairs) and (ii) a complete dataset augmented with SemFin predictions (141,669 parent-child pairs).

\item \textbf{Step 7.2.5 Enumerating License Lineages and compressing repeated operations into lineage patterns} We adapt the graph-based traversal and compression workflow established in Step 7.1.5 to construct license lineages. Each dataset variant is modeled as a directed graph where nodes correspond to models annotated with their respective licenses, and edges represent verified reuse pathways. To maintain structural consistency and prevent path duplication from multi-parent merge models, we enforce the same primary parent selection criterion described in Step 7.1.4.

We execute a root-to-leaf traversal using \Cref{lineage_enumeration} to enumerate all unique structural pathways. Following the same semantic compression logic utilized for reuse methods in step 7.1.6, consecutive duplicate license declarations are collapsed into a single step to isolate meaningful transitions (e.g., \textit{origin} $\rightarrow$ \textit{apache-2.0} $\rightarrow$ \textit{apache-2.0} $\rightarrow$ \textit{mit} is compressed to \textit{origin} $\rightarrow$ \textit{apache-2.0} $\rightarrow$ \textit{mit}). The output of this phase yields the core license lineage patterns that we use for further analysis.

\item \textbf{Step 7.2.6 Computing pattern prevalence and comparative analysis across datasets:}
For each unique license lineage pattern, we compute (i) its frequency across all root-to-leaf paths and (ii) the number of unique models associated with it. We report both absolute counts and normalized percentages to characterize the prevalence of different license sequences within the ecosystem. We apply the same pipeline to both the incomplete dataset (using only explicitly declared licenses) and the complete dataset (augmented with SemFin license predictions). We then compare the prevalence of license families and multi-step license sequences between the two datasets to assess how metadata imputation affects the visibility of license evolution patterns.

\item \textbf{Step 7.2.7 License compatibility classification:}
We define a taxonomy of license categories grounded in established open-source licensing principles and recent AI supply chain research \citep{pepe2024hugging, wang2026hidden}.

Standard software licenses are grouped into permissive, strong copyleft, and public-domain categories following Open Source Initiative definitions\footnote{opensource.org/docs/osd}. Creative Commons licenses are further separated into non-commercial, no-derivatives, attribution-only, and share-alike categories based on their legal restrictions \citep{cc_by40}. To capture AI-specific legal constraints, we additionally distinguish AI-restricted licenses (e.g., Llama and Gemma licenses) and Responsible AI Licenses (RAIL). Unlike traditional software licenses that primarily govern copyright and redistribution, RAIL licenses incorporate behavioral-use restrictions designed to promote responsible use of AI artifacts and require downstream users to preserve these restrictions when redistributing derivative works \citep{wang2026hidden}.

The resulting taxonomy consists of the following categories:

\begin{itemize}
\item \textbf{Permissive licenses} (\texttt{apache-2.0}, \texttt{mit}, \texttt{bsd-3-clause}, \texttt{bsd-2-clause}, \texttt{bsd-3-clause-clear}, \texttt{afl-3.0}, \texttt{artistic-2.0}, \texttt{isc}, \texttt{ms-pl}, \texttt{ecl-2.0}, \texttt{bsl-1.0}, \texttt{cdla-permissive-2.0}): Allow redistribution and modification with minimal conditions, without requiring derivative works to be distributed under the same license terms \citep{vendome2017license, geant_glossary}.

\item \textbf{Strong copyleft licenses} (\texttt{gpl-3.0}, \texttt{gpl-2.0}, \texttt{agpl-3.0}): Also termed "restrictive" or ``reciprocal" licenses, these require derivative works to be released under the same license terms, including source code disclosure \citep{lerner2005scope, vendome2017license}.

\item \textbf{Other copyleft licenses} (\texttt{lgpl-3.0}, \texttt{osl-3.0}, \texttt{eupl-1.1}): Encompass both weak copyleft (e.g., \texttt{LGPL}, \texttt{EUPL}) and strong copyleft (e.g., \texttt{OSL}) requirements outside the main GPL family \citep{fossa2023copyleft}.

\item \textbf{Public domain equivalents} (\texttt{cc0-1.0}, \texttt{unlicense}, \texttt{wtfpl}, \texttt{pddl}): Works dedicated to the public domain by waiving all copyright and related rights, permitting unrestricted use. In jurisdictions not recognizing waivers, these licenses include fallback permissive terms \citep{geant_glossary}.

\item \textbf{Non-commercial licenses} (\texttt{cc-by-nc-*} variants): Prohibit commercial utilization of the licensed asset \citep{geant_glossary}.

\item \textbf{No-derivatives licenses} (\texttt{cc-by-nd-4.0}, \texttt{cc-by-nc-nd-4.0}, \texttt{cc-by-nc-nd-3.0}): Permit distribution and commercial use with attribution, but forbid distribution of modified versions \citep{geant_glossary}.

\item \textbf{Attribution-only licenses} (\texttt{cc-by-*} variants excluding SA/NC/ND, \texttt{odc-by}): Licenses allowing reuse and modification, including commercial use, with mandatory attribution and no restrictions on derivatives \citep{geant_glossary}.

\item \textbf{Share-alike licenses} (\texttt{cc-by-sa-*}, \texttt{cc-by-nc-sa-4.0}, \texttt{cc-by-nc-sa-3.0}, \texttt{cc-by-nc-sa-2.0}): Require derivatives to be distributed under the same exact license terms \citep{geant_glossary}.

\item \textbf{AI-restricted licenses} (\texttt{llama2}, \texttt{llama3}, \texttt{llama3.1}, \texttt{llama3.2}, \texttt{llama3.3}, \texttt{gemma}): Impose model-specific restrictions on AI use, redistribution, and commercial deployment that extend beyond traditional open-source licensing terms \citep{mcduff2024standardization, wang2026hidden}.

\item \textbf{RAIL licenses} (\texttt{bigscience-bloom-rail-1.0}, \texttt{bigcode-openrail-m}, \texttt{creativeml-openrail-m}, \texttt{bigscience-openrail-m}, \texttt{openrail}, \texttt{openrail++}): Responsible AI Licenses that permit access and modification of AI artifacts while imposing behavioral-use restrictions intended to prevent harmful applications. These restrictions are generally propagated to downstream derivatives \citep{mcduff2024standardization, wang2026hidden}.

\item \textbf{Other and Ambiguous Identifiers} (\texttt{cdla-sharing-1.0}, \texttt{gpl}, \texttt{bsd}, \texttt{cc}, \texttt{other}): Broad or uninformative metadata tags designated as ambiguous and excluded from strict validation checks \citep{vendome2017license, pepe2024hugging}.
\end{itemize}

These classifications are operational research definitions intended for empirical analysis. They do not constitute legal advice or definitive legal determinations. Actual license compatibility may depend on jurisdiction, specific version interpretations, and legal counsel.

A key challenge during classification is that several repositories declare base license names without specifying a version number, rendering them legally ambiguous. For instance, generic declarations such as \texttt{gpl}, \texttt{bsd}, or \texttt{cc} fail to indicate which specific iteration or version restrictions apply to the model artifact. We categorize these versionless instances under the \texttt{Other and Ambiguous Identifiers} operational grouping, explicitly isolating them from adjacent license pair checks to avoid generating false alignment errors while preserving the structural integrity of the lineage paths.

For each license lineage pattern, we evaluate every adjacent pair of valid licenses along the sequence against this taxonomy to check for the following incompatibility conditions. For patterns involving permissive-to-copyleft transitions (pattern 5) and cross-copyleft family transitions (pattern 6), we draw on empirical evidence from traditional software ecosystems \citep{xu2025small}; however, the legal application of these software licensing patterns to AI models remains unsettled. We therefore identify them as potential risk indicators rather than definitive legal violations.

\begin{itemize}
\item A non-commercial license followed by a license that permits commercial deployment \citep{stalnaker2025ml}.
\item A no-derivatives license followed by any different license identifier within the lineage, which structurally implies an unpermitted modification \citep{stalnaker2025ml}.
\item A share-alike license followed by any different license type \citep{stalnaker2025ml}.
\item A strong copyleft license followed by a permissive license, which inherently violates copyleft legal obligations \citep{stalnaker2025ml}.
\item A permissive license followed by a strong copyleft license, which is compatible but introduces severe operational restrictions on subsequent lineages \citep{xu2025small}.
\item A strong copyleft license followed by a license from a different copyleft family (e.g., GPL followed by EUPL or OSL), introducing unresolved legal cross-compatibility conflicts \citep{xu2025small}.
\item An AI-restricted license followed by a non-AI license, creating severe legal uncertainty regarding downstream permissions \citep{wang2026hidden}.
\item A RAIL license followed by a non-RAIL license, where required downstream behavioral restrictions are omitted \citep{wang2026hidden}.
\end{itemize}

A license lineage pattern is classified as incompatible if it contains at least one incompatible adjacent license pair sequence. For each incompatible pattern, we record the specific incompatibility type and its exact position within the lineage.

\item \textbf{Step 7.2.8 Comparative compatibility analysis:}
Before comparing compatibility across datasets, we exclude any license lineage pattern containing the ambiguous identifier \texttt{other and ambiguous identifiers} from the analysis, as these patterns cannot be reliably evaluated for compatibility. We then apply the same compatibility assessment procedure to both the incomplete and complete datasets. For each incompatible pattern, we extract the specific compatibility reason (e.g., ``Non-commercial $\rightarrow$ Commercial'') from the \texttt{compatibility\_reason} field. Because a single lineage pattern may contain multiple incompatible license pairs (e.g., both a non-commercial to commercial violation and a share-alike violation), we aggregate individual reasons across all incompatible patterns rather than counting patterns. 

We adopt this approach rather than counting pattern instances for two reasons. First, counting by reason entries preserves the granularity of our analysis: a single lineage pattern may contain multiple distinct incompatibilities (e.g., both a non-commercial to commercial transition and a share-alike violation). Pattern-instance counting would treat such a multi-violation lineage as a single observation, discarding information about co-occurring violation types. Second, reason-entry counting better reflects the compounding compliance risk of multi-violation lineages to downstream users, as each distinct violation represents an additional legal concern. Counting by pattern instances would instead assign equal weight to lineages with a single violation and those with multiple violations, systematically underrepresenting the prevalence of co-occurring incompatibilities. For each incompatibility type, we compute its relative prevalence as:

\[
\text{Prevalence}(\text{reason}) = \frac{\text{Count of reason entries across all patterns}}{\text{Total number of reason entries across all patterns}} \times 100
\]

We then compare the distribution of incompatibility types between the incomplete and complete datasets to assess how metadata imputation affects the visibility of different licensing conflicts. This comparison quantifies the impact of metadata completeness on the visibility of legal risks within model lineages.
\end{itemize}

\head{Result}\\

\noindent\textbf{Metadata imputation significantly expands the observable license lineage ecosystem, revealing many additional license lineage chains and license lineage patterns that are not visible in incomplete metadata.} In the incomplete dataset, we identify 60,965 total license lineage chains spanning 66,825 unique models and comprising 250 unique license lineage patterns. After metadata imputation using SemFin, the complete dataset expands considerably to 131,356 total license lineage chains across 141,209 unique models, exposing 419 unique license lineage patterns. This expansion demonstrates that incomplete metadata substantially under-reports the diversity and structural complexity of license evolution across model lineages.

\noindent\textbf{Missing license metadata disproportionately obscures non-standard, Creative Commons, and AI-specific licenses, creating an incomplete view of license evolution across model lineages.} Analysis of the top-15 license lineage patterns (\Cref{license_lineage_patterns}) reveals that permissive licenses dominate the ecosystem, while metadata imputation significantly reshapes the relative prevalence of several license families and multi-step license sequences. In the incomplete dataset, \texttt{apache-2.0} is the dominant license lineage pattern, accounting for 59.02\% of all lineage chains across 38,216 unique models, followed by \texttt{mit} at 21.53\% across 13,982 models. However, after metadata imputation, the relative prevalence of \texttt{apache-2.0} decreases substantially to 34.58\%, despite increasing in absolute coverage to 48,494 unique models, while \texttt{mit} rises to 25.55\% across 35,192 models. Concurrently, license families such as \texttt{other}, \texttt{cc-by-4.0}, \texttt{cc-by-nc-4.0}, \texttt{cc-by-sa-4.0}, and \texttt{cc-by-nc-sa-4.0} become substantially more prominent in the complete dataset. This indicates that missing metadata disproportionately obscures a more diverse spectrum of custom, versionless, or less common licenses rather than standard corporate-permissive selections, showing that practitioners frequently introduce non-standard terms along a lineage that diverge from widely known benchmarks.

\noindent\textbf{Multi-step license sequences become more visible after metadata recovery but remain predominantly limited to two-step transitions.} Recovered multi-step patterns include \texttt{apache-2.0 $\rightarrow$ mit}, \texttt{apache-2.0 $\rightarrow$ other}, \texttt{apache-2.0 $\rightarrow$ llama2}, and \texttt{cc-by-sa-4.0 $\rightarrow$ mit}. However, no license lineage patterns extending beyond two distinct license steps appear among the top-15 patterns. This simple structure strongly mirrors the reuse method lineage patterns evaluated earlier, where single-step patterns dominate post-imputation and multi-step trajectories are restricted to a small minority of lineages that rarely exceed two steps. The consistent absence of deeper evolutionary sequences across both licensing and reuse modifications indicates that practitioners seldom change strategies or alter terms within a model lineage; instead, these transitions are typically executed as single steps rather than iterative transformations.

\begin{table}[t]
\caption{Union of Top-15 License Lineage Patterns Across Incomplete and Complete Datasets. Dashes (---) indicate that the lineage pattern was not observed within the top-15 ranks of that specific dataset.}
\label{license_lineage_patterns}
\centering
\begin{tabular}{lcccc}
\hline
\textbf{License Lineage Pattern} & \textbf{\begin{tabular}[c]{@{}c@{}}Incomplete\\ Prevalence (\%)\end{tabular}} & \textbf{\begin{tabular}[c]{@{}c@{}}Incomplete Unique\\ Models\end{tabular}} & \textbf{\begin{tabular}[c]{@{}c@{}}Complete\\ Prevalence (\%)\end{tabular}} & \textbf{\begin{tabular}[c]{@{}c@{}}Complete Unique\\ Models\end{tabular}} \\
\hline
\texttt{apache-2.0} & 59.02 & 38,216 & 34.58 & 48,494 \\
\texttt{mit} & 21.53 & 13,982 & 25.55 & 35,192 \\
\texttt{other} & 3.66 & 2,516 & 10.28 & 14,432 \\
\texttt{cc-by-nc-4.0} & 1.85 & 1,246 & 3.71 & 5,205 \\
\texttt{cc-by-4.0} & 1.65 & 1,050 & 3.95 & 5,476 \\
\texttt{mit $\rightarrow$ apache-2.0} & 1.39 & 986 & 0.80 & 1,316 \\
\texttt{llama2} & 1.43 & 988 & 3.61 & 5,100 \\
\texttt{cc-by-sa-4.0} & 1.00 & 665 & 3.42 & 5,888 \\
\texttt{llama3} & 0.88 & 580 & 1.30 & 1,839 \\
\texttt{cc-by-nc-sa-4.0} & 0.60 & 390 & 2.29 & 3,164 \\
\texttt{afl-3.0} & 0.51 & 316 & --- & --- \\
\texttt{apache-2.0 $\rightarrow$ mit} & 0.40 & 407 & 1.08 & 1,814 \\
\texttt{cc-by-nc-4.0 $\rightarrow$ apache-2.0} & 0.31 & 371 & --- & --- \\
\texttt{llama3.1} & 0.32 & 236 & --- & --- \\
\texttt{apache-2.0 $\rightarrow$ other} & 0.27 & 273 & 0.79 & 1,495 \\
\texttt{apache-2.0 $\rightarrow$ llama2} & --- & --- & 0.49 & 893 \\
\texttt{cc-by-sa-4.0 $\rightarrow$ mit} & --- & --- & 0.40 & 518 \\
\texttt{apache-2.0 $\rightarrow$ cc-by-nc-4.0} & --- & --- & 0.31 & 576 \\
\hline
\end{tabular}
\end{table}

\noindent\textbf{License lineage patterns often contain incompatible license transitions, and metadata imputation reveals additional incompatibilities that remain hidden in incomplete metadata.} In the incomplete dataset, 87 out of 250 license lineage patterns (34.8\%) contain at least one incompatible license pair. After metadata imputation, this increases to 154 out of 419 lineage patterns (36.8\%) in the complete dataset, indicating that missing metadata hides a substantial number of licensing conflicts embedded within lineage chains. A detailed breakdown of incompatible license types across both datasets is provided in \Cref{license_incompatibility}.

\noindent\textbf{License lineage patterns often contain incompatible license transitions, yet metadata imputation reveals that the overall proportion of conflicts remains relatively stable despite the massive increase in the number of unique lineage patterns.} In the incomplete dataset, 87 out of 250 license lineage patterns (34.8\%) contain at least one incompatible license pair. After metadata imputation, this increases to 154 out of 419 lineage patterns (36.8\%) in the complete dataset. While missing metadata hides the absolute volume of licensing conflicts embedded within lineage chains, the marginal rise of only 2.0\% indicates that the baseline probability of encountering a licensing violation does not drastically worsen upon ecosystem completion. A detailed breakdown of incompatible license types across both datasets is provided in \Cref{license_incompatibility}.

\begin{table}[t]
\centering
\caption{Distribution of License Incompatibility Reasons Across Incomplete and Complete Datasets}
\label{license_incompatibility}
\begin{tabular}{lccc}
\toprule
\multirow{2}{*}{Compatibility Reason} & \multicolumn{2}{c}{Prevalence (\%)} & \multirow{2}{*}{Sample Pattern} \\
\cmidrule(lr){2-3}
& Incomplete & Complete & \\
\midrule
Non-commercial $\rightarrow$ Commercial & 44.7\% & 40.4\% & \texttt{cc-by-nc-4.0 $\rightarrow$ apache-2.0} \\
AI-Restricted $\rightarrow$ Non-AI License & 19.3\% & 19.3\% & \texttt{llama3 $\rightarrow$ apache-2.0} \\
ShareAlike $\rightarrow$ Different License & 14.0\% & 18.7\% & \texttt{cc-by-nc-sa-4.0 $\rightarrow$ apache-2.0} \\
Ambiguous License Identifier & 10.5\% & 8.8\% & \texttt{cc-by-nc-4.0 $\rightarrow$ apache-2.0 $\rightarrow$ cc $\rightarrow$ apache-2.0} \\
RAIL $\rightarrow$ Non-RAIL & 5.3\% & 5.3\% & \texttt{apache-2.0 $\rightarrow$ openrail $\rightarrow$ cc-by-nc-4.0} \\
No-Derivatives $\rightarrow$ Modified Derivative & 3.4\% & 3.5\% & \texttt{cc-by-nc-4.0 $\rightarrow$ cc-by-nc-nd-4.0 $\rightarrow$ cc-by-nc-sa-4.0} \\
Permissive $\rightarrow$ Strong Copyleft & 4.5\% & 2.9\% & \texttt{apache-2.0 $\rightarrow$ gpl-3.0} \\
Strong Copyleft $\rightarrow$ Permissive & --- & 1.2\% & \texttt{gpl-3.0 $\rightarrow$ mit} \\
\midrule
Total patterns analyzed & 250 & 419 & \\
Incompatible patterns & 87 (34.8\%) & 154 (36.8\%) & \\
\bottomrule
\end{tabular}
\footnotesize{Note: Percentages represent the proportion of total reason entries (not patterns), as a single pattern may contain multiple incompatibility reasons. Dashes (---) indicate the reason did not appear in that dataset.}
\end{table}

\noindent\textbf{The most common incompatibilities arise when non-commercial restrictions are replaced by permissive commercial licenses, while additional incompatibility categories emerge after metadata recovery.} The most common incompatibility involves transitions from non-commercial Creative Commons licenses to permissive commercial licenses, particularly \texttt{cc-by-nc-4.0 $\rightarrow$ apache-2.0}, where non-commercial restrictions are effectively removed. This category accounts for 44.7\% of incompatibility reasons in the incomplete dataset and remains the most prevalent (40.4\%) after imputation. Additional incompatibilities emerge from AI-restricted licenses (e.g., \texttt{llama2}, \texttt{llama3}, \texttt{gemma}) transitioning to non-AI licensing schemes, remaining stable at 19.3\% after imputation, introducing uncertainty regarding downstream reuse permissions. ShareAlike conflicts (e.g., \texttt{cc-by-nc-sa-4.0 $\rightarrow$ apache-2.0}) also become substantially more visible, rising from 14.0\% to 18.7\% after metadata recovery, indicating that model reuse frequently fails to preserve required reciprocal licensing conditions. RAIL to non-RAIL incompatibilities remain stable at 5.3\%, while a previously hidden incompatibility type, strong copyleft to permissive (e.g., \texttt{gpl-3.0 $\rightarrow$ mit}), emerges only after imputation (1.2\%), demonstrating that license evolution within model ecosystems often lacks consistent governance and traceability.

\begin{Summary}
{Summary of RQ$_3$ Findings}{
\begin{itemize}
\item Metadata imputation increases traceable reuse lineage pattern occurrences from 31,795 to 131,788 and expands the number of observable reuse lineage patterns from 76 to 155.
\item Finetune is substantially under-reported in Hugging Face metadata, becoming even more dominant as the primary reuse lineage pattern (56.62\%) after metadata reconstruction, which concurrently reveals that single-step patterns dominate the top positions of the ecosystem.
\item License metadata imputation expands observable license lineage chains from 60,965 to 131,356 and increases identifiable license lineage patterns from 250 to 419.
\item The proportion of incompatible license lineage patterns rises to 36.8\%, with Non-commercial $\rightarrow$ Commercial transitions representing the most common incompatibility (44.7\% in incomplete, 40.4\% in complete).
\end{itemize}}
\end{Summary}

\section{Discussion and Implication} \label{discussion}

\subsection{Discussion}
The rapid ascendancy of Hugging Face (HF) as the central hub for PTLMs mirrors the role historically played by GitHub in the open-source software ecosystem. However, unlike traditional software repositories where version control systems inherently preserve lineage and dependency relationships, the contemporary AI ecosystem exhibits a pronounced transparency debt: a systemic lack of verifiable metadata concerning model provenance \citep{ajibode2025towards}, licenses \citep{horwitz2025we}, and reuse method transparency \citep{adekunle2025synchronization}. Addressing this deficit is a critical prerequisite for establishing a trusted AI supply chain.

Against this backdrop of missing provenance and fragmented documentation, we introduce SemFin, a novel artifact-driven approach that leverages configuration files and repository-level tags to reconstruct model identity and lineage. By analyzing 317,133 PTLMs, we uncover systematic reuse signals, recover missing metadata with high precision, and reveal a reuse ecosystem that is substantially larger, approximately five times larger and markedly deeper than what is suggested by explicit user-declared metadata alone. These results indicate that configuration files function as a model's ``structural signature'', where a core of 68 invariant configuration keys establishes the structural metadata required to preserve architectural compatibility across the ecosystem.

Yet, this reconstruction effort also exposes a deeper structural tension within the PTLM ecosystem. There exists a disconnect between a model's internal structural signals encoded in configuration files and its extrinsic repository context. Our analysis in $RQ_{2}$ demonstrates that configuration files alone perform strongly in capturing architectural constraints such as model type, but they struggle to identify pipeline tags without incorporating repository-level tags. Conversely, repository tags effectively capture user-declared intent but remain decoupled from the underlying neural implementation. This dichotomy echoes long-standing insights from software documentation research: source code explains how a system operates, whereas documentation explains why it exists \citep{lethbridge2003software}.

This structural tension motivates the core design principle of SemFin: using the presence or absence of configuration keys and repository tags as joint predictive signals. By explicitly combining configuration artifacts with repository tags, SemFin bridges the divide between implementation and intent. The result is not merely incremental performance gains but a qualitative shift in recoverability. SemFin achieves near-perfect reconstruction of model type and strong inference for reuse method while maintaining 100\% dataset coverage. In contrast, Graph Avg and Hub Avg heuristics remain inherently constrained by metadata sparsity and graph connectivity. Crucially, this universal applicability enables SemFin to recover metadata even for isolated or weakly connected models that traditional graph-based approaches cannot reach.

The governance implications of this enhanced visibility become particularly salient when examining the hidden lineages of reuse uncovered in RQ$_3$. Our reconstructed reuse method lineage patterns reveal that \textit{Finetune} becomes even more dominant (rising from 39.12\% to 56.62\% of occurrences) and that 86 previously invisible reuse method lineage patterns emerge only after imputation. While single-step patterns dominate the top positions of the ecosystem, newly uncovered multi-step trajectories remain structurally simple, including sequences such as \textit{Distillation $\rightarrow$ Finetune} and alternative multi-step pathways like \textit{Quantization $\rightarrow$ Finetune}. Furthermore, license lineage pattern imputation reveals that the overall proportion of incompatible license patterns remains relatively stable, shifting only slightly from 34.8\% to 36.8\% after metadata recovery, with common violations consistently involving Non-commercial $\rightarrow$ Commercial and AI-Restricted $\rightarrow$ Non-AI License transitions. Without systematic metadata imputation, these license violations, compliance gaps, or security vulnerabilities introduced early in a lineage may silently propagate across dozens of descendant models. By surfacing these hidden dependencies, SemFin provides one practical approach toward improving supply-chain transparency for AI models.

\subsection{From Manual Metadata to Automated AI Bill of Materials (AIBOM)}
In contemporary software engineering, the Software Bill of Materials (SBOM) is the standard for supply chain security, automated via Software Composition Analysis (SCA) tools \citep{xia2023empirical}. While recent work has attempted to map the ecosystem using graph-based propagation \citep{horwitz2025we}, our findings demonstrate that the AI ecosystem still lacks a verifiable, artifact-driven infrastructure to support an equivalent AIBOM. SemFin provides the technical foundation to close this gap by deriving provenance directly from the model itself rather than relying on external graph connectivity.

We adopt the definition of an AIBOM as a verifiable record of model provenance, licensing, training data, and intended usage \citep{rajbahadur2025building}. Currently, the PTLM ecosystem relies on manual, user-provided metadata, which our results show leads to a ``lineage mirage'' where 86 out of 155 reuse method lineage patterns (over half) remain invisible, and incompatible license lineage patterns increase from 34.8\% to 36.8\% after metadata imputation. This opacity makes compliance with emerging regulatory frameworks nearly impossible, as stakeholders cannot reliably audit the history of the models they deploy.

SemFin demonstrates that the raw materials for some of the AIBOM profiles are already present in the model artifacts, specifically, the configuration files and repository tags. Analogous to how SCA tools scan software manifests, our results show that scanning the model configuration file at ingestion time can automatically populate some of the AIBOM fields, such as type of models, hyperparameters, and metrics. By transforming metadata from a voluntary user declaration into an automated, artifact-derived property, SemFin proves that it is possible to generate AIBOMs even for models that lack explicit documentation.

Beyond linear dependencies, RQ$_3$ reveals a dominant pattern of cyclic change, characterized by recurrent, bidirectional flows between Fine-tuning and Merge. Rather than representing a one-directional refinement process, model reuse often oscillates between divergence (specialization via Fine-tuning) and convergence (integration via Merge). 

Our findings suggest that the PTLM community effectively employs Merge to integrate these specialized capabilities. However, current model-sharing practices rarely preserve a structured, machine-readable history of these integrations. As a result, much of this collaborative change remains opaque. The observation that 86 reuse method lineage patterns were entirely invisible in declared metadata indicates that the AI community is already engaging in sophisticated reuse practices, but without adequate provenance tooling to support transparency and governance. Consequently, future AIBOM standards must explicitly consider non-linear, cyclic provenance graphs that capture the iterative merging and recombination of models.

\subsection{Cyclic Change and the Need for Versioning}
Beyond linear dependencies, RQ\textsubscript{3} reveals a dominant pattern of \emph{cyclic} change, characterized by recurrent, bidirectional transitions between \emph{Finetune} and \emph{Merge}, as well as reversed pathways such as \emph{Quantization $\rightarrow$ Finetune} (1.33\% of occurrences). Rather than representing a one-directional refinement process, model reuse often oscillates between divergence (specialization via Finetune) and convergence (integration via Merge). This pattern closely resembles \emph{refactoring cycles} in collaborative software development, where developers iteratively branch to explore alternatives and subsequently reconcile changes through merging.

Our findings suggest that the PTLM community is effectively employing \textit{Merge} as a form of distributed ``community refactoring." However, in contrast to version control systems such as Git, current model-sharing practices rarely preserve a structured, machine-readable history of these merges. As a result, much of this collaborative change remains opaque. The observation that 86 out of 155 reuse method lineage patterns were entirely invisible in declared metadata indicates that the AI community is already engaging in sophisticated engineering practices, but without adequate provenance tooling to support transparency, reproducibility, and governance.

Taken together, these results suggest that, as model development practices become more complex, future AIBOM standards should explicitly consider non-linear, cyclic provenance graphs that can capture iterative merging, branching, and recombination of models.

\subsection{Metadata Imputation as an Enabler of Model Composition Analysis and Automated Versioning}
Beyond metadata imputation, \textbf{SemFin} provides a technical foundation for addressing broader challenges related to model versioning and ecosystem synchronization identified in prior empirical work. Our earlier studies document two persistent failures: (1) insufficient synchronization between training repositories (e.g., GitHub) and distribution platforms (e.g., Hugging Face), leading to inconsistent releases \citep{adekunle2025synchronization}; and (2) the absence of meaningful semantic versioning practices, wherein arbitrary naming conventions fail to convey the nature or impact of model changes \citep{ajibode2025towards}. By enabling automated, artifact-driven lineage reconstruction, SemFin directly contributes toward mitigating both challenges.

Traditional software versioning schemes such as Semantic Versioning (SemVer) rely on clear distinctions between breaking changes, additive features, and patches. However, as argued in prior work, PTLMs evolve along multiple dimensions, including architecture, data, and training procedures, rendering one-dimensional version numbers insufficient \citep{ajibode2025towards}. SemFin supplies the provenance signals necessary to support multi-dimensional versioning. For instance, configuration differentials identified in RQ\textsubscript{1} allow systematic categorization of reuse semantics: the introduction of \texttt{quantization\_config.bits} indicates a compression event that may warrant a patch-level increment, whereas the merging of distinct lineages constitutes a fundamental change in capability and provenance, justifying a major version increment. In this way, versioning transitions from an informal, manual convention to a verifiable, artifact-derived process.

The ``lineage mirages'' identified in RQ\textsubscript{3}, where 86 out of 155 reuse method lineage patterns (over half) are absent from declared metadata and incompatible license lineage patterns increase from 34.8\% to 36.8\% after imputation, are a direct manifestation of the synchronization failures previously observed \citep{adekunle2025synchronization}. When models are fine-tuned in upstream repositories but uploaded to Hugging Face without explicit lineage links, provenance chains fracture, and license violations (e.g., Non-commercial $\rightarrow$ Commercial, AI-Restricted $\rightarrow$ Non-AI License) may silently propagate. SemFin operates as a \textit{post-hoc synchronization corrective}: by mining configuration-level identifiers such as \texttt{\_name\_or\_path} alongside architectural fingerprints, many broken lineage links can be retrospectively reconstructed. This shifts provenance tracking from a purely user-dependent responsibility toward one that can be partially automated at the platform level.

Ultimately, this work motivates a transition from ad hoc metadata annotation toward comprehensive \emph{Model Composition Analysis (MCA)}, the AI analogue of Software Composition Analysis. Just as SCA tools inspect \texttt{pom.xml} or \texttt{package.json} to generate an SBOM, our results demonstrate that analyzing \texttt{config.json} in tandem with repository tags enables the construction of a \emph{Model Bill of Materials (MBOM)}. Such an MBOM captures not only immediate parentage but also inferred reuse methods, licensing compatibility, and extended lineage structure. The identification of 68 configuration keys forming an \textit{invariant core} (\Cref{invariant_keys}) further suggests a path toward standardization: modifications to this core may signal new model versions, whereas changes to peripheral configuration keys reflect iterative refinements. In summary, metadata imputation via semantic fingerprinting is not an end in itself, but a critical enabling step toward building the provenance-aware infrastructure that the PTLM ecosystem currently lacks.

\subsection{Implications}
Our findings offer practical implications for stakeholders involved in the development, deployment, and governance of PTLMs, particularly researchers, platform maintainers, and practitioners.

\subsection*{Researchers}
\begin{itemize}
\item By combining configuration files, which encode technical specifications, with repository tags, which capture social context, researchers can analyze PTLMs as part of a socio-technical ecosystem. This enables automated generation of AIBOMs by treating the fusion of technical and social signals as the unit of analysis, allowing for the study of model change using structured metadata analogous to package manifests (e.g., package.json, pom.xml) in software engineering.
\item The deep, cyclic reuse sequences we observed (with Finetune rising from 39.12\% to 56.62\% of occurrences and 86 previously invisible reuse method lineage patterns emerging after imputation) demonstrate that model change is highly iterative. Future studies must explicitly account for lineage depth and complex reuse pathways when analyzing how performance, bias, or security vulnerabilities propagate through the ecosystem.
\item SemFin provides full coverage and outperforms graph-based heuristics, offering a strong foundation for automated metadata inference and further methodological extensions.
\item Although evaluated on PTLMs, this artifact-driven approach is transferable to other model families and platforms, motivating cross-domain validation toward universal model composition analysis.
\end{itemize}

\subsection*{Platform Maintainers (e.g., Hugging Face)}
\begin{itemize}
\item Reliance on manual metadata entry creates systematic gaps and lineage mirages; platforms should transition from user-declared annotations to automated AIBOM generation by parsing configuration files at ingestion time to eliminate systematic metadata gaps.
\item Given that 86 out of 155 reuse method lineage patterns (over half) are currently invisible, and incompatible license lineage patterns increase from 34.8\% to 36.8\% after imputation, lineage reconstruction and visualization should be platform-level features rather than user-declared annotations.
\item Analogous to Software Composition Analysis in software ecosystems, model repositories should support automated detection of license conflicts, vulnerabilities, and bias propagation across reuse chains.
\item The presence of 68 invariant configuration keys suggests opportunities for standardization through structured schemas that explicitly encode provenance and reuse information.
\end{itemize}

\subsection*{Practitioners (Model Developers and Downstream Users)}
\begin{itemize}
\item Existing platform metadata significantly underestimates reuse depth; lineage-aware tools can help distinguish genuinely novel models from heavily derived reuse.
\item Deep reuse chains increase the risk of silent license or safety violations, including Non-commercial $\rightarrow$ Commercial and AI-Restricted $\rightarrow$ Non-AI License transitions, underscoring the need for provenance-aware model selection prior to deployment.
\item Configuration files should be maintained as first-class documentation artifacts, with rigor comparable to model weights and inference code.
\item Organizations managing multiple models can use SemFin to maintain internal catalogs, track reuse, and ensure compliance across reuse chains.
\end{itemize}

\section{Threats to Validity}\label{ttv}

\subsection{Internal Validity}

A primary threat to internal validity arises from our use of predicted metadata in RQ$_3$. The LightGBM classifier achieves a micro-accuracy of 0.826 for reuse methods and 0.742 for licenses (on the test set), implying that a portion of our imputed labels may be misclassified. Such errors may propagate along lineage chains. For reuse methods, these occur stochastically and do not materially affect aggregate ecosystem-level trends. For license lineage compatibility, however, prediction uncertainty carries the risk of either masking true licensing conflicts or falsely triggering hallucinated incompatibilities across adjacent reuse pathways. We mitigate this by filtering out models that no longer exist in the dataset, and by focusing our structural analysis on the most prevalent license families (e.g., permissive, non-commercial, AI-restricted) and lineage patterns rather than isolated lineage chains.

Another threat stems from the construction of the ground truth used to train the machine learning models. We derive supervision signals using keyword-based heuristics and metadata extraction rather than exhaustive manual validation of all 150{,}044 samples. We acknowledge that our ground truth for reuse methods, licenses, and parent-child relations is derived from these heuristic rules (keyword matching, priority-based resolution) rather than large-scale manual validation. While we validated on a statistically sampled subset (384 models), systematic biases may remain. Consequently, models trained on this ground truth learn the ecosystem's declared norms and our extraction heuristics, not an absolute ground truth. Readers should interpret predictive performance as recovery of declared metadata patterns, not necessarily the 'true' lineage in all cases. Nevertheless, our validation on the curated subset of 384 models and the strong performance observed on held-out test data indicate that these heuristics provide a sufficiently consistent supervision signal for learning.

Furthermore, a potential threat to internal validity concerns data leakage during feature extraction. To mitigate this, we removed all tokens from the input features that exactly matched the target labels (e.g., removing the precise string ``text-generation" when predicting pipeline tags). However, we retained compound tokens that may contain target substrings (e.g., text-generation-inference). While this could be perceived as partial leakage, we argue that these tokens constitute valid technical signals rather than artifacts. In a real-world recovery scenario, the presence of a configuration key such as text-generation-inference is a legitimate, high-fidelity indicator of the model's intended purpose. Excluding such compound tokens would artificially handicap the classifier and fail to reflect the true predictive power of the ecosystem's tooling signatures. Indeed, feature importance analysis confirms that these tooling-specific compound tokens are highly influential drivers of the classifier's decisions, which directly contributes to the strong predictive performance of SemFin \citep{jewitt2025hugging}. We acknowledge that retaining semantically adjacent tokens provides the classifier with highly predictive target substrings. Consequently, while our reported accuracies serve as a conservative lower bound compared to an in-the-wild deployment where explicit community tags are completely intact, they represent an upper bound relative to a strict zero-leakage setting where all adjacent compound indicators are entirely stripped.

A further threat concerns our comparison against the Graph Avg and Hub Avg baselines proposed by \citet{horwitz2025we}. As no replication package was publicly available, we reimplemented these methods based on the descriptions in the original paper. While we closely followed their definitions of neighbor traversal and voting logic, minor implementation differences may influence the exact performance values reported. Notably, the baselines achieve higher performance in our study than reported in the original paper. We attribute this to our evaluation on a densely connected subset of the ecosystem with strictly filtered ground-truth metadata, which provides the rich neighbor signals necessary for graph-based propagation to excel. This denser graph connectivity favors graph-based baselines, making our comparison \textit{conservative}: SemFin's relative advantages in coverage and accuracy would likely be larger in the sparser, noisier raw ecosystem.

Lastly, we treat observed user-provided metadata (e.g., repository tags) as the ground truth for training. We acknowledge that this metadata reflects community naming conventions rather than a strictly verified semantic truth. Practitioners may apply tags inconsistently or use colloquial definitions, for instance labeling a distilled model generically as fine-tuned or omitting specific license details. Because SemFin is trained on this social signal, it learns to predict metadata consistent with how the ecosystem currently describes itself, rather than enforcing a rigid taxonomic prescription. While our high cross-validation scores suggest that the model captures these community norms, the ``correctness'' of its predictions is inherently bounded by the quality of the ecosystem's self-reported data.

\subsection{External Validity}
A key threat to external validity arises from our selection criteria. We exclude models lacking a \textit{pipeline\_tag} and those without a verifiable config.json, removing 50.8\% of available models. However, we posit that this exclusion acts as a necessary quality filter rather than a source of bias. A comparison of popularity metrics confirms that the included tagged models represent the active core of the ecosystem, exhibiting a mean download count of 2{,}168 compared to just 168 for the excluded untagged cohort. This 13$\times$ disparity suggests that the excluded group largely consists of inactive or ambiguous artifacts. Consequently, while our findings are biased toward models with accessible configurations, this bias deliberately targets the functional portion of the supply chain that practitioners actually utilize.

Our study further focuses exclusively on Hugging Face. Although reuse dynamics may differ on other platforms such as GitHub or ModelScope, Hugging Face currently serves as the dominant distribution hub for PTLMs. Moreover, the SemFin approach is inherently platform-agnostic because it relies on the config.json file, a standard artifact of the transformers library, rather than proprietary platform metadata. As long as a model retains its library-standard configuration, SemFin can reconstruct its lineage regardless of whether it is hosted on Hugging Face, ModelScope, or private infrastructure. As a result, we believe our findings reasonably reflect contemporary open-source AI development practices, even if platform-specific variations exist elsewhere.

Finally, our dataset represents a snapshot of the Hugging Face ecosystem as of January 2025. While the ecosystem evolves rapidly, the scale of our analysis, covering 317,133 models, supports the stability of the structural patterns we identify, including deep lineage chains and cyclic reuse. New architectures or reuse practices emerging after this date fall outside the scope of our study.

\subsection{Construct Validity}

A construct validity threat arises from our representation of reuse history using a single categorical \textit{Reuse Method}. In practice, models may undergo multiple concurrent reuse, such as merging and quantization. To enable classification, we apply a taxonomic priority hierarchy that favors architectural transformations over secondary modifications. While necessary, this choice may under-report stacked reuse techniques when they co-occur with higher-priority methods.

Another construct threat stems from how we infer lineage links. We rely primarily on explicit metadata such as \textit{\_name\_or\_path} and \textit{base\_model}. While this avoids the ambiguity of inferring relationships based on configuration similarity, it remains susceptible to practitioner error. For instance, if a developer copies a configuration file from a sibling model without updating the provenance fields, the metadata may incorrectly reflect a parent-child relationship where a sibling relationship exists. This approach captures explicit reuse but misses implicit reuse, including local initialization or undocumented copying. Consequently, our lineage reconstruction should be interpreted as a conservative lower bound rather than a complete reconstruction of all reuse pathways.

Finally, a limitation of our artifact-driven approach is its blindness to parameter-level changes. SemFin operates exclusively on configuration files and metadata; it does not hash or inspect the model weights themselves. Consequently, if a practitioner fine-tunes a model without modifying the configuration file (e.g., retaining the exact parent config.json) and without adding repository tags, our approach may fail to distinguish the child model from its parent. This theoretically risks skewing our ecosystem analysis by under-representing ``silent" reuse (e.g., pure weight updates) while over-representing methods that force configuration changes (e.g., quantization). However, our empirical findings in $RQ_{1}$ mitigate this concern: we observed that fine-tuning systematically alters configuration fingerprints, specifically through the removal of generation parameters and the addition of task-specific keys (e.g., problem\_type). Furthermore, because SemFin fuses configuration signals with repository tags, it remains capable of detecting weight-only updates when accompanied by standard community tagging. Thus, while our lineage reconstruction represents a conservative lower bound, the multi-modal nature of our fingerprinting minimizes the impact of this parameter-blindness.

\section{Conclusion}\label{conclusion}
This study examined the fingerprinting of PTLMs on Hugging Face as a response to the growing transparency debt in the AI model ecosystem. We investigated which configuration keys signal model reuse, how effectively our proposed SemFin approach recovers missing metadata, and the depth and structure of model reuse across the platform.

This pattern closely mirrors the principle of inheritance in Object-Oriented Programming (OOP), wherein child objects retain the structural blueprint of a parent while selectively overriding behaviors. Notably, the systematic key removal phenomenon we observe, where descendant models remove generic generation configuration keys, suggests that model reuse on HF follows a largely consistent refinement pattern. In this respect, PTLM reuse resembles the optimization of a general-purpose software artifact into a production-ready deployment tailored to a specific runtime context.

Our results show that PTLM configuration files function as a form of ``structural signature", preserving a stable core of 68 invariant structural keys while encoding reuse through systematic changes in optimization settings, learning parameters, and task definitions. Crucially, SemFin exploits this structural signature using a highly lightweight approach. By relying solely on readily accessible configuration files, repository tags, and efficient TF-IDF vectorization, rather than computationally expensive model weight analysis or incomplete external graph connectivity, SemFin substantially outperforms existing graph-based and hub-based heuristics, achieving near-perfect recovery of structural attributes and complete coverage across the ecosystem.

Most notably, reconstructing the full reuse ecosystem reveals that current metadata practices drastically underrepresent both the scale and complexity of model reuse. The reconstructed reuse ecosystem is substantially larger than what visible metadata suggests, with traceable reuse method pattern occurrences expanding from 31,795 to 131,788 and license lineage chains from 60,965 to 131,356. We uncover 86 previously invisible reuse method lineage patterns and find that the overall proportion of incompatible license lineage patterns remains relatively stable, shifting only slightly from 34.8\% to 36.8\% after imputation, with common violations including Non-commercial $\rightarrow$ Commercial and AI-Restricted $\rightarrow$ Non-AI License transitions. Our evaluation uses a heuristic-derived ground truth; results reflect the ability to recover ecosystem conventions, not an absolute provenance standard. Nonetheless, these hidden lineages introduce serious governance risks, as licensing violations or safety issues introduced early in a lineage can propagate silently across dozens of undocumented reuse steps.

Taken together, these findings underscore the need for automated model composition analysis and platform-level provenance infrastructure. Replacing manual metadata entry with lightweight, automated tooling like SemFin would enable accurate, verifiable lineage tracking and reduce the transparency debt that currently obscures the realities of collaborative model development.

\section*{Data Availability}
\label{sec:availability}
The datasets generated and analyzed during this study are available in the replication package~\citep{SemFin}.
\section*{Funding} 
This research was supported by the NSERC Discovery Grant RGPIN-2025-04654.
\section*{Ethical Approval} This study does not involve human participants or animals.
\section*{Informed Consent} No human subjects were involved in this study.
\section*{Conflicts of Interests/Competing Interests}
The authors declare that they have no known competing interests or personal relationships that could have (appeared to) influenced the work reported in this article.
\section*{Author Contributions}
\begin{itemize}
    \item Adekunle Ajibode: Conceptualization, Data Collection, Methodology, Data Analysis, Writing – Original Draft.
    \item Oussama Ben Sghaier: Methodology, Data Validation, Writing – Review \& Editing.
    \item Keheliya Gallaba: Research Direction - Review \& Editing.
    \item Bram Adams: Supervision, Writing – Review \& Editing, Conceptual Guidance, Research Direction.
    \item Ahmed E. Hassan: Supervision, Research Direction.
\end{itemize}

\bibliographystyle{plainnat}
\bibliography{main_draft}

\newpage
\titleformat{\section}
  {\normalfont\Large\bfseries}
  {Appendix \thesection:}{1em}{}

\appendix

\end{document}